\documentclass[useAMS,usenatbib,letterpaper]{mn2e}
\usepackage{journal_abbreviations, custom_commands}
\usepackage{graphicx, amsmath, amssymb}
\usepackage{bold-extra} % For bold \sc fonts
\usepackage{multirow} % For spanning columns in tables

\usepackage[totalwidth=500pt,totalheight=650pt,centering,layoutvoffset=-0.3cm]{geometry}

\def\ngaltotal{854}
\def\ngaluv{473}
\def\ngalneb{381}
\def\ngalabs{244}
\def\ngalem{50}
\def\ngalboth{179}
\def\ngalns{41}
\def\ngalmf{340}
\def\ngalnebpercent{45}
\def\medz{2.34}
\def\medzuv{2.35}
\def\medzneb{2.33}
\def\ngalinner{13}
\def\ngalinnertwo{24}
\def\ngalinnertwoneb{19}

\def\ngalbothnebanduv{238}
\title[Metal-line absorption around galaxies in the KBSS]{Metal-line absorption
 around $z\approx2.4$ star-forming galaxies in the Keck Baryonic 
Structure Survey\thanks{Based on data obtained
at the W.M. Keck Observatory, which is operated as 
a scientific partnership among the California Institute of 
Technology, the University of California, and NASA, and was made possible
by the generous financial support of the W.M. Keck Foundation.}}
\author[Turner et al.]{Monica L. Turner,$^{1}$\thanks{E-mail: turnerm@strw.leidenuniv.nl}
Joop Schaye$^{1}$,
Charles C. Steidel$^{2}$,\newauthor
Gwen C. Rudie$^{3}$,
and Allison L. Strom$^{2}$\\
$^{1}$Leiden Observatory, Leiden University, PO Box 9513, 2300 RA Leiden, The Netherlands\\
$^{2}$California Institute of Technology, MS 249-17, Pasadena,
\def\ngalbothnebanduv{230} CA 91125, USA\\
$^{3}$Carnegie Observatories, 813 Santa Barbara Street, Pasadena, CA 91101, USA\\}

\begin{document}

%\date{Accepted yyyy Month dd. Received yyyy Month dd; in original form yyyy Month dd}
%\date{2015 Feb 9}

\pagerange{\pageref{firstpage}--\pageref{lastpage}}% \pubyear{2013}

\maketitle

\label{firstpage}

\begin{abstract}
We study metal absorption around
\ngaltotal\ $z\approx2.4$ star-forming  
galaxies taken from the Keck Baryonic Structure Survey (KBSS). 
The galaxies examined in this work lie in the fields of 
15 hyper-luminous background QSOs, with galaxy impact parameters ranging from
35~proper kpc (pkpc) to 2~proper Mpc (pMpc). Using the pixel optical depth
technique, we present the first galaxy-centred 2-D maps of the median absorption by \osix,
\nfive, \cfour, \cthree, and \sifour, as well as updated results for
\hone. At small galactocentric radii 
we detect a strong enhancement of the absorption
relative to randomly located regions
that extend out to at least 180~pkpc in the transverse direction, 
and $\pm240$~\kmps\ along the line-of-sight (LOS, $\sim1$~pMpc in 
the case of pure Hubble flow) for all ions except \nfive. 
For \cfour\ (and \hone) we detect a significant enhancement of the
absorption signal out to 2~pMpc in the transverse direction,
corresponding to the maximum impact parameter in our sample. 
After normalising the median absorption profiles to account for
variations in line strengths and detection limits, in the transverse direction
we find no evidence for a sharp drop-off in metals distinct from
that of \hone.  We argue instead that non-detection of some metal line
species in the extended circumgalactic medium is consistent with differences in
the detection sensitivity. Along the LOS, the normalised profiles
reveal that the enhancement in the absorption is more 
extended for  \osix, \cfour, and \sifour\ than for \hone.
 We also present measurements of the scatter in the pixel optical depths, covering
fractions, and equivalent widths as a function of projected galaxy distance. 
Limiting the sample to the \ngalmf\ galaxies with redshifts 
measured from nebular emission lines 
does not decrease the extent of the enhancement along the LOS compared to that 
in the transverse direction. This rules out redshift errors as the source of the observed 
redshift-space anisotropy and thus implies that we have detected the signature of gas 
peculiar velocities from infall, outflows, or virial motions for \hone, \osix, \cfour, 
\cthree, and \sifour.
\end{abstract}


%%%%%%%%%
% INTRO %
%%%%%%%%%

\section{Introduction}
\label{sec:intro}

The exchange of baryons between galaxies and their surroundings
remains a poorly understood problem in galaxy formation theory.
Currently, hydrodynamical cosmological simulations 
suffer from large uncertainties in their implementations of sub-grid 
physics, particularly those related to feedback from
star formation and active galactic nuclei (AGN). Variations within
these recipes can create vast differences in the resulting
galaxy properties \citep[e.g.,][]{haas13a, haas13b}, which manifest themselves also
in the distribution of cosmic metals \citep[e.g.,][]{wiersma11}. 
Metals in different ionisation states hold clues to the structure, 
kinematics, temperature, and composition of the gas in which they 
reside, and they are therefore an important observational tool for comparison
with simulations. 

Observations have shown that the movements of these metals through the 
circumgalactic and intergalactic medium (CGM and IGM, respectively) 
are influenced by galactic-scale outflows,
which are commonly observed in star-forming 
galaxies at $z\sim2\text{--}3$
\citep[e.g.,][]{steidel96,  pettini00, quider09, steidel10}; but also
in nearby galaxies $(z<2)$ \citep[e.g.,][]{heckman90, heckman00, martin05, rupke05, tremonti07, weiner09}
as well as in those at higher redshifts \citep[e.g.,][]{franx97, steidel99, ajiki02, shapley03, jones12}. 
These outflows are often metal-rich, with velocities
of up to 800--1000~\kmps. 

\citet{aguirre01}, \citet{oppenheimer08}, and 
\citet{oppenheimer10} studied such winds in cosmological simulations,
and found that they are likely responsible for metal pollution in the IGM.
Furthermore, \citet{oppenheimer10} determined that these outflows are often bound to the galaxies
and fall back in, with most star forming gas at $z\sim1$ coming 
from such recycled material. 
Simulations by \citet{booth12} suggest that galaxies residing in low-mass 
($M_{\rm tot}<10^{10}$~\msol) haloes are
required to account for the observed \cfour\ absorption associated with relatively
weak \hone\ absorption \citep{schaye03}. 
However, \citet{wiersma10} found that only half 
of intergalactic metals originated from $M_{\rm tot}\lesssim10^{11}$~\msol\ haloes.
The authors also noted that in their simulations, half of the intergalactic metals at $z=2$ 
were ejected between $2<z<3$. This prediction 
is consistent with observations by \citet{simcoe11} that suggested
that 50\% of metals observed in the IGM at $z\sim2.4$ have been there since $z\sim4.3$ (1.3~Gyrs). 
In general, many questions still remain about the  masses of the galaxies responsible 
for metal pollution, as well as the epoch(s) at which the bulk of it may have occurred. 

To search for intergalactic metals, observations 
of absorption-line systems are often used.
\citet{bahcall69} first suggested that 
intervening absorbers could be associated with galaxy haloes;
indeed, it has has been shown that
\mgtwo\ \citep[e.g.,][]{bergeron91, zibetti05, nielsen13},
\cfour\ \citep[e.g.,][]{chen01}, and 
\osix\ \citep[e.g.,][]{stocke06, chen09, prochaska11,tumlinson11}
absorbers are found to occur near galaxies. 
\citet{adelberger03, adelberger05a} found evidence for metals out to
 300~proper~kpc using \cfour-galaxy cross-correlation studies at $z\sim3$, while
\citet{steidel10} used galaxy pairs to observe the equivalent widths (EWs) of
various ions near $\sim L_{*}$ galaxies at $z\sim$2--3, galaxies, and showed that these 
galaxies have metal-enriched gas out to $\sim125$~proper~kpc.

Theoretical studies of metals around simulated 
galaxies have also been undertaken. For example, \citet{ford13} examined the 
distribution of the absorption of various ions around 
$z=0.25$ galaxies with halo masses $10^{11}$--$10^{13}$~\msol,
and found that all ions examined showed enhanced absorption near
the galaxy centres, with lower ions arising in denser gas 
closer to the galaxies and higher ions found further away. 
A similar effect was seen by \citet{shen13}, who analysed
a zoom-in simulation of a $z\sim3$, $\sim10^{11}\msolm$ 
galaxy. They found that low ions were predominantly tracing
cool ($T<10^5$~K) gas at radii less than the galaxy
virial radius ${\rm R}_{vir}$, while higher ions trace hotter gas  
out to beyond 2$\times {\rm R}_{vir}$.

In this work, we study the distribution of metals around galaxies
using data from the Keck Baryonic Structure Survey
\citep[KBSS,][]{rudie12, steidel14}.
 This survey consists of a combination of
high-quality quasi-stellar object (QSO)  spectra, and a survey 
focused on redshifts below those of the QSOs.
We combine these two components in order to use a galaxy-centred 
approach to study metals in the CGM, by examining
metal optical depths and EWs in the QSO spectra as a function 
of galaxy impact parameter and line-of-sight (LOS) distances. The metal ions
examined here are, ordered by decreasing ionisation energy,
\osix, \nfive, \cfour, \cthree, and \sifour. 
Additionally, we revisit the results for \hone\ previously
examined in \citet{rakic12} using the updated KBSS galaxy sample.
Thanks to observations with MOSFIRE \citep{mclean12}, not only has the total sub-sample 
size grown from 679 to \ngaltotal\ galaxies, but both the fraction and total number
of galaxies with redshifts measured using nebular emission lines 
have increased substantially from 10\% (71 galaxies)
to \ngalnebpercent\% (\ngalneb\ galaxies).

The structure of the paper is as follows:
we first describe our galaxy and QSO samples in \S~\ref{sec:gal_sample},
while in \S~\ref{sec:od_recovery} we discuss the pixel optical depth recovery
for the various ions.  The results are presented in \S~\ref{sec:results},
where in particular we examine optical depth maps in \S~\ref{sec:maps};
cuts through these maps in \S~\ref{sec:profiles};
optical depths as a function of 3-dimensional (3-D)
 Hubble distance in \S~\ref{sec:hubble}; 
the optical depth distributions in \S~\ref{sec:odhist};
EWs in \S~\ref{sec:ew};
covering fractions in \S~\ref{sec:fcover}; and the effects of 
the galaxy redshift measurement errors in \S~\ref{sec:var_gal_sample}.
Finally, we present our conclusions in \S~\ref{sec:conclusion}.
Throughout the paper, we quote proper rather than comoving units
(denoted as pkpc and pMpc), and have used cosmological parameters
determined from the Planck mission \citep{planck13}, i.e.
$H_{\rm 0}=67.1$~\kmps~Mpc$^{-1}$, $\Omega_{\rm m} = 0.318$, and $\Omega_{\Lambda} = 0.683$.

\begin{figure*}
\includegraphics[width=0.8\textwidth, clip=true, trim = 0mm 0 0 0]{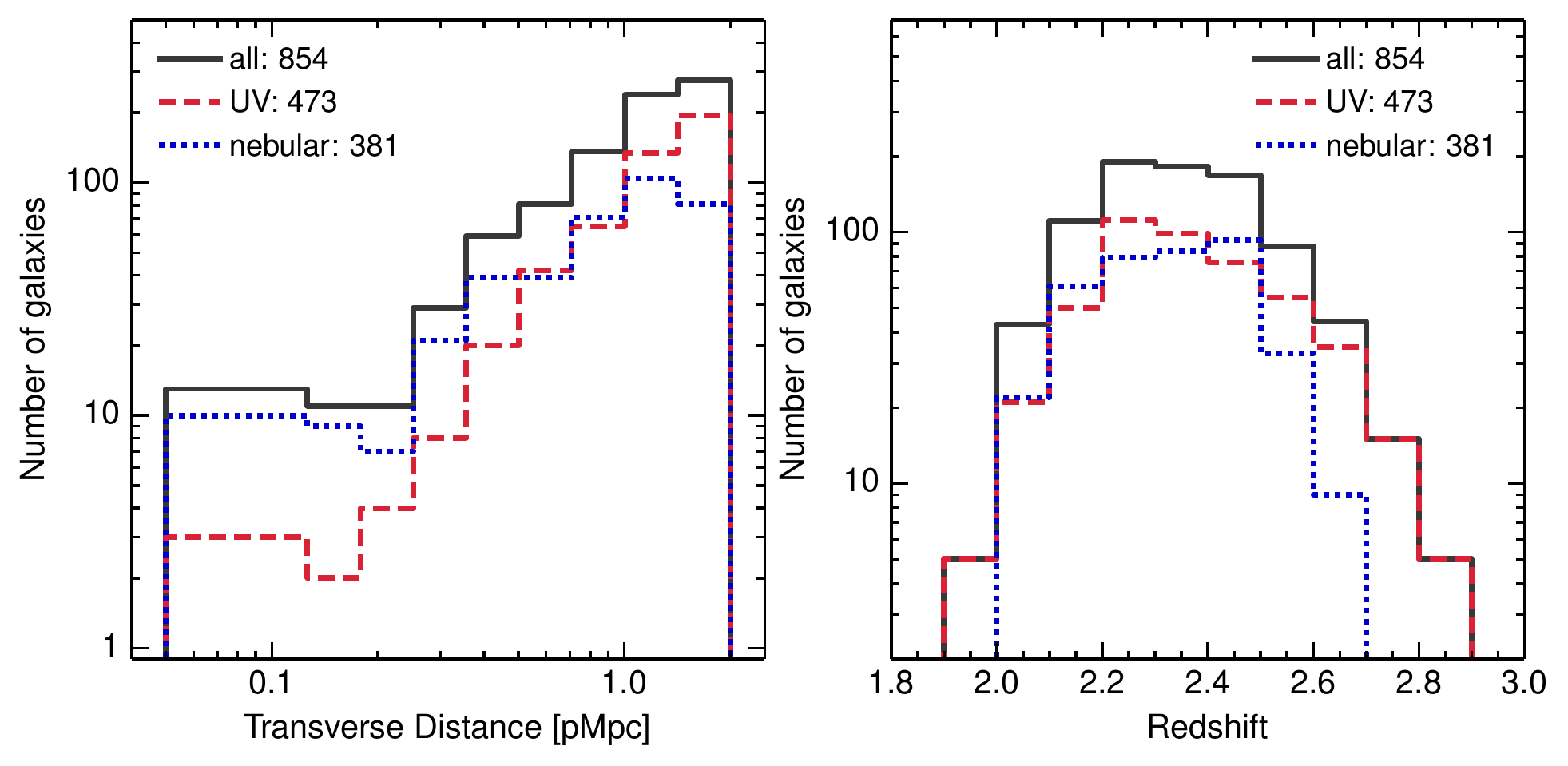}
\caption{Histograms of the sub-sample of KBSS galaxies used in this work, 
selected to have redshifts 
in the \lya\ forest and impact parameters $\leq2$~pMpc. The left panel
shows the distribution as a function of impact parameter, and the right
as a function of galaxy redshift. The values for the impact parameter
histogram are given in Table~\ref{tab:bins}.}
\label{fig:galaxy_sample_hist}
\end{figure*}

\begin{table}
\caption{Number of galaxies per impact parameter bin. Except for the innermost
bin (which has been extended in order to include the smallest impact parameter
galaxies), the bins are spaced logarithmically and are 0.15~dex in size.}
\label{tab:bins}
\begin{tabular}{@{}rrrrr}
\hline
\hline
Bin 	& Lower edge	& Upper edge 	& Total \#  	& \# with  \\
\# 	&  (pMpc)    	& (pMpc)	  	& of galaxies	& nebular $z$ \\
\hline
1 & 0.04 & 0.13 & 13 & 10 \\ 
2 & 0.13 & 0.18 & 11 & 9 \\ 
3 & 0.18 & 0.25 & 11 & 7 \\ 
4 & 0.25 & 0.36 & 29 & 21 \\ 
5 & 0.36 & 0.50 & 59 & 39 \\ 
6 & 0.50 & 0.71 & 81 & 39 \\ 
7 & 0.71 & 1.00 & 136 & 71 \\ 
8 & 1.00 & 1.42 & 238 & 104 \\ 
9 & 1.42 & 2.00 & 276 & 81 \\ 
\hline
\end{tabular}
\end{table}

\section{Galaxy Sample}
  \label{sec:gal_sample}

The sample of galaxies used in this work comes from a subset of the KBSS, 
which consists of 
$\sim2550$  $\langle z\rangle\sim2.3$ galaxies selected to lie in the fields 
of 15 hyper-luminous ($L_{\rm bol} \gtrsim 10^{14} L_{\sun}$) redshift 
$2.5$--$2.85$ QSOs which all have extremely high-quality, i.e.,
high resolution and signal-to-noise (S/N) Keck/HIRES spectra.
The galaxies in each QSO field were chosen primarily using ultraviolet (UV) colour selection techniques 
\citep{steidel03,steidel04, adelberger04} with the purpose of tuning the galaxy redshift selection
functions to optimise overlap with the range probed by the QSO spectra. 
Galaxies with apparent magnitude $m_R\leq25.5$ were then 
followed up spectroscopically using Keck/LRIS, NIRSPEC, and/or MOSFIRE, with
priority given to those likely to have redshifts in the QSO \lya\ forest 
and those near the QSO sightline. For more details on the observations,
see \S~2.1 of \citet{rudie12}.

Typically, the galaxies in the full survey sample
have dynamical masses of
$\sim7\times10^{10}$~\msol \citep{erb06c}, and 
reside in halos with masses $\sim10^{12}$~\msol
\citep{adelberger05a, conroy08, trainor12, rakic13}.
This corresponds to virial radii and circular velocities of 
$\approx90$~pkpc and $\approx217$~\kmps, respectively.
They tend to have median star formation rates $\sim25$~\msol~yr$^{-1}$
\citep{erb06b, steidel14},
gas-phase metallicites $\sim0.5\,Z_\odot$ \citep{erb06a}
and  stellar ages $\sim0.7$~Gyr \citep{erb06c}. 

The sub-sample that we use satisfies two constraints. Firstly, we require 
that the galaxies have impact parameters $\leq2$~pMpc 
(or $\simeq4\arcmin$ at $z\sim2.4$), so that
the range in common is covered in all 15 KBSS fields. 
Secondly, we only use galaxies that have a redshift within
the range of the \lya\ forest, since our pixel optical depth 
recovery is limited to this region (see \S~\ref{sec:od_recovery}). 
We define the redshift of the \lya\ forest as follows:
\begin{equation}
  (1 + \zqsom) \dfrac{\lambda_{\lybm} }{ \lambda_{\lyam}}- 1  \leq z \leq \zqsom - (1+\zqsom) \dfrac{3000\,\kmpsm}{c}
 \label{eq:zlim}
\end{equation}
where $\lambda_{\lyam}=1215.7$~\AA\ and $\lambda_{\lybm}=1025.7$~\AA\
are the \hone\ \lya\ and \lyb\ rest wavelengths, respectively. 
The lower limit on the galaxy redshifts is set by the beginning of the \lyb\ forest in \hone, and a cut is made
3000~\kmps\ bluewards of the redshift $\zqsom$ to avoid proximity effects (i.e., these regions
can be affected by ejecta and/or the ionising radiation field
originating from the QSO). 
We note that in our analysis (\S~\ref{sec:results}), we 
search for absorption within $\pm1350$~\kmps\ of galaxies. 
Therefore, in practice, our sample contains galaxies
that have redshifts extending 1350~\kmps\ above or below
the \lya\ forest limits given above.

Figure~\ref{fig:galaxy_sample_hist} shows histograms of the
galaxy impact parameters (left panel) and redshifts (right panel)
for the \ngaltotal\ galaxies that satisfy the above constraints.
We also show the distributions separately for the galaxies whose redshifts were 
measured from rest-frame UV features (using LRIS; \ngaluv\ galaxies)
and from rest-frame optical nebular emission lines (using NIRSPEC and MOSFIRE; \ngalneb\ galaxies).
Since the impact parameter binning shown in this figure is used throughout this 
paper, we have included the bin edge values as well as the number of galaxies
in each bin in Table~\ref{tab:bins}.
The two smallest impact parameter bins exhibit 
the strongest optical depth enhancement
for the metals studied in this work, so many of our results are based on the nearest
\ngalinnertwo\ galaxies, of which \ngalinnertwoneb\ have nebular redshifts.
As the volume sampled in these inner bins is comparably small, 
the number of galaxies at small impact parameters is somewhat reduced compared
to the larger bins.
The median redshifts of the three galaxy samples shown (all galaxies, UV-only,
and nebular-only) are $\langle z \rangle = \medz$, $\medzuv$, and $\medzneb$, respectively.

\subsection{QSO Spectra}

The 15 quasars that are part of the KBSS were all observed 
with Keck/HIRES, and their spectra have a typical resolution of $R\approx36000$.
A detailed description of the data is given in \citet{rudie12}; 
briefly, the spectra were reduced using T. Barlow's MAKEE package where
each spectral order was continuum normalised using low-order spline interpolation,
and the final spectra were rebinned to pixels of 2.8~\kmps. 
The final continua were fitted by hand, with an automated iterative correction procedure
applied redwards of the quasar's \lya\ emission line
(described in Appendix~\ref{sec:pod_details}). Based on tests
done in \citet{aguirre02}, we expect the errors induced due to continuum fits
to be $\lesssim1$\%. We also test the effects of the automated continuum fit
that we apply redwards of \lya\ in Appendix~\ref{sec:var_pod}.

Six of the spectra contain damped \lya\ systems (DLAs) in the \lya\ forest region, 
which have been fitted with  Voigt profiles and had their damping wings divided out 
(as described in \citealt{rudie12}). The saturated portions of the
six \lya\ forest region DLAs were masked and not used for the recovery of
optical depths for ions in this region (our masking procedure is described more 
fully in Appendix~\ref{sec:pod_details}). For more information about the QSOs, see Table~1
of \citet{rakic12} and also \citet{trainor12}.

\subsection{Galaxy Redshifts}

Redshifts for KBSS galaxies are measured from features in their rest-frame 
far-UV and optical spectra. The strongest features in the rest-frame UV
spectra of these galaxies are the \hone\ \lya\ emission line (when present) and a series 
of metallic interstellar absorption lines.  All of these strong features 
have been empirically determined  to lie at significant velocity offsets 
with respect to the systemic velocity of the galaxy 
\citep{shapley03, adelberger03, steidel10, rakic12}
which is generally interpreted as evidence for strong mass outflows from such systems. 
The rest-frame optical spectra of KBSS galaxies consist primarily of nebular emission lines 
which arise in \htwo\ regions within these galaxies and therefore trace the systemic
velocity of the system to much higher fidelity. For this reason, we prefer redshifts 
measured from nebular emission lines, when they are available.

Many galaxies in the KBSS have been observed with the near-IR Keck instruments 
NIRSPEC and MOSFIRE
which have spectral resolutions $R\approx1200$ and $\approx3600$, respectively.
Such observations allowed measurement of galaxy redshifts using
their nebular emission lines H$\alpha$, H$\beta$, and 
[\othree]~$\lambda\lambda4959,5007$ for 
\ngalneb\ galaxies in our KBSS subsample.
Since the nebular emission lines are good tracers of the systemic galaxy velocity, 
for those galaxies with nebular redshifts 
we take $z_{\rm gal} = z_{\rm neb}$. The measurement uncertainties 
for the two instruments, which were estimated by comparing
multiple observations of the same galaxy either on two separate occasions 
or in two different bands, are determined to be $\approx\pm60$~\kmps\ and 
$\approx\pm18$~\kmps, respectively 
We emphasise that most of the galaxies in the 
three smallest impact parameter bins
have nebular redshifts (see Table~\ref{tab:bins}). 

The remaining \ngaluv\ galaxies in our KBSS subsample
lack rest-frame optical spectra and therefore have 
redshifts measured
from rest-frame UV lines observed with Keck/LRIS
($R\approx800$--1300).
To account for the offset of rest-frame UV features
from the systemic galaxy velocity, we apply a correction to those galaxy redshifts 
estimated from interstellar absorption lines, $z_{\rm ISM}$, 
and \lya\ emission lines, $z_{\lyam}$.

\citet{rakic11} used
the fact that the mean foreground galaxy \lya\ absorption profiles
seen in QSO spectra should be symmetric around the true galaxy redshift to determine
the corrections needed to infer the systemic galaxy redshifts.
Another approach, which is the one that we use in this work, was first applied by 
\citet{adelberger05b} and \citet{steidel10}. They used a sample of galaxies 
having redshifts measured from both rest-frame UV and nebular emission lines
to estimate the typical offset between the two measurement techniques.

To calculate the velocity offset values, we use all galaxies with $2 <z_{\rm neb}<3$ 
that have both rest-frame UV and optical spectra, where 
the errors on the mean have been determined from bootstrap resampling the galaxies
1000 times with replacement, and taking the $1\sigma$ confidence intervals.
The resulting offset values are implemented as follows:
\begin{itemize}
\item For galaxies with only \lya\ emission lines detected (\ngalem\ galaxies),
\begin{equation}
 z_{\rm gal,\lyam} = z_{\lyam}-220^{+30}_{-29}\kmpsm.
\end{equation}
\item For galaxies with only interstellar absorption lines (\ngalabs\ galaxies),
\begin{equation}
 z_{\rm gal,ISM} = z_{\rm ISM}+110^{+19}_{-16}\kmpsm.
\end{equation}
\item For galaxies with both \lya\ emission and interstellar
 absorption (\ngalboth\ galaxies), first the following corrections are made 
 to the measured redshifts:
 \begin{equation}
 \begin{split}
 z_{\rm gal,\lyam} &= z_{\lyam}-370^{+17}_{-16}\kmpsm. \\
  z_{\rm gal,ISM} &= z_{\rm ISM}+200^{+18}_{-16}\kmpsm.
 \end{split}
\end{equation}
The arithmetic mean of the corrected redshifts, 
\begin{equation}
 z_{\rm gal} = \dfrac{z_{\rm gal,\lyam} + z_{\rm gal,ISM}}{2},
\end{equation}
is then used as the final corrected value, 
unless it does not satisfy
\begin{equation}
  z_{\rm ISM} < z_{\rm gal}<  z_{\lyam},
\end{equation}
in which case we use the arithmetic mean of the uncorrected values,
\begin{equation}
 z_{\rm gal} = \dfrac{z_{\lyam} + z_{\rm ISM}}{2}.
\end{equation}
\end{itemize}
Based on the sample of \ngalbothnebanduv\ galaxies with both nebular
and UV redshift estimates, we find that the rest-frame UV redshifts exhibit a $1$-$\sigma$ scatter 
of $\pm150$~\kmps. This value corresponds to the error for individual systemic
redshift estimates, rather than the error on the mean velocity offset.

\begin{figure}
\includegraphics[width=0.5\textwidth, clip=true, trim = 0mm 0mm 0mm 0]{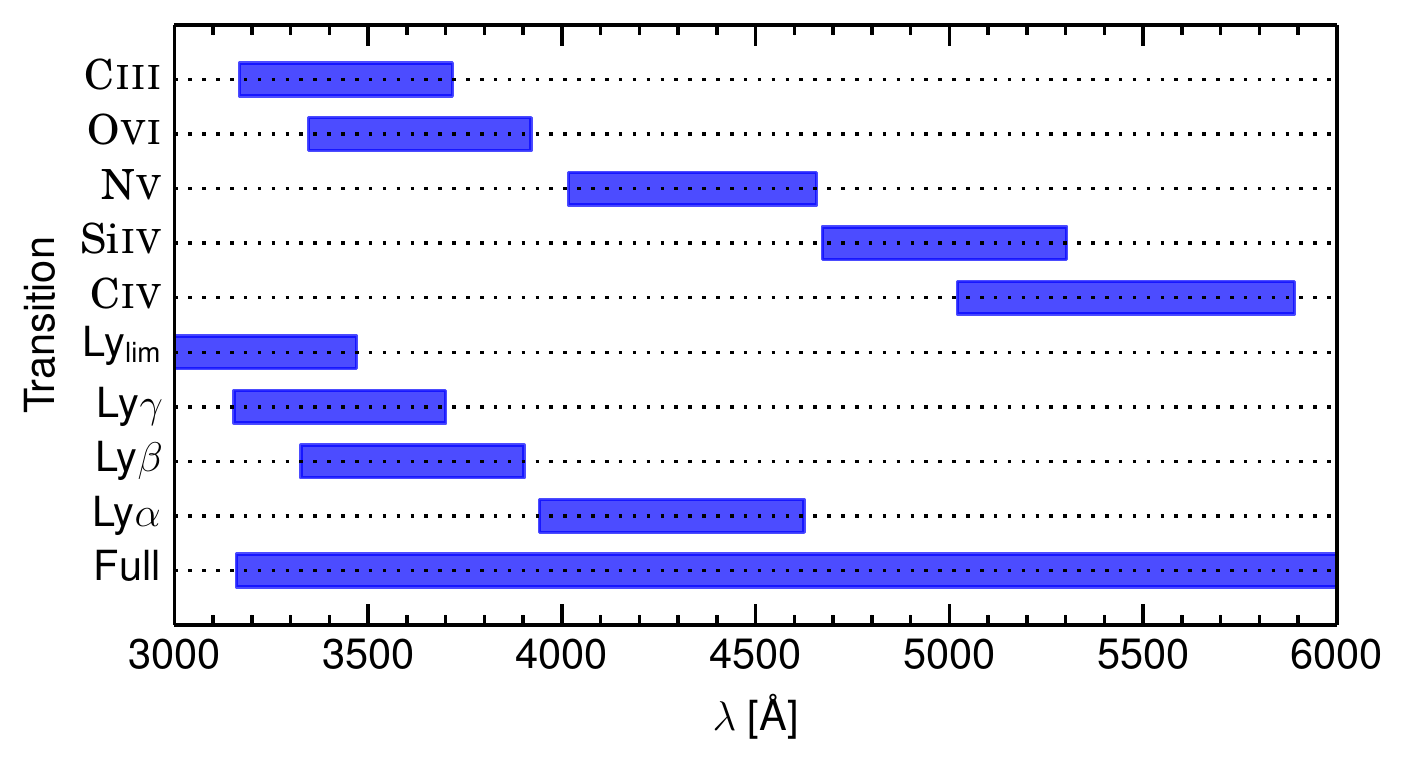}
\caption{Wavelength ranges of various transitions in one of our QSOs
(Q1549+1933, $\zqsom=2.8443$), when their redshifts are restricted to that of 
the \hone\ \lya\ forest ($z =2.24$--$2.81$) plus the additional ion-dependent constraints outlined in
\S~\ref{sec:od_recovery}. The bottom row (marked ``Full'') shows the wavelength
range covered by the observed spectrum.}
\label{fig:coverage}
\end{figure}

\begin{table*}
\caption{Rest wavelengths of both strong and weak components and their separation 
  (if applicable), as well as the pixel optical depth recovery implementation, for
  the different metal ions studied in this work.}
\label{tab:pod}
\begin{tabular}{@{}lrrrccc}
\hline
Metal &	 \multicolumn{2}{c}{\lrest (\AA)}&$\Delta v$	& Subtr. higher-& Doublet 	& Self- \\
ion   &	strong   & weak	& (\kmps)	&  order \hone 	& min. 		& contam. \\
\hline
\hline
\osix 	& 1031.927 &1037.616	& 1650		& $\checkmark$	& $\checkmark$ 	&  \\
\hline
\nfive 	& 1238.821 &1242.804	& 962		&		& $\checkmark$	&  \\
\hline
\cfour 	& 1548.195 &1550.770	& 498		&		&  		& $\checkmark$ \\
\hline
\cthree & 977.020  &\ldots	& \ldots	& $\checkmark$ 	&  		&  \\
\hline
\sifour & 1393.755 &1402.770	& 1930		&		& $\checkmark$	&  \\
\hline
\end{tabular}
\end{table*}

\section{Optical depth recovery}
\label{sec:od_recovery}

To study the absorption of metals in the vicinity of galaxies, 
we have used the pixel optical depth method 
\citep{cowie98, songaila98, ellison00, schaye00, aguirre02, schaye03}
rather than the complementary technique of fitting Voigt profiles to absorption lines.
The pixel optical depth approach is advantageous in the sense that it allows one to 
quickly and objectively measure absorption strengths
in a statistical sense, even for weak signals in highly contaminated regions.
On the other hand, with the standard pixel optical depth method, information 
about line widths is lost, and 
the interpretation of optical depths is not 
always as straightforward as that of column densities.
In this section, we give a brief description of our
specific implementation which is taken
largely from \citet{aguirre02} with some minor improvements; 
more details can be found in Appendix~\ref{sec:pod_details}.

\subsection{Redshift ranges}

For each metal transition, the so-called fiducial redshift range that we use for the recovery
is first set by that of the \lya\ forest, which was described in \S~\ref{sec:gal_sample}
and is given by Equation~\ref{eq:zlim}. As for \hone, for all ions considered we set the 
upper limit to be $\zqsom-3000$~\kmps to avoid proximity effects.
Additional redshift range modifications are made based on the transition in question,
the reasons for which can be seen more clearly by examining Figure~\ref{fig:coverage}.

Firstly, the \osix\ region overlaps mainly with the \lyb\ forest, 
but extends marginally into the \lya\ forest since its rest wavelength 
is slightly higher than that of \lyb. 
In an effort to make the contamination across the recovery area uniform, 
we limit \osix\ to overlap only with the \lyb\ forest, and take 
$z_{\rm max} = (1 + \zqsom) \lambda_{\honem,\lybm} / \lambda_{\osixm,2} - 1$
where $\lambda_{Z,k}$ is the rest wavelength of multiplet component $k$ 
of the ion $Z$. 

% \sithree\ ($\lrestm=1206.6$~\AA) and \nfive\ ($\lrestm=[1238.8,1242.8]$~\AA) 
% have rest wavelengths which are, respectively, 
% slightly below and above that of \hone\ \lya\ ($\lrestm=1215.8$]~\AA), 
% and therefore the recovery regions of these metals lie mainly 
% in the \lya\ forest. Although no correction is done for contamination by \lya, to avoid
% contamination from higher-order Lyman lines and to keep the recovery 
% region uniform, we impose the constraint 
% that the recovered optical depth region must not extend to redshifts outside of the \lya\ forest, 
% namely $z_{\rm min} =(1 + \zqsom) \lambda_{\lybm} / \lambda_{\sithreem}$ 
% for \sithree\ and 
%  $z_{\rm max} = (1 + \zqsom) \lambda_{\lyam} / \lambda_{\nfivem}$ for \nfive.

The \nfive\ doublet ($\lrestm=[1238.8,1242.8]$~\AA) 
has rest-wavelengths which are
slightly above that of \hone\ \lya\ ($\lrestm=1215.8$\AA), 
and therefore the recovery region of this species lies mainly 
in the \lya\ forest. Although no correction is done for contamination by \lya,
to keep the level of contamination in the recovery region uniform, we impose the constraint 
that the recovered optical depth region must not extend to redshifts outside of the \lya\ forest, 
namely 
 $z_{\rm max} = (1 + \zqsom) \lambda_{\lyam} / \lambda_{\nfivem} - 1$.
 
The recovery regions of \cfour\ ($\lrestm = [1548.2, 1550.8]$~\AA) and 
\sifour\ ($\lrestm = [1393.8, 1402.8]$~\AA) both lie mainly redwards of the \lya\ forest.
In the case of \sifour, we avoid any contamination from \lya\ by excluding wavelengths bluewards
of the quasar's \lya\ emission, corresponding to a minimum redshift of
$z_{\rm min} = (1+\zqsom)\lambda_{\lyam}/\lambda_{\sifourm}-1$. 
For \cfour, the limits given in Equation~\ref{eq:zlim} 
are used; however, it is possible to expand the range to lower redshifts until
the quasar's \lya\ emission limit is reached 
($z_{\rm min} = [1+\zqsom]\lambda_{\lyam}/\lambda_{\cfourm}-1$), which
can significantly increase the number of galaxies in the sample.
We experiment with this modification in Appendix~\ref{sec:c4_var_pod}.

\begin{figure*}
\includegraphics[width=\textwidth]{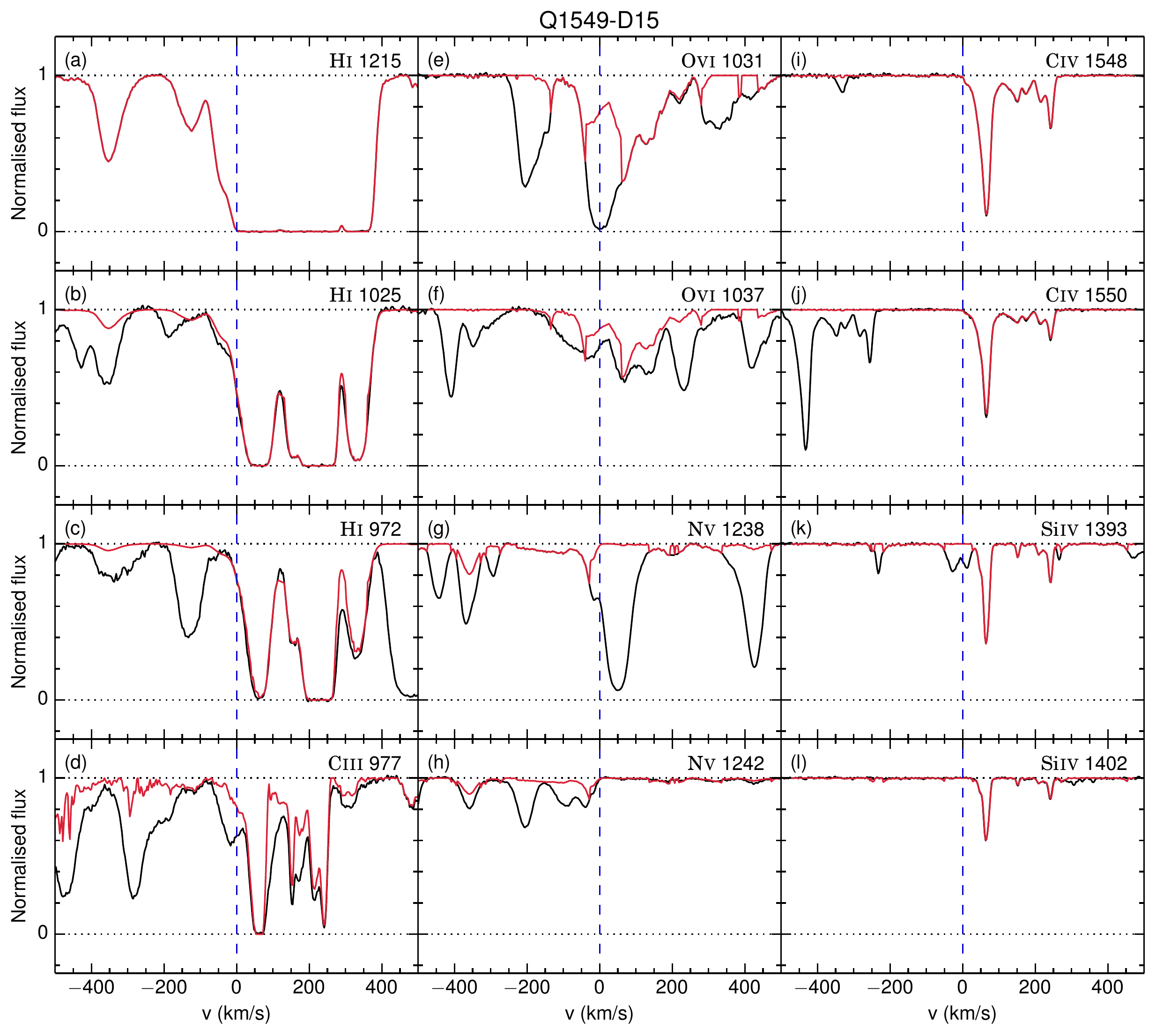}
\caption{Normalised flux (black lines) and recovered optical depths 
 converted to normalised flux (red lines) for regions $\pm500$~\kmps\ around
 the ions studied in this work at the redshift of the galaxy Q1549-D15,
which has an impact parameter of 58~pkpc.
 $v = 0$~\kmps\ corresponds to the galaxy redshift, and is marked by the 
 vertical dashed blue line. 
 We stress here that, particularly for those metal ions bluewards of \lya\ (\osix, 
 \nfive, and \cthree), the derived optical depths are {\it apparent} 
 rather than true optical depths, since we have no way of fully correcting
 for all possible contamination. 
 Panels (a), (b) and (c) show \lya, \lyb\ and \lyg, respectively,
 and demonstrate how higher-order \hone\ components often have unsaturated pixels 
 which can be used to correct pixels which are saturated in \lya\ 
 (such as in the region shown here from 0--400~\kmps). Panel (d) makes clear 
 that much of the absorption in the \cthree\ region can be attributed to 
 higher order \hone\ absorption, and is at least partially corrected for 
 in the \hone\ subtraction 
 procedure. We note that the noise from the subtracted \hone\ is responsible for 
 the fine structure seen at negative velocities for \cthree.  Both \nfive\ (panels (g) and (h)) 
 and \sifour\ (panels (k) and (l)) have their optical depths corrected 
 for contamination by taking the minimum optical depth between the two 
 double components. This procedure allowed us to identify and correct for the 
 contaminating absorption systems located at $\sim50$ and 
 $\sim450$~\kmps for \nfive\ (panel (g)), and at $\sim-250$ and 
 $\sim0$~\kmps\ for \sifour\ (panel (k)). 
 The recovery procedure for the \osix\ doublet, shown in panels 
 (e) and (f), uses both \hone\ subtraction as well as taking the minimum
 optical depth value between the two doublet components. 
 Finally, the $\sim-350$~\kmps absorption line
 in panel (i) was identified and removed by the \cfour\ self-contamination
 correction.  }
\label{fig:linechart_a}
\end{figure*}

\subsection{Corrections for contamination}

Below we briefly outline the optical depth recovery method used
(see Appendix~\ref{sec:pod_details} for the full
description). A summary of the metal ion rest 
wavelengths, the doublet
separation in \kmps\ (if applicable), 
and pixel optical depth implementations (to be explained below)
 is given in Table~\ref{tab:pod}. 
Here we also note that the separation between the strong \osix\ transition 
($\lrestm=1031.927$~\AA) and \hone\ \lyb\ ($\lrestm=1025.7223$~\AA)  is 
1810~\kmps. It is important to keep the above transition separations in mind
when examining the optical depths, since
contamination can be expected on these characteristic scales. 
To visualise the corrections for contamination, in Figure~\ref{fig:linechart_a} we have 
plotted $\pm500$~\kmps\ regions of the spectrum around the galaxy
Q1549-D15, for each of the ions studied. The black lines denote the original flux,
while the red lines are calculated from the recovered optical depths 
derived using the following procedure. 

Implementing the above redshift ranges, we define the optical depth for each ion $Z$
and multiplet component $k$
as $\tau_{Z,k}(z) = -\ln(F)$, where $F(\lambda)$ is the normalised flux at
$\lambda = \lambda_{k}(1+z)$.
Beginning with \hone\ \lya, the main source of error is the saturation of the absorption;
to account for this, for every saturated pixel we search for unsaturated 
higher-order lines at the same redshift and take the minimum of the optical depths (corrected
for differences in oscillator strengths and rest-frame wavelengths, and accounting for noise). 
An example of a saturated region with corresponding unsaturated higher-order
lines can be seen in panels (a)--(c) of Figure~\ref{fig:linechart_a}. 
We also use the higher-order transitions to search for and flag \lya\ pixels
contaminated by metal line absorption,
something which was not implemented in \citet{aguirre02}.

The recovered \hone\ optical depths are then used to clean the \cthree\ and 
\osix\ regions by subtracting the optical depths of five higher 
order \hone\ lines starting from \lyb. In both \citet{aguirre02} and this work,
the \hone\ subtraction is performed on all unsaturated metal ion pixels.
However, the subtraction cannot be done reliably for saturated  metal ion pixels since their 
optical depths are not well defined. In \citet{aguirre02}, 
such pixels remain unchanged by the subtraction procedure, which could result
in recovered optical depths being biased high. 
To combat this, we have made another addition to the recovery method,
where for saturated metal line pixels 
we sum the optical depths from the higher-order \hone\ components. If this value
is consistent with saturation, we flag the pixel as contaminated and discard it. 
Panel (d) of Figure~\ref{fig:linechart_a}
shows the effect of \hone\ subtraction on the \cthree\ region, where any absorption
that is seen in the black spectrum but not in the red is due to known higher-order
\hone\ lines. 

Furthermore, since \osix\ is a doublet,
we perform the \hone\ subtraction on both doublet components and 
further correct for contamination by taking the minimum of the
optical depths between the two components at each redshift,
taking into account relative oscillator strengths and rest
wavelengths. 
This doublet minimum correction is also made for \nfive\ and 
\sifour, which are mainly contaminated by \hone\ \lya\ and \cfour,
respectively. We note that for metals that have a doublet component,
we always scale the weaker transition optical depths to the strongest 
transition (in this work, all doublets considered have
an optical depth ratio of 2). Unlike in \citet{aguirre02}, 
we allow pixels from the weaker transition to be used in the doublet 
minimum correction even if they have been flagged as having flux $>1$
(or a negative optical depth). Panels (g)--(h) (\nfive) and (k)--(l) (\sifour) of 
Figure~\ref{fig:linechart_a} demonstrate how this technique can
remove contaminating absorption lines. 

Finally, \cfour\ shows relatively strong absorption with a recovery
region that lies redwards of the \lya\ forest; this means that the main
source of contamination comes from its own doublet. To correct for this, 
we iteratively subtract the expected optical depth of the contaminating
doublet from each pixel. The result of doing so is shown in panels (i)--(j) 
of Figure~\ref{fig:linechart_a}.

How important is it to correct for contamination? In Appendix~\ref{sec:var_pod}
we examine the recovery of each metal ion in turn to see how the above procedure 
affects our final result. In general, we find that performing the correction increases
the dynamic range of recovered optical depths and decreases the scatter
when they are binned as a function of galactocentric distance. 
However, we note that even without any corrections, the main conclusions
of this work still hold. 

It is important to note that the corrections for contamination are not
perfect. For example, because we can only reliably recover \hone\ \lya\ 
in the range of the \lya\ forest, we cannot correct contamination of
\osix\ and \cthree\ (which lie in the \lyb\ forest) by \lya. 
Hence, the optical depths quoted should not be taken at face value. 
This is particularly relevant for \cthree, \nfive\ and \osix, which all suffer from
substantial residual contamination from \hone. However, because the 
contamination is due to gas at very different redshifts, it does not 
vary systematically with the distance to galaxies will therefore not 
give rise to spurious trends of absorption strength with separation 
from galaxies. It can only compromise our ability to detect such a trend.

%%%%%%%%%%%
% Results %
%%%%%%%%%%%

\section{Results}
\label{sec:results}

We can now use the recovered optical depths from \S~\ref{sec:od_recovery}
and combine them with the redshifts and impact parameters of the galaxies 
from \S~\ref{sec:gal_sample} to investigate how \hone\ and metal 
ions are distributed around galaxies.
In \S~\ref{sec:maps}, we bin the pixel optical depths by galaxy impact
parameter and along the LOS to create 2-dimensional (2-D) optical depth maps,
while in \S~\ref{sec:profiles} we make the cuts through these maps
to look for redshift space distortions 
in order to compare the optical depth distributions for different ions. 
We bin the pixels by their 3-D 
distance to the galaxy in \S~\ref{sec:hubble}, where distances are estimated assuming 
pure Hubble flow, and in  \S~\ref{sec:odhist} we show how the distribution of 
optical depths for each ion varies as a function of impact parameter. 
Covering fractions are examined in \S~\ref{sec:fcover},
and lastly we explore the effect of galaxy redshift errors on 
our results in \S~\ref{sec:var_gal_sample}.
Unless specified otherwise, all errors are calculated 
by bootstrap resampling (with replacement) the galaxy sample 1000 times within 
each impact parameter bin, and taking the $1\sigma$ confidence intervals. 

\subsection{2-D Optical depth maps}
\label{sec:maps}

\def\wa{0.44\textwidth}
\begin{figure*}
 \includegraphics[width=\wa]{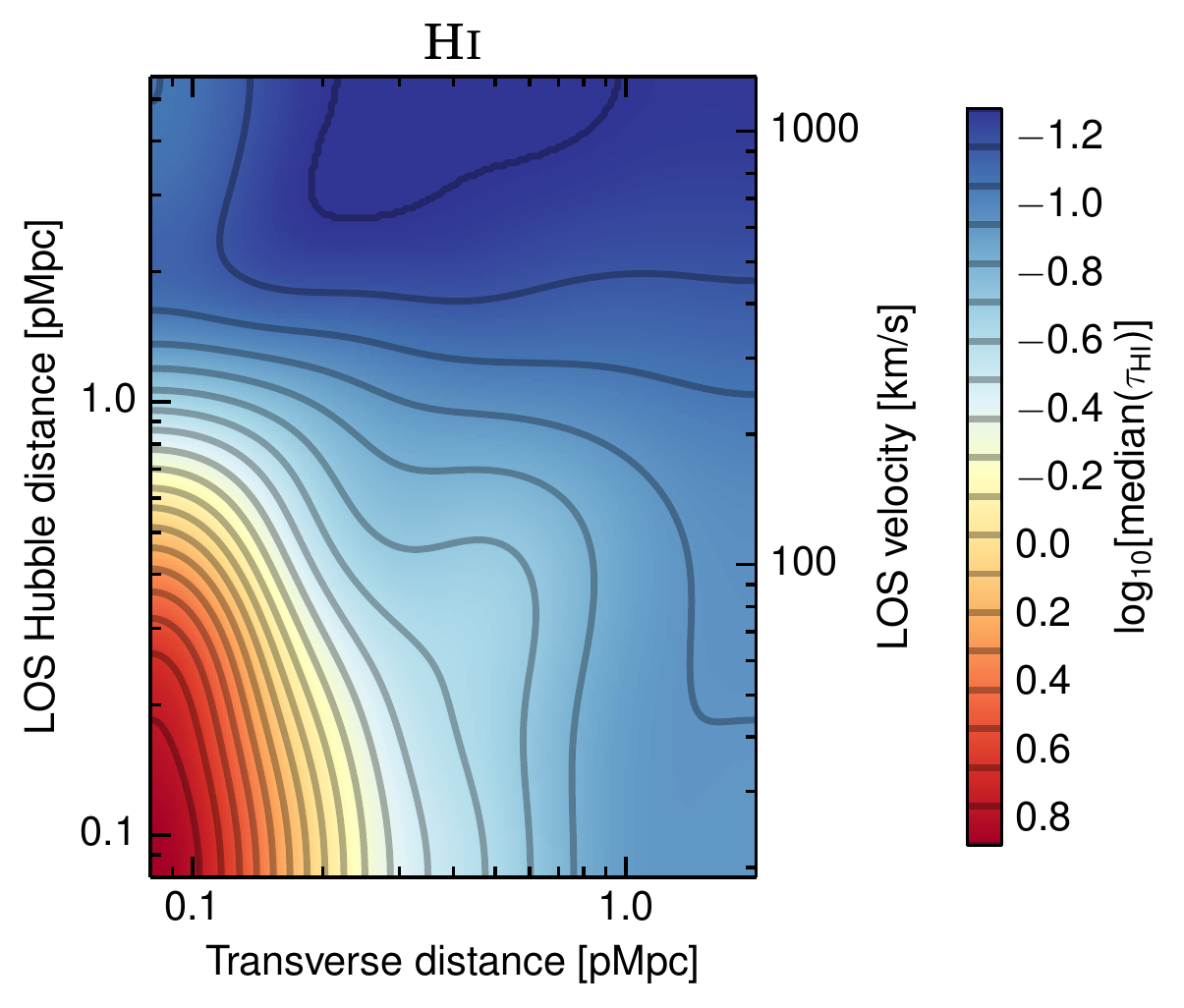}
 \includegraphics[width=\wa]{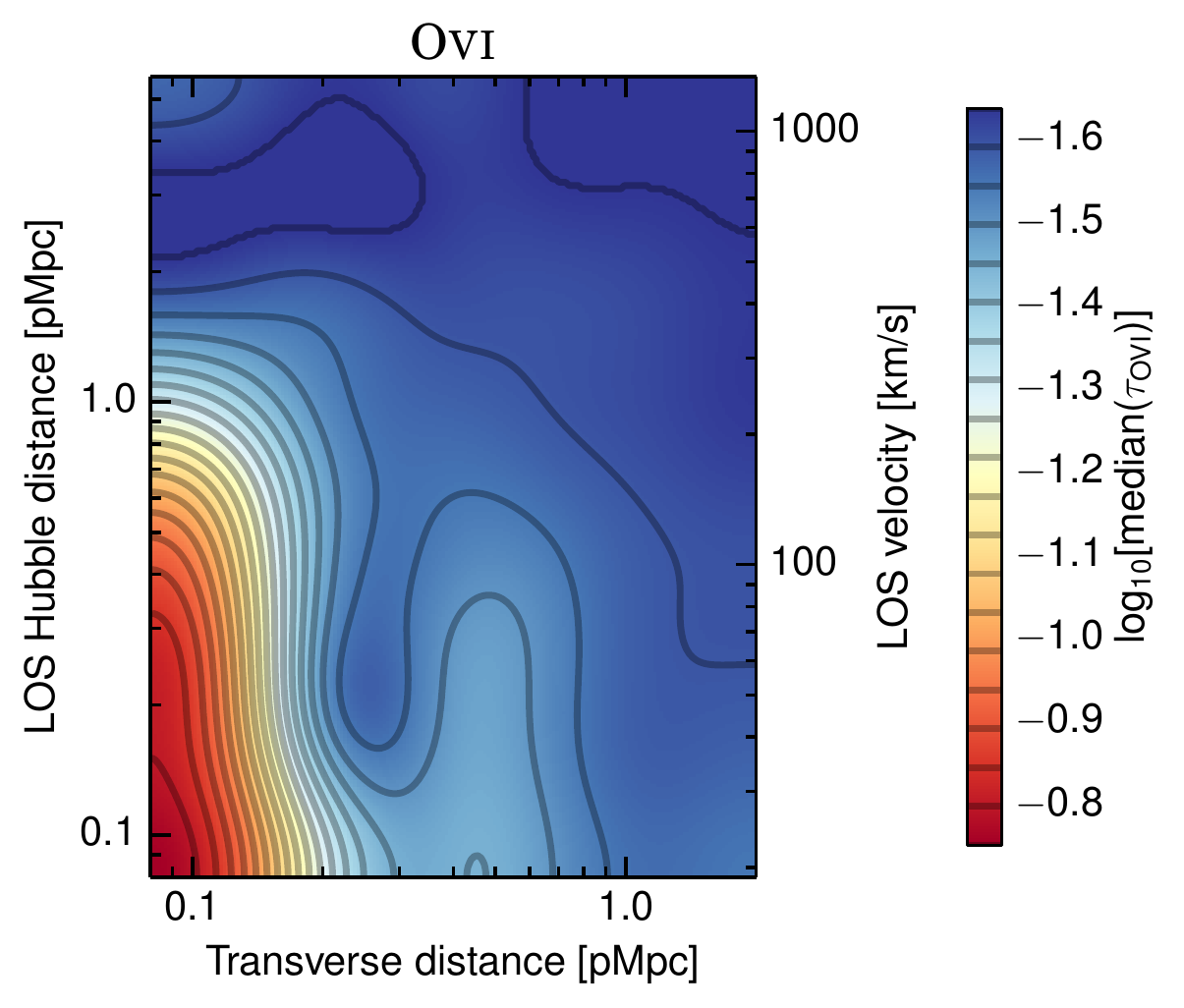}\\
 \includegraphics[width=\wa]{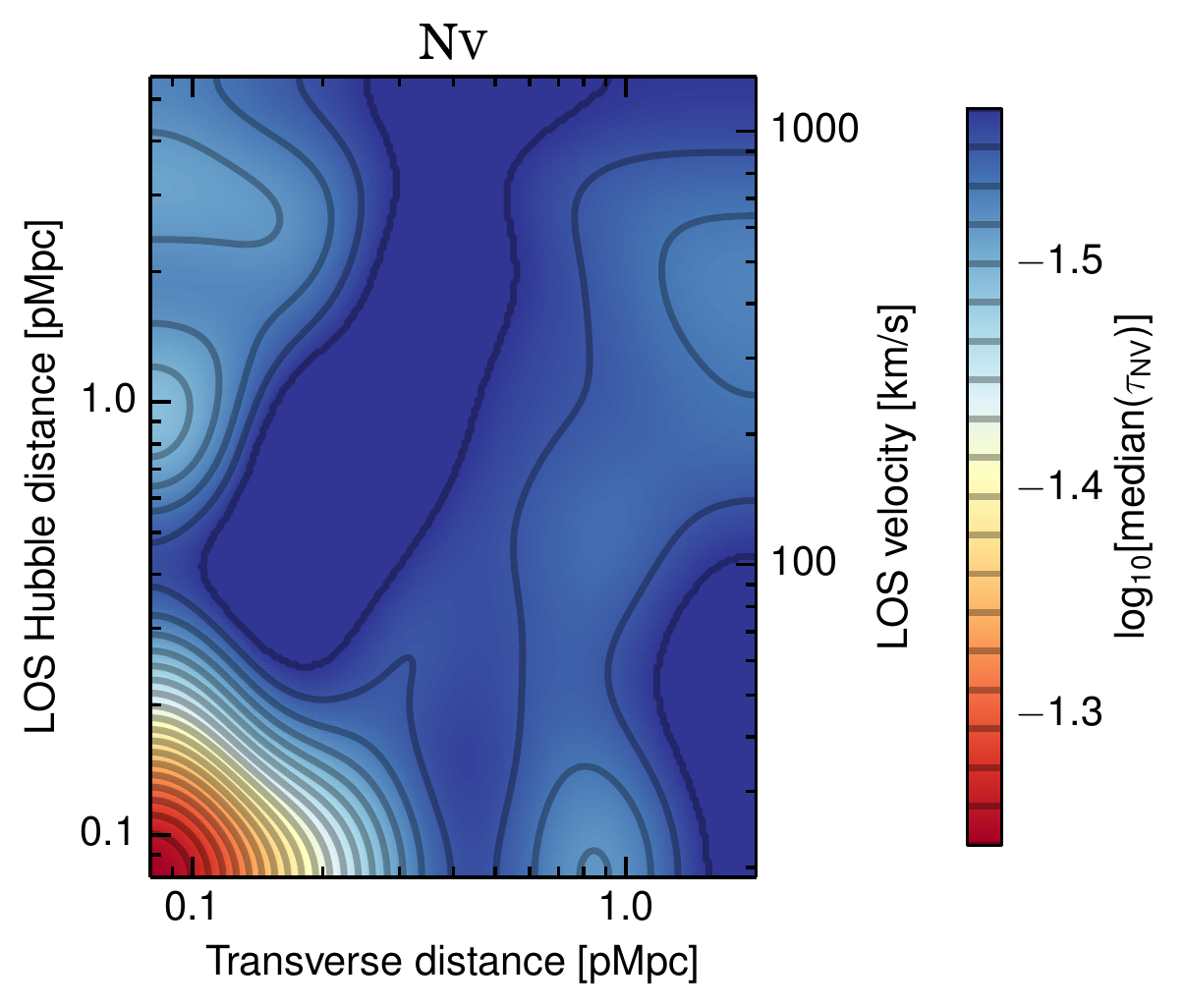}
 \includegraphics[width=\wa]{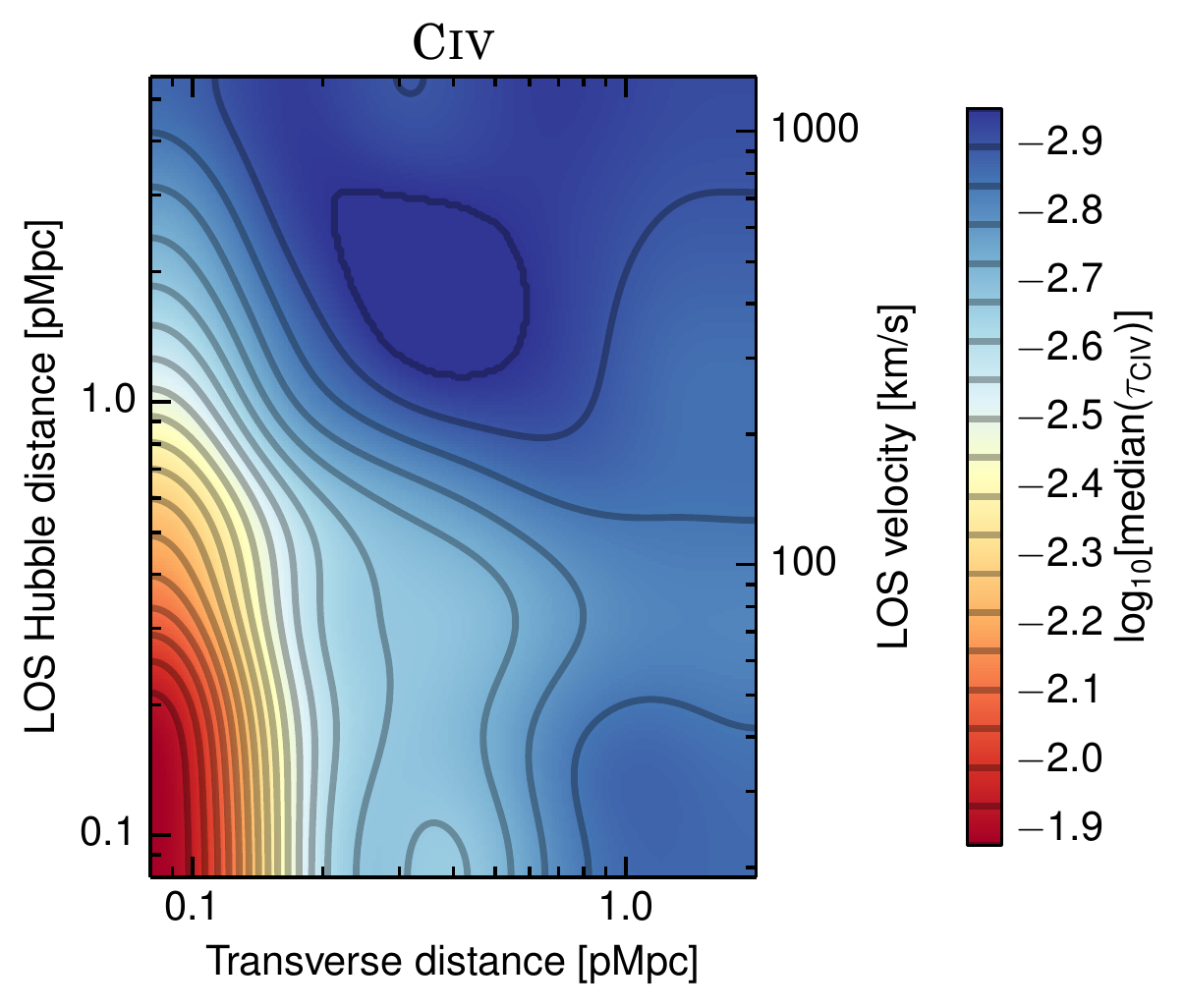}\\
 \includegraphics[width=\wa]{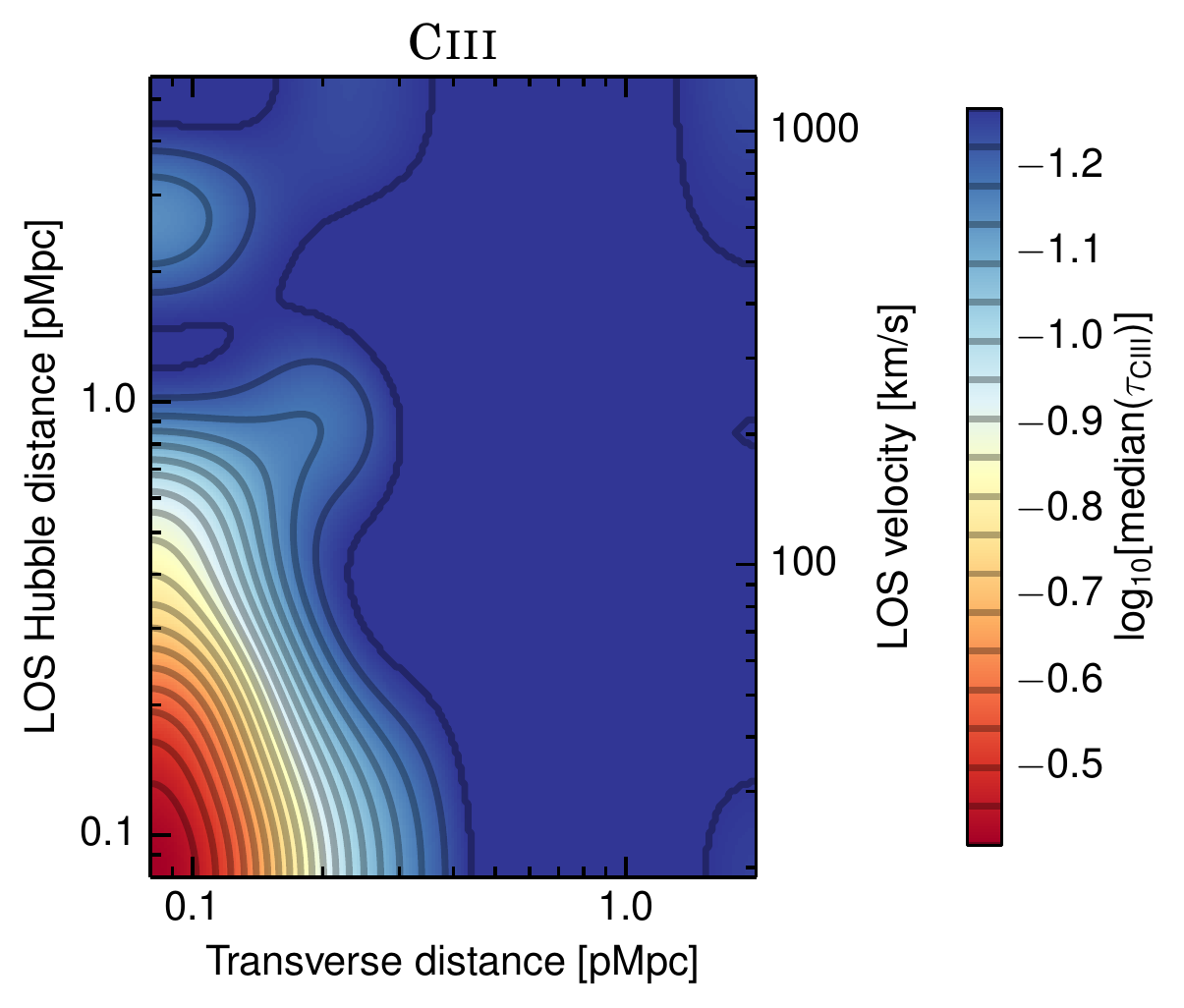}
 \includegraphics[width=\wa]{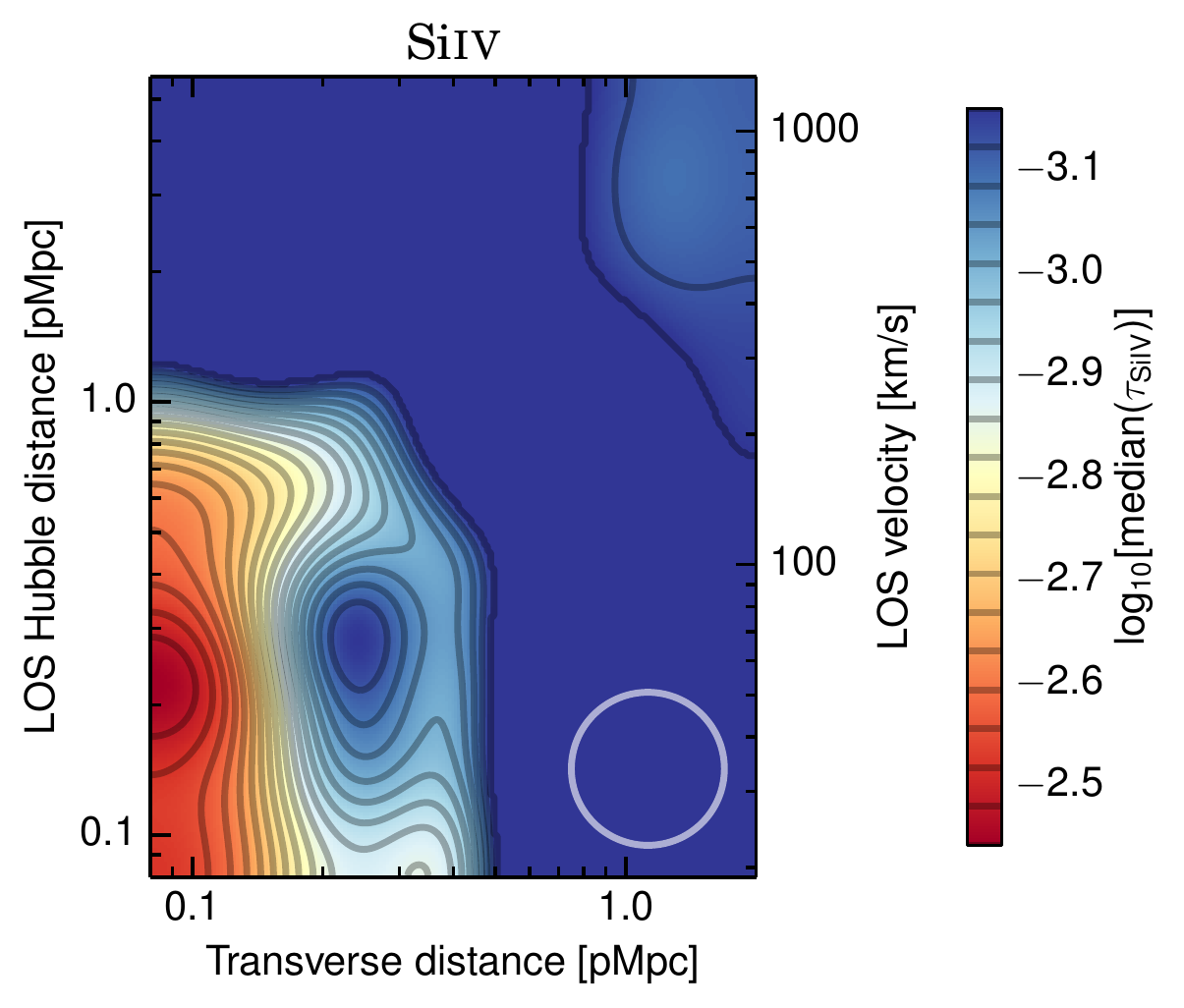}
\caption{2-D optical depth maps constructed by binning the galaxy impact parameters and
LOS distance (assuming pure Hubble flow) logarithmically, taking the absolute LOS distance, 
and calculating the median optical depth in each bin.
The bins are 0.15~dex wide except for the innermost bin, which runs from 0 to 130~pkpc. 
The maps have been smoothed using a Gaussian that has $\sigma$ equal to the bin size, and 
the FWHM of the smoothing kernel is indicated by open the white circle in the bottom right corner
of the \sifour\ map. 
For each ion, the range of optical depths is set
to run from the median of all pixels in the considered redshift range
 to the maximum value in the smoothed image.
Note that the minimum optical depth shown typically reflects the median level of noise and 
contamination rather than the detection of true metal absorption. 
The ions considered and their rest frame wavelengths are, from left to right:
\hone\ (1215.7~\AA),
\osix\	(1031.9~\AA),
\nfive\ (1238.8~\AA),
\cfour\ (1548.2~\AA),
\cthree\ (977.0~\AA), and
\sifour\ (1393.8~\AA). 
In every case, there is a region of very strongly enhanced absorption at impact parameters of $\lesssim2$~pkpc,
which is smeared along the LOS direction. This elongation likely originates 
primarily from gas peculiar velocities, as discussed in \S~\ref{sec:var_gal_sample}.}
\label{fig:all_maps_b}
\end{figure*}

\begin{table*}
 \caption{The median $\log_{10}\tau_{\honem}$ and $1\sigma$ errors  as a function 
   of impact parameter (rows) and distance along the LOS (columns), used to construct Figure~\ref{fig:all_maps_b}. The values
    from the innermost transverse  and LOS distance bins are plotted in Figure~\ref{fig:zspacedist}. Tables for 
     \osix, \nfive, \cfour, \cthree, and \sifour\ are available online at 
  {\tt http://www.strw.leidenuniv.nl/\mytilde turnerm/kbss\_metals\_table.pdf}.}
\label{tab:maps}
\hone
\begin{tabular}{@{}lrrrrrrrrr}
\hline\hline
$D$ (pMpc) & 	0.04--0.13 & 	0.13--0.18 & 	0.18--0.25 & 	0.25--0.36 & 	0.36--0.50 & 	0.50--0.71 & 	0.71--1.00 & 	1.00--1.42 & 	1.42--2.00\\
\hline
0.00--0.13	& $1.06_{-0.29}^{+0.94}$	& $0.13_{-0.37}^{+0.25}$	& $0.08_{-1.37}^{+0.39}$	& $-0.83_{-0.14}^{+0.29}$	& $-0.28_{-0.17}^{+0.15}$	& $-0.78_{-0.17}^{+0.18}$	& $-0.98_{-0.09}^{+0.09}$	& $-0.97_{-0.08}^{+0.08}$	& $-0.92_{-0.08}^{+0.09}$ \\ 
0.13--0.18	& $0.95_{-0.31}^{+0.32}$	& $-0.12_{-0.24}^{+0.18}$	& $0.11_{-0.99}^{+0.24}$	& $-0.94_{-0.21}^{+0.27}$	& $-0.40_{-0.17}^{+0.12}$	& $-0.81_{-0.15}^{+0.11}$	& $-0.95_{-0.11}^{+0.12}$	& $-0.98_{-0.09}^{+0.08}$	& $-0.92_{-0.09}^{+0.10}$ \\ 
0.18--0.25	& $0.99_{-0.21}^{+0.17}$	& $-0.40_{-0.11}^{+0.63}$	& $-0.01_{-0.65}^{+0.23}$	& $-0.90_{-0.27}^{+0.42}$	& $-0.49_{-0.14}^{+0.15}$	& $-0.72_{-0.15}^{+0.13}$	& $-1.00_{-0.07}^{+0.12}$	& $-0.93_{-0.07}^{+0.08}$	& $-0.97_{-0.08}^{+0.09}$ \\ 
0.25--0.36	& $0.96_{-0.70}^{+0.20}$	& $-0.31_{-0.51}^{+0.28}$	& $-0.55_{-0.16}^{+0.25}$	& $-0.71_{-0.38}^{+0.21}$	& $-0.50_{-0.16}^{+0.12}$	& $-0.55_{-0.13}^{+0.09}$	& $-0.96_{-0.07}^{+0.08}$	& $-0.95_{-0.07}^{+0.09}$	& $-0.96_{-0.07}^{+0.06}$ \\ 
0.36--0.50	& $0.43_{-0.72}^{+0.43}$	& $-0.68_{-0.25}^{+0.76}$	& $-0.65_{-0.10}^{+0.17}$	& $-0.91_{-0.20}^{+0.24}$	& $-0.52_{-0.18}^{+0.16}$	& $-0.63_{-0.09}^{+0.11}$	& $-0.86_{-0.09}^{+0.11}$	& $-0.98_{-0.06}^{+0.06}$	& $-0.96_{-0.06}^{+0.07}$ \\ 
0.50--0.71	& $0.05_{-0.47}^{+0.37}$	& $-0.53_{-0.41}^{+0.79}$	& $-0.88_{-0.22}^{+0.32}$	& $-0.87_{-0.28}^{+0.24}$	& $-0.75_{-0.19}^{+0.24}$	& $-0.70_{-0.10}^{+0.07}$	& $-0.85_{-0.08}^{+0.11}$	& $-0.93_{-0.06}^{+0.06}$	& $-0.99_{-0.07}^{+0.06}$ \\ 
0.71--1.00	& $-0.51_{-0.12}^{+0.09}$	& $-0.57_{-0.23}^{+0.32}$	& $-0.68_{-0.29}^{+0.27}$	& $-0.98_{-0.18}^{+0.17}$	& $-0.79_{-0.13}^{+0.12}$	& $-0.96_{-0.08}^{+0.10}$	& $-0.92_{-0.08}^{+0.06}$	& $-1.00_{-0.06}^{+0.05}$	& $-1.02_{-0.05}^{+0.05}$ \\ 
1.00--1.42	& $-0.81_{-0.22}^{+0.18}$	& $-0.89_{-0.11}^{+0.13}$	& $-0.94_{-0.19}^{+0.47}$	& $-1.09_{-0.12}^{+0.14}$	& $-1.04_{-0.11}^{+0.10}$	& $-1.13_{-0.09}^{+0.09}$	& $-1.06_{-0.06}^{+0.07}$	& $-1.09_{-0.05}^{+0.05}$	& $-1.10_{-0.05}^{+0.05}$ \\ 
1.42--2.00	& $-1.36_{-0.14}^{+0.19}$	& $-1.45_{-0.18}^{+0.23}$	& $-1.04_{-0.27}^{+0.33}$	& $-1.14_{-0.11}^{+0.09}$	& $-1.26_{-0.07}^{+0.07}$	& $-1.17_{-0.08}^{+0.08}$	& $-1.09_{-0.06}^{+0.07}$	& $-1.16_{-0.04}^{+0.04}$	& $-1.16_{-0.04}^{+0.04}$ \\ 
2.00--2.83	& $-0.98_{-0.22}^{+0.27}$	& $-1.43_{-0.15}^{+0.11}$	& $-1.32_{-0.10}^{+0.09}$	& $-1.39_{-0.13}^{+0.13}$	& $-1.36_{-0.08}^{+0.07}$	& $-1.16_{-0.06}^{+0.06}$	& $-1.29_{-0.04}^{+0.04}$	& $-1.17_{-0.04}^{+0.03}$	& $-1.25_{-0.04}^{+0.04}$ \\ 
2.83--3.99	& $-1.01_{-0.14}^{+0.11}$	& $-1.34_{-0.09}^{+0.10}$	& $-1.43_{-0.14}^{+0.21}$	& $-1.24_{-0.09}^{+0.09}$	& $-1.27_{-0.07}^{+0.06}$	& $-1.28_{-0.05}^{+0.05}$	& $-1.33_{-0.05}^{+0.05}$	& $-1.22_{-0.03}^{+0.03}$	& $-1.28_{-0.03}^{+0.03}$ \\ 
3.99--5.64	& $-1.02_{-0.14}^{+0.12}$	& $-1.20_{-0.12}^{+0.11}$	& $-1.38_{-0.12}^{+0.17}$	& $-1.21_{-0.10}^{+0.10}$	& $-1.34_{-0.06}^{+0.06}$	& $-1.38_{-0.06}^{+0.05}$	& $-1.27_{-0.04}^{+0.04}$	& $-1.29_{-0.02}^{+0.03}$	& $-1.28_{-0.03}^{+0.03}$ \\ 
\hline
\end{tabular}
\end{table*}

\begin{table}
\caption{The log of the median optical depth, and the median 
  continuum S/N of all pixels (with normalized flux $>0.7$)
   in the redshift range considered for 
  the particular ion and recovery method.}
\label{tab:taumed}
\begin{tabular}{@{}lcc}
\hline\hline
Ion & $\log_{10}$\taumedk  & S/N$_{\rm med}$\\
\hline
\hone & $-1.29$ & $70.1$ \\ 
\osix & $-1.64$ & $45.7$ \\ 
\nfive & $-1.57$ & $72.4$ \\ 
\cfour & $-2.95$ & $83.6$ \\ 
\cthree & $-1.27$ & $37.8$ \\ 
\sifour & $-3.16$ & $84.4$ \\ 
\hline
\end{tabular}
\end{table}

\begin{figure*}
 \includegraphics[width=0.9\textwidth]{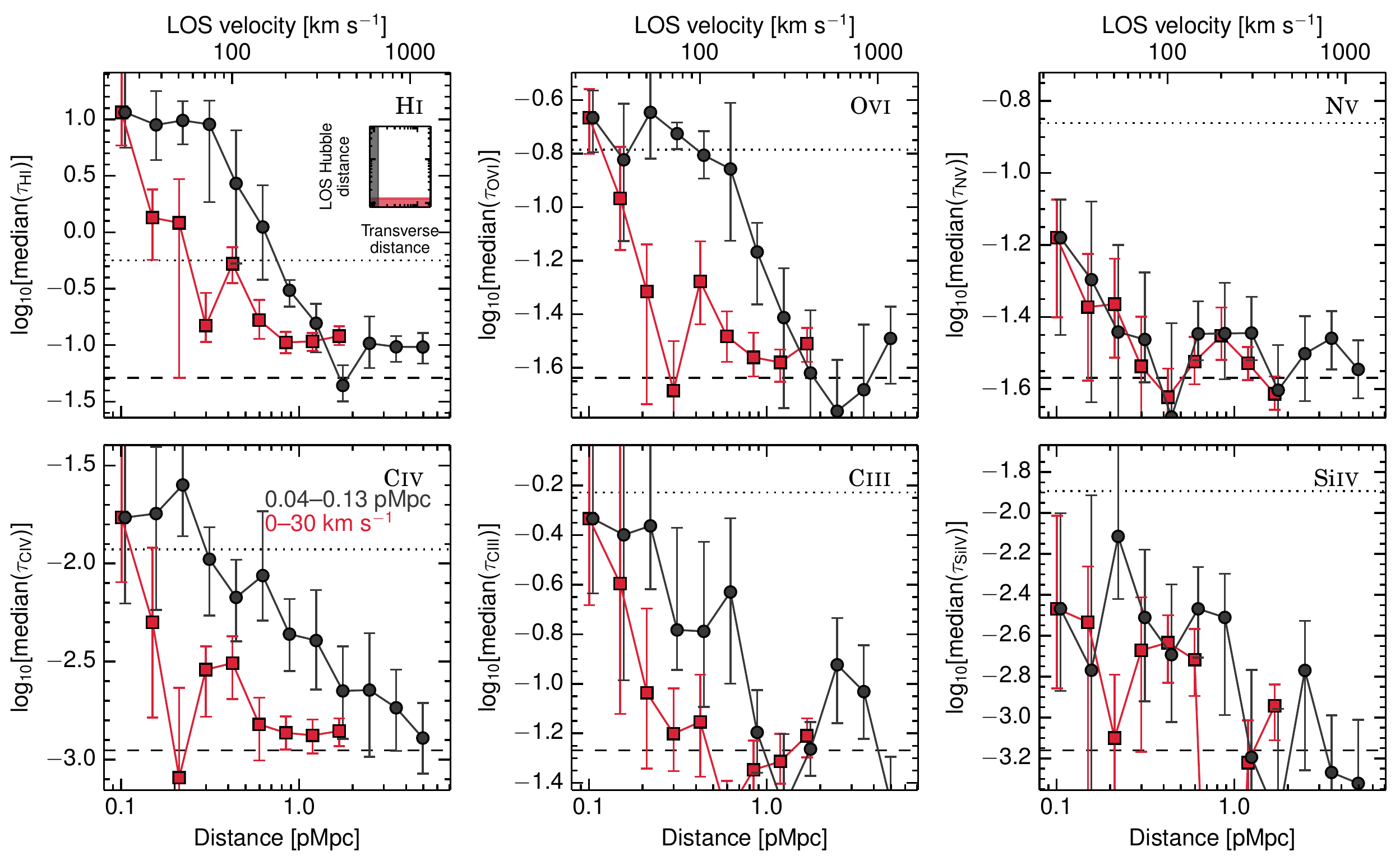}
\caption{Cuts through the median optical depth 
maps shown in Figure~\ref{fig:all_maps_b} along the 
innermost LOS and transverse distance bins,
 where each panel represents a different ion.
 Contrary to Figure~\ref{fig:all_maps_b}, the data have not been smoothed. 
 The black circles run along the LOS and are 
 taken from a 0.04--0.13~pMpc cut in impact parameter, while the red squares run along the 
  transverse direction using a a $\pm30.0$~\kmps\ LOS cut size, which corresponds to 
  0.13~pMpc at $z=\medz$ in the case of pure Hubble flow. The insets show the scaled down
 areas plotted in Figure~\ref{fig:all_maps_b}, where the coloured strips denote the cut sizes. 
 The horizontal dashed lines represent the median optical depths of all pixels in the redshift range
 considered for each ion, while the horizontal dotted line shows the $1\sigma$ scatter. 
In every case except for \nfive, the median optical depth values along the LOS are greater
 than than those in the transverse direction for bins 2--7 (i.e. out to $\sim1$~pMpc or 240~\kmps). 
  }
\label{fig:zspacedist}
\end{figure*}

For each ion we have constructed a galaxy-centred map of the median optical depth as a function of the 
transverse and LOS separation from the galaxies. The maps extend 2~pMpc in the transverse direction and 
5.64~pMpc along the LOS (i.e. $\pm1350$~\kmps), where velocity differences were converted into proper  
distances given each galaxy's redshift and assuming pure Hubble flow. The pixels contributing to the maps 
were binned logarithmically in the LOS 
and transverse directions, where the first bin runs from 0.04--0.13 pMpc and subsequent bins are 0.15~dex 
wide. We note that each bin contains pixels coming from several different galaxy sightlines, where the number of 
contributing galaxies depends on the impact parameter (see Table~\ref{tab:bins}). The median pixel optical depth 
in each bin was then taken to construct Figure~\ref{fig:all_maps_b}, where the images were smoothed by a 
Gaussian  with a $\sigma$ equal to the bin size. The right $y$-axis indicates the LOS velocity difference 
assuming the median redshift of the galaxies in our sample, $z=2.34$. The (unsmoothed) optical depth values
 in each bin, plus their $1\sigma$ errors, are given in Table~\ref{tab:maps}.  

We note that the optical depth scales on the maps presented in Figure~\ref{fig:all_maps_b} 
run from the median optical depth of all 
 pixels  in the redshift range considered for the particular ion and recovery method
(which we denote as \taumedk\ for each ion $Z$), up to the
maximum value of all smoothed pixels in the image, in order
to maximise the dynamic range. It is therefore advised that care be taken in the
interpretation of these maps, as the true dynamic range of any one species may not be properly 
captured since the minimum of the range (set by \taumedk) is set by the 
contamination and/or shot noise. 
We list \taumedk\ (as well as the median S/N for all considered pixels)
for each ion in Table~\ref{tab:taumed}. 

One should also keep in mind that
although these maps are constructed by situating the galaxies at the origin,
at large distances the median absorption is likely affected by gas
near neighbouring galaxies, most of which will be undetected. 
These maps should be interpreted as the average properties
of gas around galaxies, which include the effects of clustering. 
We elaborate further on this point in \S~\ref{sec:fcover}. 

Before further discussing Figure~\ref{fig:all_maps_b}, we 
first summarise the results from a similar analysis of 
\hone\ from \citet{rakic12}, using an earlier sample from the KBSS
(see also \citealt{rudie12}). We note that the 2-D \hone\ median optical
depth map presented in Figure~6 of \citet{rakic12} is analogous 
to the top left panel of Figure~\ref{fig:all_maps_b} in this work 
(but created with a slightly different galaxy sample). 
As described in \citet{rakic12}, the first point to note from these figures is the
strongly enhanced absorption extending  $\sim10^2$~pkpc in the transverse direction
and $\sim1$~pMpc ($\sim240$~\kmps) along the LOS. They concluded that this redshift
space distortion, often called the ``finger of God'' effect, could have two origins. 
Firstly, the redshift estimates of the galaxies have associated errors 
that smear the signal along the LOS. Specifically, 
the errors are roughly $\Delta v \approx150$~\kmps\ for LRIS, 
$\Delta v \approx60$~\kmps\ for NIRSPEC, and $\Delta v \approx18$~\kmps\ for 
MOSFIRE redshifts. Additionally, peculiar velocities
of the gas arising from infall, outflows or virial motions 
may be responsible for this effect, particularly 
because the extent of the elongation is greater 
than that which one may expect purely from redshift errors.

The second main result from the 2-D \hone\ median optical
depth map studied in \citet{rakic12} (again, see their Figure~6 or
the top left panel of our Figure~\ref{fig:all_maps_b})
was the presence of an anisotropy on large scales. They noticed that while
\hone\ absorption was enhanced out to the maximum considered impact 
parameter of 2~pMpc, along the LOS it already dropped off at 
$\approx1.5$~pMpc ($\approx300$~\kmps). Such a feature could be
attributed to the \citet{kaiser87} effect, originally defined as
Doppler shifts in galaxy redshifts caused by the large-scale coherent
motions of the galaxies towards cluster centres, which  
manifest as LOS distortions. Although in this work we are not specifically 
looking at clusters, any large-scale coherent motions should produce a similar effect. 
In the case of \citet{rakic12}, since one would expect
redshift space distortions caused by measurement errors
to elongate the signal, the observed compression is likely due to
peculiar velocities of infalling gas. Indeed, \citet{rakic13}
used simulations to show that the observed anisotropy is consistent with 
being due to the Kaiser effect. However, we note that a
similar study at $z\sim3$ by \citet{tummuangpak13} examining the \lya-galaxy
correlation function found the observed infall to be 
smaller than would be predicted by simulations.

We emphasise here that the typical galaxy redshift error of our sample
has changed substantially since \citet{rakic12}. Most galaxies in the innermost
bins now have redshifts measured using nebular lines, while the converse was true
for \citet[see their Figure~1]{rakic12}. However,
as we show in Appendix~\ref{sec:galsample_appendix},
 the extent of the elongation of 
the optical depth signal along the LOS direction has not changed significantly 
between the two samples, even though the redshift errors are now considerably smaller.
This suggests that the main origin of the small-scale 
anisotropy is gas peculiar velocities
(see \S~\ref{sec:var_gal_sample} for further discussion). Note also that 
the large-scale anisotropy detected by \citet{rakic12} is also present here,
as can be seen most clearly from the top left panel of Figure~\ref{fig:zspacedist}
(compare the last red point with the black point at the same distance).

While the goal of \citet{rakic12} was to study \hone, in this work
we are extending the analysis to metal ions.
In Figure~\ref{fig:all_maps_b} we compare the 2-D \hone\ optical 
depth distribution to those of the metals. 
In all cases we see a strong central enhancement of absorption, 
which is elongated along the LOS. For a more quantitative picture,
we have made ``cuts'' through the maps, 
and plotted the (unsmoothed) optical depth values from the nearest
LOS bin (red squares) and the innermost transverse distance bin 
(black circles) for each ion in Figure~\ref{fig:zspacedist} 
(see also Table~\ref{tab:maps}). Note that the first bin is 
identical for the transverse and LOS directions. 
For every ion except for \nfive\ (and one data point in \sifour), the optical depths 
of the data points in the LOS direction are greater than those in the transverse 
direction (i.e., the black points are above the red points) 
for bins 2--7, or out to $\sim1$~pMpc.

The discrepancy between \nfive\ and the other ions 
is likely due to the fact that \nfive\ is both relatively weak and
has \hone\ contamination that is difficult to correct for, 
and not necessarily due to intrinsically 
different redshift-space structure. This is supported by the fact that the
dynamic range of optical depths probed for \nfive\ is much smaller than 
those for the other ions. 

In Figure~\ref{fig:zspacedist} we show the median optical depth of 
a random region (horizontal dashed line) as well as the $1\sigma$ scatter
(horizontal dotted line). From this, it is clear that the scatter is 
quite large relative to many of the detected optical depth enhancements. 
The full distribution of pixel optical depths is discussed
in more detail in \S~\ref{sec:odhist}.

We estimate the confidence levels associated with these measurements
by bootstrap resampling the galaxies in each impact parameter bin.
In each bootstrap realisation we 
randomise the galaxy redshifts and compute the 
median optical depth as a function of transverse or LOS distance for both the original
and randomised galaxy redshifts. Then, in each transverse or LOS distance bin, we calculate 
the fraction of bootstrap realisations for which the median optical depth computed 
using galaxies with randomised redshifts is higher than the median optical depth 
computed using the original galaxy redshifts. 
From this, we determine that for \nfive, only in the innermost bin
is the optical depth enhancement detected with a confidence level  
$\geq68$\% ($1\sigma$). 

\begin{figure*}
 \includegraphics[width=0.7\textwidth]{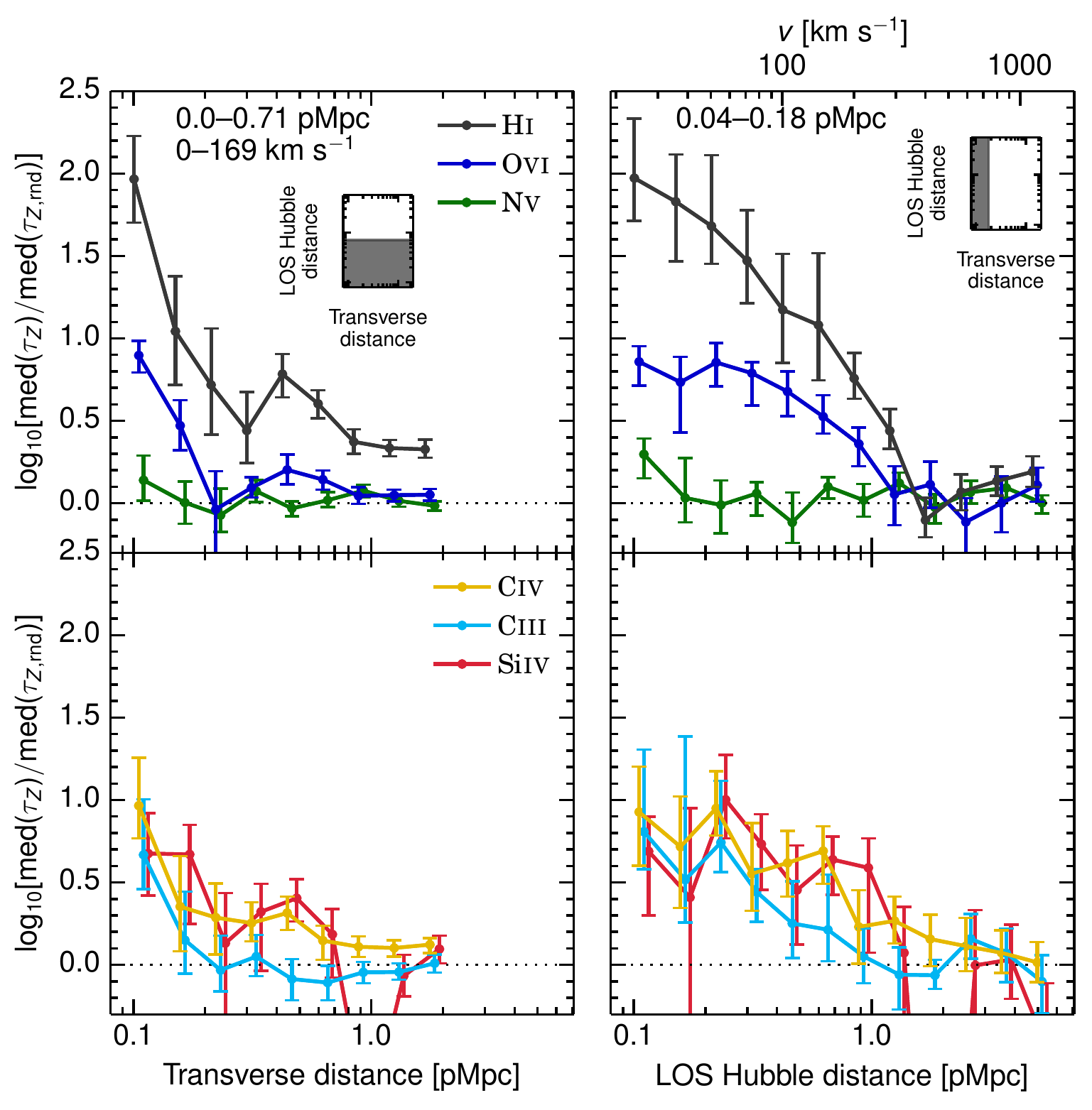}
\caption{Cuts through the (unsmoothed) median
optical depth maps from Figure~\ref{fig:all_maps_b}
along the transverse (left) and LOS (right) directions. All optical
depths were divided by their corresponding \taumedk\ so that the medians of 
all ions are aligned, and denoted by the horizontal dotted black line. 
For clarity, the different ions have been divided 
between the top and bottom panels, and the points have been offset from 
each other horizontally by 0.02~dex. The shaded strip in the inset shows the 
region included in each cut. Except for \nfive, the optical
depths of all ions are enhanced out to at least 180~pkpc  
in the transverse direction and $\sim1$~pMpc ($\sim240$~\kmps) along the LOS.
 Like \hone, the optical depth of \cfour\ is
significantly enhanced along the entire extent of the transverse direction
(bottom left panel).}
\label{fig:cuts_a}
\end{figure*}

\begin{table*}
\caption{The median $\log_{10}$\tauk\ and $1\sigma$ errors determined by taking cuts through Figure~\ref{fig:all_maps_b}. 
Unlike in Figure~\ref{fig:zspacedist}, bins have now been combined in order to reduce the amount of noise and 
facilitate comparisons between different ions, and the results are plotted in Figure~\ref{fig:cuts_a}. Specifically, 
the values along the transverse direction (top portion of the table) are measured from 
a  170~\kmps\ (0.71~pMpc, or combining the first six bins) cut along the LOS, while the values 
along the LOS (bottom portion of the table)
are calculated from a 0.18~pMpc (combining the first two bins) transverse direction cut.
For comparison, the bottom row of the table lists the median optical depths measured 
over the full redshift range of each ion (\taumedk). We note that 
$\log_{10}\tau_{\sifourm}=-6.00$ is a flag value which indicate that the median
optical depth in that bin is negative. 
}
\label{tab:cuts}
\begin{tabular}{@{}rlrrrrrr}
\hline\hline
 & $D_{\rm trans}$ (pMpc) & \multicolumn{1}{c}{\hone} & \multicolumn{1}{c}{\osix} & \multicolumn{1}{c}{\nfive} & \multicolumn{1}{c}{\cfour} & \multicolumn{1}{c}{\cthree} & \multicolumn{1}{c}{\sifour} \\
\hline
 & 0.04--0.13	& $0.68_{-0.26}^{+0.26}$	& $-0.74_{-0.10}^{+0.09}$	& $-1.43_{-0.12}^{+0.15}$	& $-1.99_{-0.20}^{+0.29}$	& $-0.60_{-0.21}^{+0.34}$	& $-2.49_{-0.26}^{+0.24}$ \\ 
 & 0.13--0.18	& $-0.25_{-0.33}^{+0.33}$	& $-1.17_{-0.15}^{+0.15}$	& $-1.56_{-0.13}^{+0.13}$	& $-2.60_{-0.27}^{+0.31}$	& $-1.12_{-0.20}^{+0.30}$	& $-2.49_{-0.42}^{+0.18}$ \\ 
 & 0.18--0.25	& $-0.57_{-0.30}^{+0.34}$	& $-1.67_{-0.34}^{+0.23}$	& $-1.64_{-0.10}^{+0.16}$	& $-2.67_{-0.22}^{+0.21}$	& $-1.30_{-0.13}^{+0.21}$	& $-3.03_{-0.51}^{+0.30}$ \\ 
 & 0.25--0.36	& $-0.85_{-0.20}^{+0.23}$	& $-1.54_{-0.06}^{+0.06}$	& $-1.50_{-0.07}^{+0.07}$	& $-2.70_{-0.11}^{+0.13}$	& $-1.22_{-0.12}^{+0.13}$	& $-2.84_{-0.36}^{+0.17}$ \\ 
 & 0.36--0.50	& $-0.51_{-0.14}^{+0.12}$	& $-1.44_{-0.09}^{+0.09}$	& $-1.60_{-0.05}^{+0.04}$	& $-2.64_{-0.10}^{+0.10}$	& $-1.35_{-0.13}^{+0.12}$	& $-2.76_{-0.14}^{+0.12}$ \\ 
 & 0.50--0.71	& $-0.69_{-0.09}^{+0.08}$	& $-1.49_{-0.06}^{+0.05}$	& $-1.55_{-0.04}^{+0.05}$	& $-2.81_{-0.12}^{+0.09}$	& $-1.38_{-0.11}^{+0.10}$	& $-2.98_{-0.26}^{+0.15}$ \\ 
 & 0.71--1.00	& $-0.92_{-0.07}^{+0.08}$	& $-1.59_{-0.05}^{+0.05}$	& $-1.49_{-0.04}^{+0.04}$	& $-2.84_{-0.06}^{+0.06}$	& $-1.31_{-0.07}^{+0.06}$	& $-4.13_{-1.87}^{+0.55}$ \\ 
 & 1.00--1.42	& $-0.95_{-0.05}^{+0.05}$	& $-1.59_{-0.04}^{+0.03}$	& $-1.55_{-0.04}^{+0.03}$	& $-2.85_{-0.05}^{+0.05}$	& $-1.31_{-0.05}^{+0.05}$	& $-3.23_{-0.12}^{+0.13}$ \\ 
 & 1.42--2.00	& $-0.96_{-0.05}^{+0.06}$	& $-1.59_{-0.04}^{+0.04}$	& $-1.58_{-0.03}^{+0.03}$	& $-2.83_{-0.05}^{+0.04}$	& $-1.26_{-0.06}^{+0.05}$	& $-3.06_{-0.09}^{+0.08}$ \\ 
\hline\hline
$v_{\rm LOS}$ (\kmps) & $D_{\rm LOS}$ (pMpc)
 & \multicolumn{1}{c}{\hone} & \multicolumn{1}{c}{\osix} & \multicolumn{1}{c}{\nfive} & \multicolumn{1}{c}{\cfour} & \multicolumn{1}{c}{\cthree} & \multicolumn{1}{c}{\sifour}\\
\hline
  0--30 & 0.00--0.13	& $0.68_{-0.26}^{+0.36}$	& $-0.78_{-0.15}^{+0.09}$	& $-1.27_{-0.15}^{+0.10}$	& $-2.03_{-0.33}^{+0.28}$	& $-0.46_{-0.23}^{+0.50}$	& $-2.47_{-0.39}^{+0.21}$ \\ 
 30--42 & 0.13--0.18	& $0.54_{-0.36}^{+0.29}$	& $-0.90_{-0.31}^{+0.15}$	& $-1.54_{-0.15}^{+0.24}$	& $-2.24_{-0.37}^{+0.31}$	& $-0.75_{-0.26}^{+0.87}$	& $-2.75_{-3.25}^{+0.54}$ \\ 
 42--60 & 0.18--0.25	& $0.39_{-0.23}^{+0.43}$	& $-0.78_{-0.15}^{+0.12}$	& $-1.58_{-0.17}^{+0.15}$	& $-2.00_{-0.16}^{+0.22}$	& $-0.53_{-0.18}^{+0.37}$	& $-2.16_{-0.23}^{+0.27}$ \\ 
 60--84 & 0.25--0.36	& $0.18_{-0.26}^{+0.31}$	& $-0.85_{-0.20}^{+0.07}$	& $-1.51_{-0.13}^{+0.07}$	& $-2.40_{-0.23}^{+0.31}$	& $-0.83_{-0.18}^{+0.14}$	& $-2.43_{-0.28}^{+0.18}$ \\ 
 84--119 & 0.36--0.50	& $-0.12_{-0.32}^{+0.34}$	& $-0.96_{-0.15}^{+0.12}$	& $-1.69_{-0.12}^{+0.18}$	& $-2.33_{-0.20}^{+0.19}$	& $-1.02_{-0.21}^{+0.26}$	& $-2.71_{-0.33}^{+0.27}$ \\ 
119--169 & 0.50--0.71	& $-0.21_{-0.34}^{+0.43}$	& $-1.11_{-0.10}^{+0.13}$	& $-1.47_{-0.07}^{+0.06}$	& $-2.26_{-0.20}^{+0.15}$	& $-1.06_{-0.19}^{+0.33}$	& $-2.52_{-0.21}^{+0.14}$ \\ 
169--238 & 0.71--1.00	& $-0.53_{-0.12}^{+0.15}$	& $-1.28_{-0.14}^{+0.10}$	& $-1.55_{-0.10}^{+0.10}$	& $-2.72_{-0.22}^{+0.22}$	& $-1.22_{-0.16}^{+0.17}$	& $-2.57_{-0.51}^{+0.18}$ \\ 
238--336 & 1.00--1.42	& $-0.85_{-0.11}^{+0.13}$	& $-1.58_{-0.19}^{+0.17}$	& $-1.45_{-0.06}^{+0.06}$	& $-2.69_{-0.14}^{+0.15}$	& $-1.33_{-0.21}^{+0.17}$	& $-3.09_{-0.72}^{+0.28}$ \\ 
336--475 & 1.42--2.00	& $-1.39_{-0.10}^{+0.13}$	& $-1.52_{-0.14}^{+0.14}$	& $-1.59_{-0.10}^{+0.08}$	& $-2.80_{-0.16}^{+0.15}$	& $-1.33_{-0.08}^{+0.09}$	& $-6.00_{-0.00}^{+2.41}$ \\ 
475--671 & 2.00--2.83	& $-1.22_{-0.11}^{+0.11}$	& $-1.75_{-0.21}^{+0.14}$	& $-1.50_{-0.07}^{+0.07}$	& $-2.84_{-0.15}^{+0.17}$	& $-1.11_{-0.12}^{+0.15}$	& $-3.16_{-1.35}^{+0.34}$ \\ 
671--948 & 2.83--3.99	& $-1.15_{-0.09}^{+0.09}$	& $-1.64_{-0.18}^{+0.16}$	& $-1.48_{-0.06}^{+0.07}$	& $-2.88_{-0.12}^{+0.14}$	& $-1.20_{-0.17}^{+0.15}$	& $-3.13_{-0.24}^{+0.21}$ \\ 
948--1339 & 3.99--5.64	& $-1.10_{-0.09}^{+0.09}$	& $-1.53_{-0.10}^{+0.11}$	& $-1.57_{-0.06}^{+0.05}$	& $-2.94_{-0.12}^{+0.12}$	& $-1.37_{-0.19}^{+0.16}$	& $-3.67_{-2.33}^{+0.40}$ \\ 
\hline\hline
& $\log_{10}(\tau_{Z,{\rm rnd}})$	& \multicolumn{1}{l}{$-1.29$}	& \multicolumn{1}{l}{$-1.64$}	& \multicolumn{1}{l}{$-1.57$}	& \multicolumn{1}{l}{$-2.95$}	& \multicolumn{1}{l}{$-1.27$}	& \multicolumn{1}{l}{$-3.16$} \\ 
\hline
\end{tabular}
\end{table*}

In Figure~\ref{fig:cuts_a} (Table~\ref{tab:cuts}) we now make larger cuts through the 
maps from Figure~\ref{fig:all_maps_b}, and divide the results for each ion by 
\taumedk\ in order to compare the ions with each other in the same figure.
Since the innermost impact parameter bin contains only \ngalinner\ galaxies, and because 
we see an enhanced optical depth signal out to larger distances, 
for the cut along the LOS direction we combine the two innermost impact parameter bins
to create a single bin out to 0.18~pMpc. 
In the transverse direction, we combine the six smallest bins, which leads to a cut size 
of $\pm170$~\kmps\ (or 0.71~pMpc). A similar bin size was previously motivated by 
\citet{rakic11}\footnote{Due to a slightly different cosmology and median galaxy redshift 
used in \citet{rakic12}, the velocity interval corresponding to 0.71~pMpc was closer to 
$\pm165$~\kmps.}
because it is the scale over which $\tau_{\honem}$ is smooth in 
the LOS direction, and because 
they found errors between points along LOS direction (but not in the 
transverse direction) to be correlated on smaller scales (i.e. for separations 
$\lesssim100$~\kmps; see Appendix~B of \citealt{rakic12}). 
We note that the conclusions from this section are not affected by the precise size of the cut.

From the right panels of Figure~\ref{fig:cuts_a} we can see that in general, along the LOS, 
the optical depth enhancement is very small or undetectable beyond 
$\sim240$~\kmps\ ($\sim1$~pMpc) for both \hone\ and the metals.
In the transverse direction only \hone\ and \cfour\ remain
significantly enhanced above the noise level for all 
impact parameters. Calculating confidence intervals as before, we find
a $\geq84$\% confidence level for \hone\ in every bin, and a $\geq68$\% (i.e., $1\sigma$)
confidence level for \cfour\ at every point except for the sixth bin (0.50--0.71~pMpc) 
which has a confidence level of 55\%.

It is important to mention that the non-detection of this
large-scale enhancement for the other the metal ions may be due to sensitivity
limits rather than a true paucity of these species at larger impact parameters.
Due either to contamination (for ions
with rest wavelengths shortward of \lya) or shot noise and 
continuum fitting errors (for \sifour), we are likely 
unable to probe optical depths down to the true median level
for these ions.

\begin{figure*}
\includegraphics[width=0.7\textwidth]{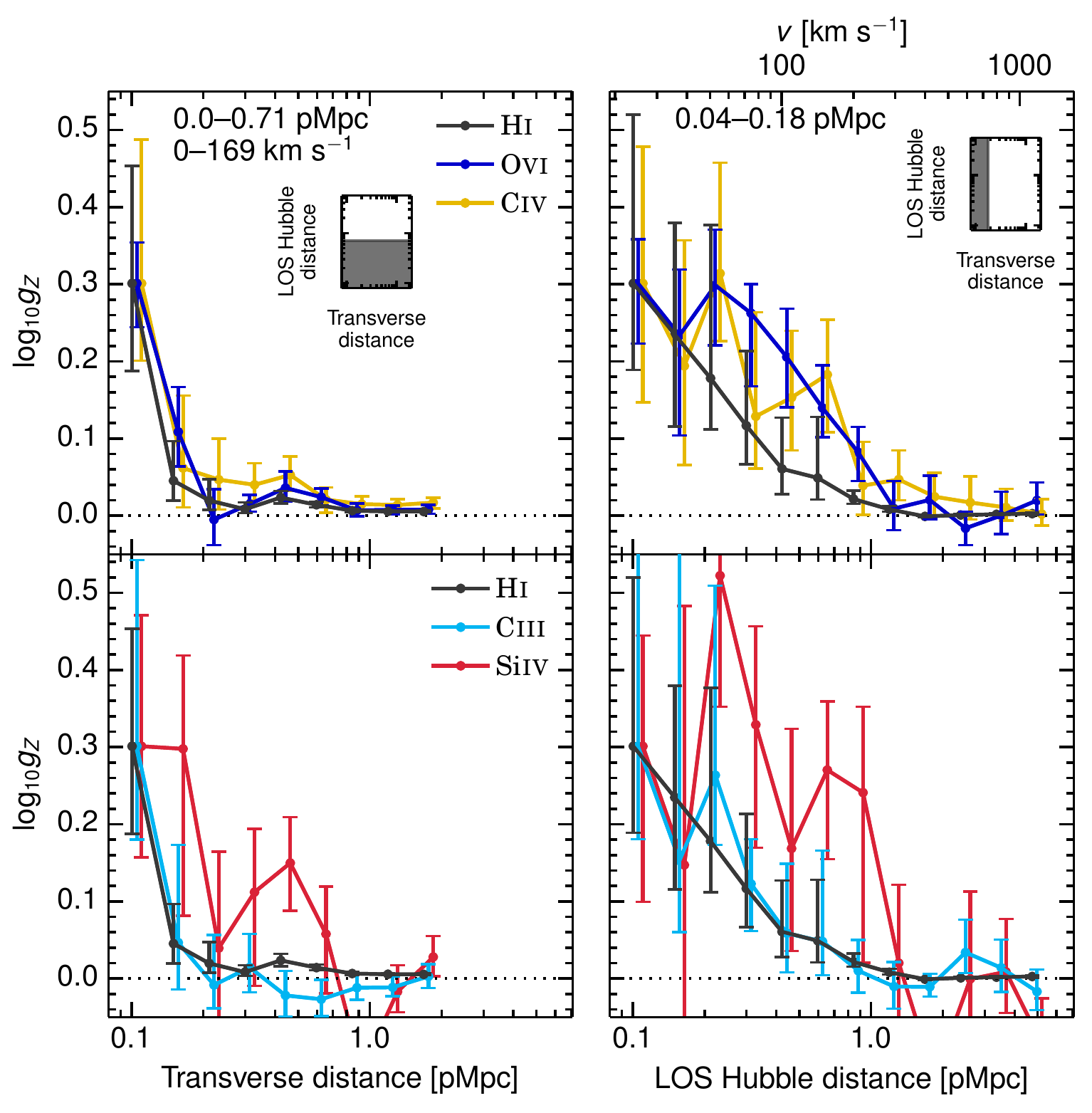}
\caption{The same as Figure~\ref{fig:cuts_a}, but after re-scaling
the optical depths using Equation~\ref{eq:norm2} to account for differences
in transition strengths and detection limits. Note that \hone\ is shown
in both the top and bottom panels, while \nfive\ has been removed. In general, the shapes
of the rescaled metal optical depth profiles appear similar to that of \hone,
although \osix, \cfour\ and \sifour\ are somewhat enhanced relative to \hone\ along 
the LOS direction.}
\label{fig:cuts_b}
\end{figure*}

\subsection{Rescaled profiles}
\label{sec:profiles}
Next we consider the distribution of the metals 
presented in our optical depth maps and compare them to 
\hone\ and to each other, in an effort to explore whether the measured
absorption signals trace the same gaseous structure.
Of course, the metal optical depths
are set not only by the metal abundance, but
are also determined by the varying degree of ionisation and hence
by the density and temperature of the gas in which they reside.

With this point in mind, we explore whether the 
optical depth profiles of the different ions arise from the same intrinsic 
functional form, i.e. whether their spatial (in the transverse direction)
or kinematic (in the LOS direction) profiles have the same shape. 
However, as 
Figures~\ref{fig:all_maps_b}, \ref{fig:zspacedist}, and \ref{fig:cuts_a} suggest, the comparison
is not straightforward, as some metals have 
a very small dynamic range where the median optical depth (set by the detection limit) 
lies close to the maximum binned optical depth value. We therefore propose a
method to normalise the curves to alleviate this problem. 
 
If all ions follow the same intrinsic profile, but appear to vary 
because the transitions through which 
they are observed have different strengths and because they 
suffer from different levels of contamination and noise, 
then for each ion $Z$, the median optical depth profile can be written as
\begin{equation}
 f_{Z}(x) = a_Z f(x) + {\rm med}_Z
\label{eq:norm1}
\end{equation}
where $f_{Z}(x)$ is the observed, apparent optical depth of species $Z$,
$f(x)$ is the intrinsic profile (which varies from 
$\max(f)$ at $x=0$ to $0$ as $x\rightarrow\infty$),
$a_Z$ is a scale factor that sets the relative strengths of different transitions, 
and med$_Z$ is the median optical depth to which the curve asymptotes 
(i.e. the detection limit set by noise and contamination).
For \cfour\ and \sifour, this value is set mostly by shot noise and continuum 
fitting errors, while
for the other transitions it is mainly unrelated absorption (i.e. contamination).

Although the values of $a_Z$ and med$_Z$ will certainly depend on 
the ion in question, it is possible that the intrinsic 
profile $f(x)$ does not vary between different ions. 
To examine this, we can try and normalise the curves such that, if
indeed $f(x)$ does not depend on the ion $Z$,
then the normalised curves should also be equivalent for different ions.
This can be achieved using the following transformation:
\begin{equation}
 g_Z(x) = \dfrac{f_{Z}(x)-{\rm med}_Z}{\max_Z - {\rm med}_Z} + 1
\label{eq:norm2}
\end{equation}
 where $\max_Z = f_{Z}(x_{\max})$ and $x_{\max}$ is the value of 
 $x$ at which the maximum value of $f_{Z}(x)$ occurs. 
 For the optical depth profiles, 
 $x_{\max}$ is usually the innermost transverse or LOS distance bin. 
 Combining Equations~\ref{eq:norm1} and \ref{eq:norm2}, we obtain the expression
  $g_Z(x) = f(x) / f(x_{\max}) +1$, which is independent of $Z$ 
  and varies between $1$ and $2$. 
In summary, if the observed spatial and/or kinematic optical depth profiles 
of different ions are all the same apart from a multiplicative 
factor reflecting the strength of the transition and an additional
constant reflecting the levels of contamination and noise, then
by using the transformation in Equation~\ref{eq:norm2}, the resulting
curves should all overlap each other. 

We implemented the above by taking med$_k$ to be \taumedk,
and max$_k$ to be the optical depth of the innermost point of the curve in question
(the values differ along the LOS and transverse directions, due to the different
cut sizes taken).
The resulting curves are shown in Figure~\ref{fig:cuts_b}.
Note that the normalisation method will amplify any small point-to-point
variations inversely with dynamic range. Since \nfive\ has a relatively
small dynamic range, variations which appear small in the top left panel
(particularly the first few points along the transverse direction) of 
Figure~\ref{fig:cuts_a} manifest themselves as large deviations
compared to the other curves, and for this reason we have chosen to omit
\nfive\ from Figure~\ref{fig:cuts_b}.

The curves resulting from cuts along the transverse direction 
(left panels of Figure~\ref{fig:cuts_b}) appear quite similar 
for all species observed. In every case
(including \hone), the optical depth drops sharply after the first impact parameter bin, 
and quickly asymptotes to the median. 
The plateau in the absorption at impact parameters 180~pkpc--2~pMpc that 
was clearly detected for \hone\ and \cfour\ in Figures~\ref{fig:all_maps_b} and \ref{fig:cuts_a} may
therefore also be present for the other species, but cannot be detected due to the smaller 
dynamic range in their recovered optical depths. 
Along the LOS (right panels of  Figure~\ref{fig:cuts_b}), 
the situation is less uniform.
While \cthree\ tends to trace the \hone\ profile,  
\osix, \cfour\ and \sifour\ appear to be
\textit{more enhanced} than \hone\ out to $\sim1$~pMpc or 240~\kmps. 
This result suggests that we might be seeing two different gas phases,
i.e. a relatively compact one traced mainly by \hone\ and \cthree,
and a phase that is more extented in real and/or velocity space and that is 
traced by \osix, \cfour\ and \sifour.

We again point out that any fluctuations in the normalised optical 
depth values for the metal ions are magnified in these figures, due to the
lower dynamic range of optical depths. Although this makes it difficult to 
draw secure conclusions about any observed differences in the distributions of 
the metals with respect to each other and to \hone,
we can at least say that the strong \hone\ enhancement seen along the LOS
direction is also present for the metal ions \osix, \cfour, \cthree\ and \sifour.

\begin{figure*}
 \includegraphics[width=0.9\textwidth]{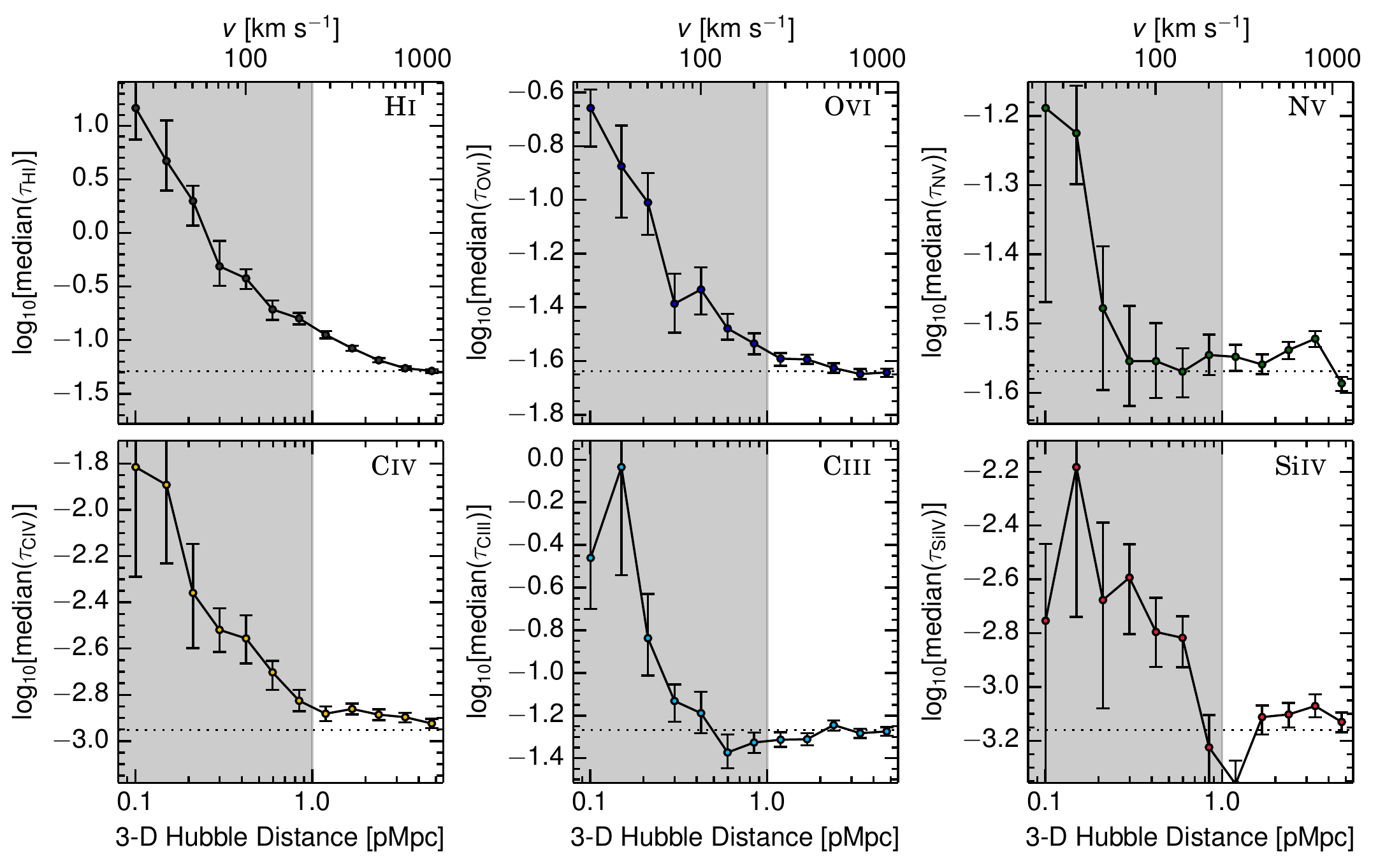}
 \caption{The median optical depth as a function of 3-D Hubble distance for all galaxies 
in our sample, assuming that velocity differences between the galaxies 
and associated absorption regions are purely
due to Hubble flow. This assumption is certainly incorrect for $\lesssim1$~pMpc, and 
we have therefore greyed out the regions that we believe are effected by redshift-space 
distortions. The horizontal dotted black lines indicate the median optical depth 
for all pixels of the ion shown. One advantage of examining 3-D Hubble distance is that
it allows us to decrease the noise by combining large numbers of pixels, particularly at 
large distances. As a result, the enhancement in the median optical 
depth of \cfour\ can now be detected out to $\sim4$~pMpc. }
\label{fig:hubble_a}
\end{figure*}

\begin{table*}
\caption{The median $\log_{10}$\tauk\ and $1\sigma$ errors as a function of 3-D Hubble distance (left column). Columns 2--7
 display values for the different ions $Z$ studied in this work, and the data are plotted in
 Figure~\ref{fig:hubble_a}.
 For comparison, the bottom row of the table lists the median optical depths measured 
over the full redshift range of each ion (\taumedk).
 }
\label{tab:hubble}
\begin{tabular}{@{}rlrrrrrr}
\hline\hline
$v_{\rm Hubble}$ (\kmps) & $D_{\rm Hubble}$ (pMpc) & \multicolumn{1}{c}{\hone} & \multicolumn{1}{c}{\osix} & \multicolumn{1}{c}{\nfive} & \multicolumn{1}{c}{\cfour} & \multicolumn{1}{c}{\cthree} & \multicolumn{1}{c}{\sifour} \\
\hline
  0--30 & 0.00--0.13	& $1.16_{-0.29}^{+1.39}$	& $-0.66_{-0.14}^{+0.07}$	& $-1.19_{-0.28}^{+0.11}$	& $-1.81_{-0.47}^{+0.63}$	& $-0.46_{-0.24}^{+4.46}$	& $-2.75_{-3.25}^{+0.29}$ \\ 
 30--42 & 0.13--0.18	& $0.67_{-0.28}^{+0.38}$	& $-0.88_{-0.19}^{+0.15}$	& $-1.22_{-0.07}^{+0.07}$	& $-1.89_{-0.34}^{+0.26}$	& $-0.03_{-0.51}^{+0.38}$	& $-2.18_{-0.56}^{+0.30}$ \\ 
 42--60 & 0.18--0.25	& $0.30_{-0.23}^{+0.14}$	& $-1.01_{-0.12}^{+0.11}$	& $-1.48_{-0.12}^{+0.09}$	& $-2.36_{-0.24}^{+0.21}$	& $-0.84_{-0.17}^{+0.21}$	& $-2.68_{-0.40}^{+0.29}$ \\ 
 60--84 & 0.25--0.36	& $-0.31_{-0.18}^{+0.24}$	& $-1.39_{-0.11}^{+0.11}$	& $-1.55_{-0.06}^{+0.08}$	& $-2.52_{-0.10}^{+0.09}$	& $-1.13_{-0.10}^{+0.08}$	& $-2.59_{-0.21}^{+0.12}$ \\ 
 84--119 & 0.36--0.50	& $-0.42_{-0.10}^{+0.08}$	& $-1.33_{-0.09}^{+0.08}$	& $-1.55_{-0.05}^{+0.06}$	& $-2.56_{-0.11}^{+0.10}$	& $-1.19_{-0.09}^{+0.10}$	& $-2.80_{-0.13}^{+0.13}$ \\ 
119--169 & 0.50--0.71	& $-0.71_{-0.10}^{+0.08}$	& $-1.48_{-0.04}^{+0.05}$	& $-1.57_{-0.04}^{+0.03}$	& $-2.70_{-0.08}^{+0.05}$	& $-1.37_{-0.07}^{+0.09}$	& $-2.82_{-0.11}^{+0.08}$ \\ 
169--238 & 0.71--1.00	& $-0.80_{-0.06}^{+0.05}$	& $-1.53_{-0.04}^{+0.04}$	& $-1.55_{-0.03}^{+0.03}$	& $-2.83_{-0.05}^{+0.05}$	& $-1.33_{-0.05}^{+0.05}$	& $-3.22_{-0.17}^{+0.12}$ \\ 
238--336 & 1.00--1.42	& $-0.95_{-0.03}^{+0.03}$	& $-1.59_{-0.03}^{+0.02}$	& $-1.55_{-0.02}^{+0.02}$	& $-2.88_{-0.03}^{+0.03}$	& $-1.31_{-0.03}^{+0.03}$	& $-3.36_{-0.17}^{+0.09}$ \\ 
336--475 & 1.42--2.00	& $-1.08_{-0.02}^{+0.03}$	& $-1.59_{-0.02}^{+0.02}$	& $-1.56_{-0.01}^{+0.01}$	& $-2.86_{-0.02}^{+0.02}$	& $-1.31_{-0.03}^{+0.03}$	& $-3.11_{-0.07}^{+0.04}$ \\ 
475--671 & 2.00--2.83	& $-1.19_{-0.02}^{+0.02}$	& $-1.63_{-0.02}^{+0.02}$	& $-1.54_{-0.01}^{+0.01}$	& $-2.89_{-0.02}^{+0.02}$	& $-1.24_{-0.03}^{+0.02}$	& $-3.10_{-0.05}^{+0.04}$ \\ 
671--948 & 2.83--3.99	& $-1.26_{-0.02}^{+0.02}$	& $-1.65_{-0.02}^{+0.02}$	& $-1.52_{-0.01}^{+0.01}$	& $-2.90_{-0.02}^{+0.02}$	& $-1.28_{-0.02}^{+0.02}$	& $-3.07_{-0.04}^{+0.04}$ \\ 
948--1339 & 3.99--5.64	& $-1.29_{-0.02}^{+0.02}$	& $-1.64_{-0.02}^{+0.01}$	& $-1.59_{-0.01}^{+0.01}$	& $-2.92_{-0.02}^{+0.02}$	& $-1.27_{-0.02}^{+0.02}$	& $-3.13_{-0.04}^{+0.04}$ \\ 
\hline\hline
& $\log_{10}(\tau_{Z,{\rm rnd}})$	& \multicolumn{1}{l}{$-1.29$}	& \multicolumn{1}{l}{$-1.64$}	& \multicolumn{1}{l}{$-1.57$}	& \multicolumn{1}{l}{$-2.95$}	& \multicolumn{1}{l}{$-1.27$}	& \multicolumn{1}{l}{$-3.16$} \\ 
\hline
\end{tabular}
\end{table*}

\subsection{3-D Hubble distance}
 \label{sec:hubble}

If we assume that the velocity differences between galaxies and the nearby
absorption regions in the quasar are due only to Hubble flow, then we can calculate
the 3-D Hubble distance as $\sqrt{b^2 + (H(z)\Delta v)^2}$, where $b$ is the 
galaxy impact parameter, $H(z)$ is the Hubble parameter, and $\Delta v$ is the
LOS separation between the absorber and the galaxy. 
We have computed the 
3-D Hubble distance for every galaxy-pixel pair, divided them into 0.15~dex
distance bins as in Figure~\ref{fig:all_maps_b}, and taken the median optical 
depth in every bin. The result is plotted in Figure~\ref{fig:hubble_a},
and the data are given in Table~\ref{tab:hubble}.
We stress that this metric is a poor approximation for 
distances $\lesssim1$~pMpc since Figure~\ref{fig:zspacedist} 
revealed strong anisotropies along the LOS on this scale, and we have 
shaded the poor-approximation region grey in Figure~\ref{fig:hubble_a}.

Although using the Hubble distance results in the loss of some
important information, by using this projection
we are able to achieve higher S/N ratios because each 3-D 
bin contains many more pixels than those constructed from
cuts through the median optical depth maps.
Specifically, we are able see a significant enhancement of 
optical depth above the median for \hone\ out 
to $\sim2.8$~pMpc (99.5\% confidence)  and for \cfour\ out to $\sim4.0$~pMpc
(90\% confidence)\footnote{We note that in the eighth 3-D Hubble distance bin,
 which runs from 1.0--1.4~pMpc, we only detect the \cfour\ enhancement with a 
confidence level of 80\%.}.
The latter result suggests that the enhancement of \cfour\ above the median 
seen throughout the 2-D map in Figure~\ref{fig:all_maps_b} is a real effect.
We note that for \nfive, the enhancement seen at $\sim900$~\kmps\ is likely 
due to residual contamination from its own doublet.

\subsection{Covering fraction}
\label{sec:fcover}

\begin{table*}
\caption{Covering fraction and $1\sigma$ errors as a function of transverse distance (top row),
 which is defined as the fraction of galaxies in each impact parameter bin that
  have a median optical depth within $\pm170$~\kmps\ above some threshold value. 
  The threshold values are set by the optical depth at which the covering fraction 
  of 1000 random $\pm170$~\kmps\, regions, $f_{\rm IGM}$,  are equal to 
   0.25, 0.05, and 0.01 (second column),
    and the results are plotted in Figure~\ref{fig:fcover}.}
\label{tab:fcover}
\begin{tabular}{@{}lcccccccccc}
\hline\hline
\multicolumn{2}{r}{$D_{\rm trans}$ (pMpc)} & 	0.04--0.13 & 	0.13--0.18 & 	0.18--0.25 & 	0.25--0.36 & 	0.36--0.50 & 	0.50--0.71 & 	0.71--1.00 & 	1.00--1.42 & 	1.42--2.00\\
Ion & $f_{\rm IGM}$ & & & & & & & & & \\\hline
  & 0.25	& $1.00_{-0.00}^{+0.00}$	& $0.80_{-0.10}^{+0.10}$	& $0.64_{-0.18}^{+0.18}$	& $0.54_{-0.12}^{+0.08}$	& $0.61_{-0.07}^{+0.07}$	& $0.57_{-0.05}^{+0.05}$	& $0.43_{-0.04}^{+0.05}$	& $0.45_{-0.03}^{+0.03}$	& $0.45_{-0.03}^{+0.03}$ \\ 
\hone & 0.05	& $0.83_{-0.08}^{+0.08}$	& $0.50_{-0.20}^{+0.10}$	& $0.27_{-0.09}^{+0.09}$	& $0.19_{-0.08}^{+0.08}$	& $0.25_{-0.05}^{+0.05}$	& $0.19_{-0.04}^{+0.04}$	& $0.14_{-0.03}^{+0.03}$	& $0.13_{-0.02}^{+0.02}$	& $0.14_{-0.02}^{+0.02}$ \\ 
  & 0.01	& $0.42_{-0.17}^{+0.17}$	& $0.20_{-0.10}^{+0.10}$	& $0.09_{-0.09}^{+0.09}$	& $0.04_{-0.04}^{+0.04}$	& $0.14_{-0.05}^{+0.05}$	& $0.07_{-0.03}^{+0.03}$	& $0.02_{-0.02}^{+0.02}$	& $0.04_{-0.01}^{+0.01}$	& $0.04_{-0.01}^{+0.01}$ \\ 
\hline 
  & 0.25	& $0.83_{-0.08}^{+0.08}$	& $0.64_{-0.18}^{+0.18}$	& $0.36_{-0.18}^{+0.18}$	& $0.27_{-0.08}^{+0.08}$	& $0.35_{-0.05}^{+0.05}$	& $0.36_{-0.06}^{+0.06}$	& $0.31_{-0.04}^{+0.05}$	& $0.32_{-0.03}^{+0.03}$	& $0.31_{-0.03}^{+0.03}$ \\ 
\osix & 0.05	& $0.58_{-0.17}^{+0.17}$	& $0.27_{-0.18}^{+0.18}$	& $0.09_{-0.09}^{+0.09}$	& $0.04_{-0.04}^{+0.04}$	& $0.16_{-0.05}^{+0.05}$	& $0.04_{-0.03}^{+0.01}$	& $0.09_{-0.02}^{+0.02}$	& $0.08_{-0.02}^{+0.02}$	& $0.07_{-0.02}^{+0.02}$ \\ 
  & 0.01	& $0.08_{-0.08}^{+0.08}$	& $0.09_{-0.09}^{+0.09}$	& $0.00_{-0.00}^{+0.00}$	& $0.00_{-0.00}^{+0.00}$	& $0.04_{-0.02}^{+0.02}$	& $0.01_{-0.01}^{+0.01}$	& $0.03_{-0.02}^{+0.02}$	& $0.03_{-0.01}^{+0.01}$	& $0.02_{-0.01}^{+0.01}$ \\ 
\hline 
  & 0.25	& $0.42_{-0.17}^{+0.17}$	& $0.27_{-0.09}^{+0.11}$	& $0.20_{-0.10}^{+0.10}$	& $0.32_{-0.11}^{+0.07}$	& $0.25_{-0.05}^{+0.05}$	& $0.26_{-0.06}^{+0.06}$	& $0.34_{-0.04}^{+0.04}$	& $0.33_{-0.03}^{+0.03}$	& $0.27_{-0.03}^{+0.03}$ \\ 
\nfive & 0.05	& $0.00_{-0.00}^{+0.00}$	& $0.09_{-0.09}^{+0.09}$	& $0.10_{-0.10}^{+0.10}$	& $0.04_{-0.04}^{+0.04}$	& $0.04_{-0.02}^{+0.02}$	& $0.06_{-0.03}^{+0.03}$	& $0.08_{-0.02}^{+0.02}$	& $0.06_{-0.01}^{+0.01}$	& $0.05_{-0.01}^{+0.01}$ \\ 
  & 0.01	& $0.00_{-0.00}^{+0.00}$	& $0.00_{-0.00}^{+0.00}$	& $0.00_{-0.00}^{+0.00}$	& $0.00_{-0.00}^{+0.00}$	& $0.00_{-0.00}^{+0.00}$	& $0.00_{-0.00}^{+0.00}$	& $0.00_{-0.00}^{+0.00}$	& $0.00_{-0.00}^{+0.00}$	& $0.00_{-0.00}^{+0.00}$ \\ 
\hline 
  & 0.25	& $0.77_{-0.09}^{+0.08}$	& $0.45_{-0.18}^{+0.18}$	& $0.55_{-0.18}^{+0.18}$	& $0.46_{-0.11}^{+0.07}$	& $0.46_{-0.07}^{+0.07}$	& $0.37_{-0.05}^{+0.05}$	& $0.24_{-0.04}^{+0.04}$	& $0.34_{-0.03}^{+0.04}$	& $0.35_{-0.03}^{+0.03}$ \\ 
\cfour & 0.05	& $0.62_{-0.15}^{+0.15}$	& $0.18_{-0.09}^{+0.09}$	& $0.18_{-0.09}^{+0.09}$	& $0.14_{-0.07}^{+0.07}$	& $0.24_{-0.05}^{+0.05}$	& $0.09_{-0.04}^{+0.03}$	& $0.11_{-0.03}^{+0.03}$	& $0.09_{-0.02}^{+0.02}$	& $0.12_{-0.02}^{+0.02}$ \\ 
  & 0.01	& $0.15_{-0.08}^{+0.08}$	& $0.09_{-0.09}^{+0.09}$	& $0.00_{-0.00}^{+0.00}$	& $0.00_{-0.00}^{+0.00}$	& $0.03_{-0.02}^{+0.03}$	& $0.01_{-0.01}^{+0.01}$	& $0.01_{-0.01}^{+0.01}$	& $0.01_{-0.01}^{+0.01}$	& $0.01_{-0.01}^{+0.01}$ \\ 
\hline 
  & 0.25	& $0.60_{-0.20}^{+0.20}$	& $0.43_{-0.14}^{+0.14}$	& $0.29_{-0.14}^{+0.14}$	& $0.29_{-0.10}^{+0.10}$	& $0.27_{-0.07}^{+0.07}$	& $0.24_{-0.06}^{+0.06}$	& $0.25_{-0.04}^{+0.04}$	& $0.20_{-0.03}^{+0.03}$	& $0.31_{-0.03}^{+0.03}$ \\ 
\cthree & 0.05	& $0.30_{-0.10}^{+0.10}$	& $0.14_{-0.14}^{+0.14}$	& $0.00_{-0.00}^{+0.00}$	& $0.05_{-0.05}^{+0.05}$	& $0.05_{-0.02}^{+0.02}$	& $0.08_{-0.04}^{+0.04}$	& $0.02_{-0.01}^{+0.01}$	& $0.03_{-0.01}^{+0.01}$	& $0.05_{-0.01}^{+0.01}$ \\ 
  & 0.01	& $0.00_{-0.00}^{+0.00}$	& $0.00_{-0.00}^{+0.00}$	& $0.00_{-0.00}^{+0.00}$	& $0.00_{-0.00}^{+0.00}$	& $0.00_{-0.00}^{+0.00}$	& $0.00_{-0.00}^{+0.00}$	& $0.00_{-0.00}^{+0.00}$	& $0.00_{-0.00}^{+0.00}$	& $0.00_{-0.00}^{+0.00}$ \\ 
\hline 
  & 0.25	& $0.50_{-0.17}^{+0.17}$	& $0.56_{-0.11}^{+0.22}$	& $0.38_{-0.12}^{+0.12}$	& $0.35_{-0.12}^{+0.12}$	& $0.37_{-0.06}^{+0.06}$	& $0.30_{-0.07}^{+0.05}$	& $0.20_{-0.04}^{+0.04}$	& $0.25_{-0.03}^{+0.03}$	& $0.26_{-0.03}^{+0.03}$ \\ 
\sifour & 0.05	& $0.25_{-0.08}^{+0.17}$	& $0.11_{-0.11}^{+0.11}$	& $0.00_{-0.00}^{+0.00}$	& $0.12_{-0.06}^{+0.06}$	& $0.10_{-0.04}^{+0.04}$	& $0.07_{-0.03}^{+0.03}$	& $0.04_{-0.02}^{+0.02}$	& $0.06_{-0.02}^{+0.02}$	& $0.06_{-0.02}^{+0.02}$ \\ 
  & 0.01	& $0.17_{-0.08}^{+0.08}$	& $0.11_{-0.11}^{+0.11}$	& $0.00_{-0.00}^{+0.00}$	& $0.06_{-0.06}^{+0.06}$	& $0.06_{-0.04}^{+0.04}$	& $0.03_{-0.02}^{+0.02}$	& $0.00_{-0.00}^{+0.00}$	& $0.01_{-0.01}^{+0.01}$	& $0.01_{-0.01}^{+0.01}$ \\ 
\hline 
\hline
\end{tabular}
  \end{table*}

\begin{figure*}
 \includegraphics[width=0.9\textwidth]{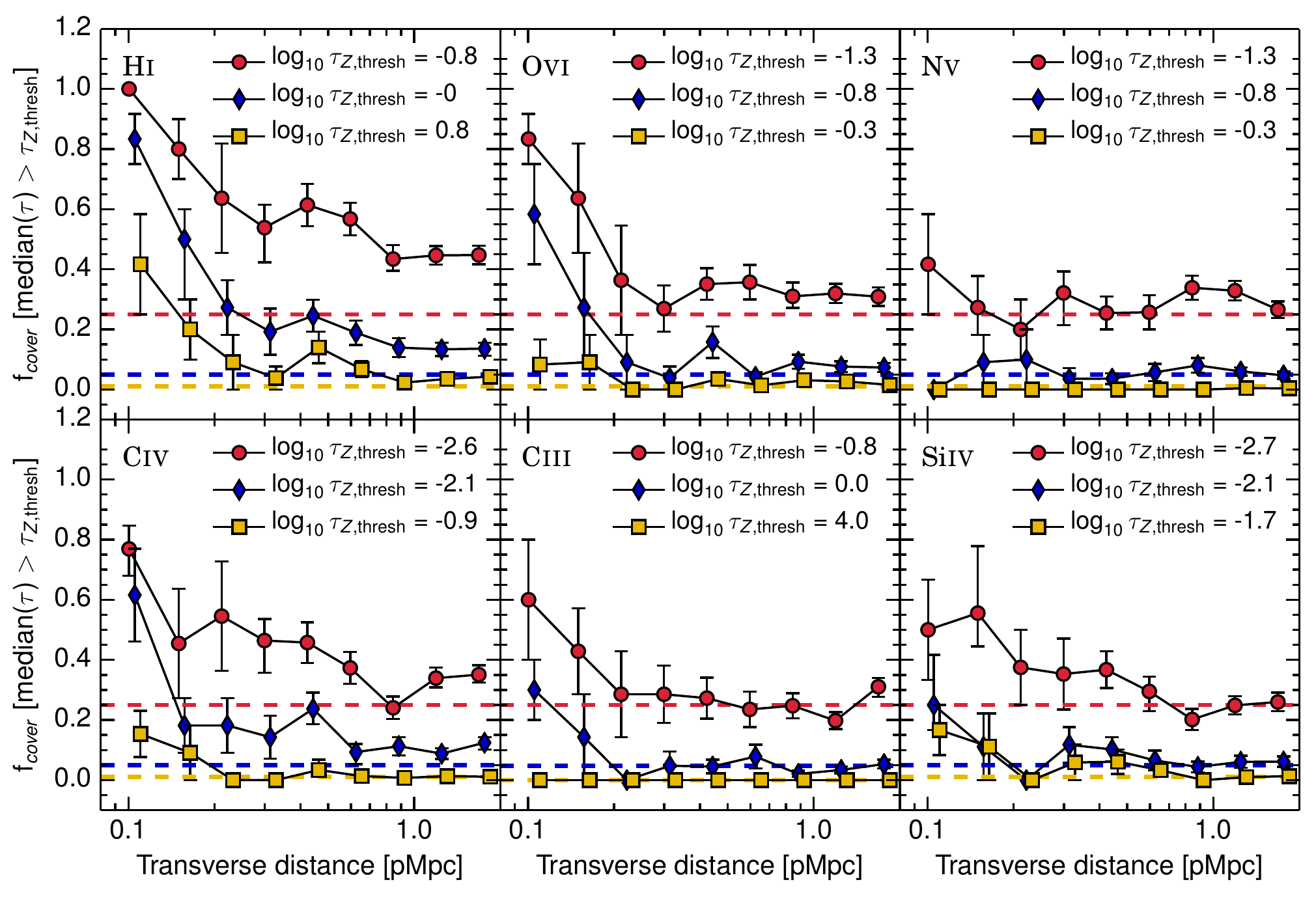}
\caption{Covering fraction for each ion, defined as the fraction of galaxies within some 
impact parameter bin that have a median optical depth 
within $\pm170$~\kmps\ above some threshold value. These values are chosen to be the optical depths
 for which the covering fraction computed for 1000 random
regions in the IGM is equal to 0.25 (red circles), 0.05 (blue diamonds), and 0.01 (yellow squares).
The IGM covering fraction values are indicated by the horizontal dashed lines of corresponding colour.
Points determined using different \tauthresh\ values have been offset horizontally by 0.02~dex for clarity.
We note that points where the covering fraction is 0 or 1 have no errorbar because
bootstrapping galaxies from the same sample cannot change the covering fraction within that
particular impact parameter bin.
}
\label{fig:fcover}
\end{figure*}

The results from \S~\ref{sec:maps} tell us about the median
optical depths as a function of distance from galaxies. However, we are also 
interested in the rate of occurrence of relatively high optical depth 
(and hence more rare) systems. To quantify this, we use the covering fraction,
which we define as the fraction 
of galaxies in a particular impact parameter bin
for which the median pixel optical depth within $\pm170$~\kmps\  of the galaxy 
exceeds some threshold value \tauthresh. We have checked that
the results remain unchanged using a larger velocity interval of $\pm300$~\kmps. 

To see which values of \tauthresh\ may be informative, 
for each ion $Z$ we calculate the median optical depth for 1000 random 
$\pm170$~\kmps\ regions selected from the full recovered redshift range,
which are taken to represent typical IGM values. We 
then select the \tauthresh\ values for which the
covering fractions of the 1000 random regions are 0.25, 0.05, and 0.01 
(the exact values of \tauthresh\ for each ion are displayed in Figure~\ref{fig:fcover}). 
In Appendix~\ref{sec:fcover_ew} we instead use an EW 
threshold and find that our conclusions still hold.
The covering fractions for \hone\ and each metal, as a function of impact parameter 
and for different values of \tauthresh\ are shown in 
Figure~\ref{fig:fcover} and Table~\ref{tab:fcover}, with the associated 
random region covering fraction denoted by the horizontal dashed lines.

First we examine values of \tauthresh\ for which
the IGM covering fraction is 0.25 and 0.05 (red circles and blue diamonds respectively).
In every case except for \nfive, we observe elevated covering fractions
within the two smallest impact parameter bins.
For larger impact parameters, the \hone\ covering
fraction is higher than the IGM value out to 2~pMpc, 
which is in agreement with \S~5 of \citet{rudie12}. 
Additionally, \osix\ and in particular \cfour\ both have covering fractions 
significantly above that of the IGM along the full range of impact parameters. 
This is consistent with the finding that we are able to detect levels of  
\cfour\ optical depth enhancement above the median IGM value
at all impact parameters probed (Figures~\ref{fig:all_maps_b} 
and \ref{fig:cuts_a}). 
Again, the fact that for impact parameters $\gtrsim180$~pkpc we only see 
significantly enhanced covering fractions for \hone, \osix, and \cfour\ does 
not necessarily mean that the other metal ions are not present, but could
rather be due to contamination and noise preventing us from
probing low enough optical depths, and could also be due to changes in 
the ionisation level of the gas. 

Finally, turning our attention to the values of \tauthresh\ 
for which the IGM covering fraction is 0.01 (yellow squares), 
only \hone, \cfour, and \sifour\ show significant enhancement at 
small impact parameters.
This suggests that for the ions bluewards of \lya, the highest optical
depth values are largely the result of contamination.
Furthermore, \hone\ and \cfour\ no longer have enhanced covering
fractions at large impact parameters, which suggests that the rare,
high optical depth systems are preferentially located very near galaxies. 

Here we address the two major transverse direction 
enhancement extents seen in \hone\ and the metal ions: a strong enhancement 
(out to $\approx180$~pkpc) and a weak enhancement
(out to $\approx2$~pMpc). We have observed the strong enhancement 
for all ions studied out to $\approx180$~pkpc, in both the transverse cuts 
(left panels of Figure~\ref{fig:cuts_a}) and the covering fraction (Figure~\ref{fig:fcover}). 
To put this distance into context, we take the comoving number density of identically 
selected galaxies in this survey from \citet{reddy08},
which is $\Phi = 3.7\times10^{-3}$~cMpc$^{-3}$ (for $\mathcal{R}\leq25.5$). 
From this, we can infer that the regions  within $180$~pkpc around galaxies 
comprise only $\sim0.4$\% of the total volume of the universe. In spite of
being present in a very small volume, the high optical depths and covering 
fractions within this distance indicate that these regions contain substantial 
amounts of \hone\ and metals ions. 

The second, weaker enhancement out to $\approx2$~pMpc 
can be seen for \hone\ and \cfour\ (and marginally for \osix) 
in the transverse-direction cuts (left panels of Figure~\ref{fig:cuts_a}), 
3-D Hubble distance (Figure~\ref{fig:hubble_a}), 
and covering fraction (Figure~\ref{fig:fcover}). 
 
Although the enhanced absorption observed on these scales is statistically significant, 
we note that 2~pMpc corresponds to $\sim20\times$ the virial radii of 
the galaxies in our sample, far beyond the physical sphere of influence 
of the galaxies centred at the origins of these figures. Performing the 
same calculation as above, to determine the fraction of the volume 
of the Universe within 2~pMpc of these galaxies results in a value that exceeds 
unity -- i.e., if these galaxies were uniformly distributed, all of space would 
lie within 2~pMpc of a galaxies that meets these selection criteria. 
 However, it is well established that galaxies
are clustered \citep[e.g.,][]{adelberger05a}, so we would
expect many of the ``spheres'' from our simple calculation to be overlapping.

Indeed, \citet{rudie12} noted that the galaxy-galaxy 
autocorrelation scale-length for 
this same sample of galaxies has been measured to be 
$\approx2.8$~pMpc \citep{trainor12}, which certainly suggests 
that the 2-halo term plays a role in the enhancement at large scales.
Another possibility is that we are seeing metals in the IGM; 
in reality we are likely observing a combination of both CGM metals from 
other clustered galaxies, as well as some truly intergalactic metals. 

Studies by \citet{adelberger03} at $z\sim3$ and 
\citet{adelberger05b} at $z\sim2.5$, using decrements in transmitted flux, 
found evidence for elevated (with respect to 
random IGM positions)
\hone\ absorption out to impact parameters of $\sim2$~pMpc.
The authors also measured a strong correlation between the 
positions of \cfour\ absorbers and galaxies, with the strength 
of the correlation increasing with absorber strength. 
Furthermore, at $z\sim0.5$ \citet{zhu14} found \mgtwo\ absorber
and galaxy positions to be correlated out to $\sim20$~pMpc. 
The above results are certainly consistent with the fact that
we are finding enhanced covering fractions for \hone\ and 
\cfour\ out to impact parameters of 2~pMpc.

\begin{figure*}
 \includegraphics[width=0.9\textwidth]{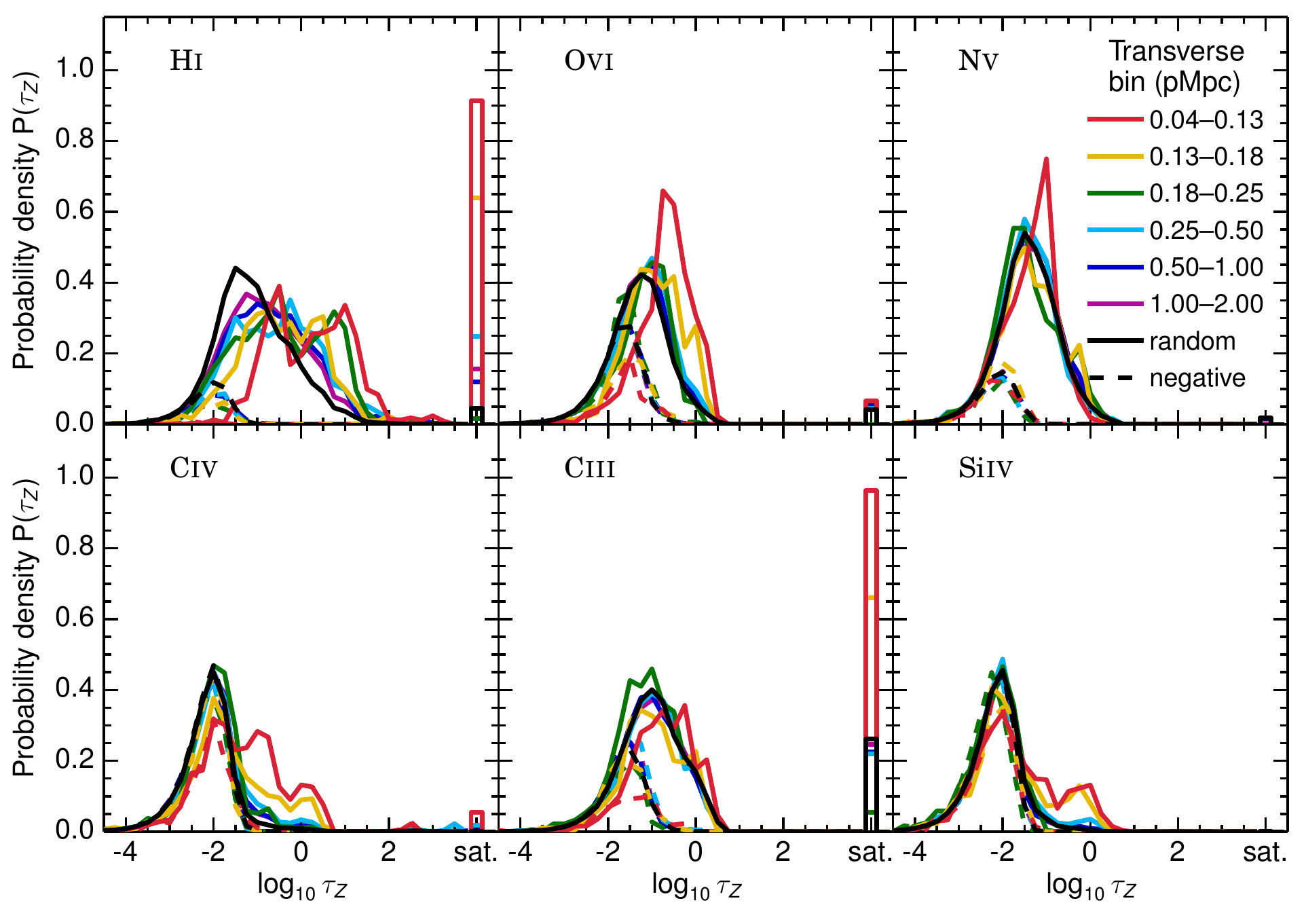}
 \caption{Pixel optical depth PDFs within $\pm170$~\kmps\ of galaxies divided by 
  impact parameter bin (coloured lines), where each panel displays a different ion. 
  We have used the usual binning scheme, but combined the last six bins to create three
  in order to reduce the number of lines plotted. The black line shows the PDF for 
  1000 random $\pm170$~\kmps\ regions within the full redshift range corresponding to each
  ion. The portion marked ``sat'' represents pixels that were found to 
  be saturated, and for which the optical depth could not be recovered. The dashed lines show the PDFs for
  pixels which had negative optical depths, where we have taken the log of their absolute values.
  Except for \nfive, the PDFs for the two smallest impact parameter bins
  (and more for \hone) tend to higher optical depth values 
  compared to the PDF of random regions. }
\label{fig:odhist}
\end{figure*}

\subsection{Optical depth distributions}
\label{sec:odhist}

Up until this point we have only considered median optical depths.
To acquire a sense of how individual pixel optical depth values are distributed, we have plotted 
their probability density functions (PDFs) for $\pm170$~\kmps\ regions
around galaxies in Figure~\ref{fig:odhist}, where each panel shows a different ion. 
The galaxies are divided into the usual 
impact parameter bins (coloured lines), except for the final six bins which have been combined 
into three (with sizes of 0.30~dex for clarity). The values marked ``sat.'' are for pixels which
we found to be saturated, and, in the case of \hone, whose optical depths 
could not be recovered from higher order Lyman series lines. We set the 
optical depth values of such pixels to $10^4$. 
We have also determined the PDFs for 1000 random 
regions in the IGM, which are shown by the black lines. Finally, for pixels which have negative
optical depths, we have taken the log of their absolute values and plotted their PDFs using  
dashed lines.

First we consider the source of the positive optical depths,
by comparing their distributions 
(solid lines) to those of negative optical depths (dashed lines),
which we expect to reflect the level of shot-noise  in the spectrum. 
Aside from the bins with impact parameters $<0.18$~pMpc, for
\cfour\ and \sifour\ the distributions of
 positive and negative optical depths are similar to each other,
suggesting that most of the positive optical depths arise from shot noise 
around the level of the continuum. This is not surprising for ions 
with rest-frame wavelengths redwards of the QSO \lya\ emission,
i.e. regions that do not suffer from \hone\ contamination. 
On the other hand,  for the remaining ions, the negative optical depths
contribute minimally to the full distribution, which means that
the positive optical depths should primarily reflect 
real absorption (including contamination).

Figure~\ref{fig:odhist} demonstrates that the scatter in optical depths is quite large. 
Even for regions at small galactocentric distances, 
many optical depths reach values typical of random IGM regions
(which may of course be close to undetected galaxies).
However, except perhaps for \nfive, 
the optical depth PDFs of the 2 or 3 smallest impact parameter bins 
for each ion appear substantially different from the 
random region PDF, and in particular tend
towards higher optical depth values.\footnote{Unfortunately, because of spatial 
correlations between the optical depth pixels,
the use of the Kolmogorov-Smirnov test to compare the PDFs
is not appropriate for this data.}

\begin{figure*}
\centering
\includegraphics[width=\textwidth, clip=true, trim = 0mm 0 0mm 0]{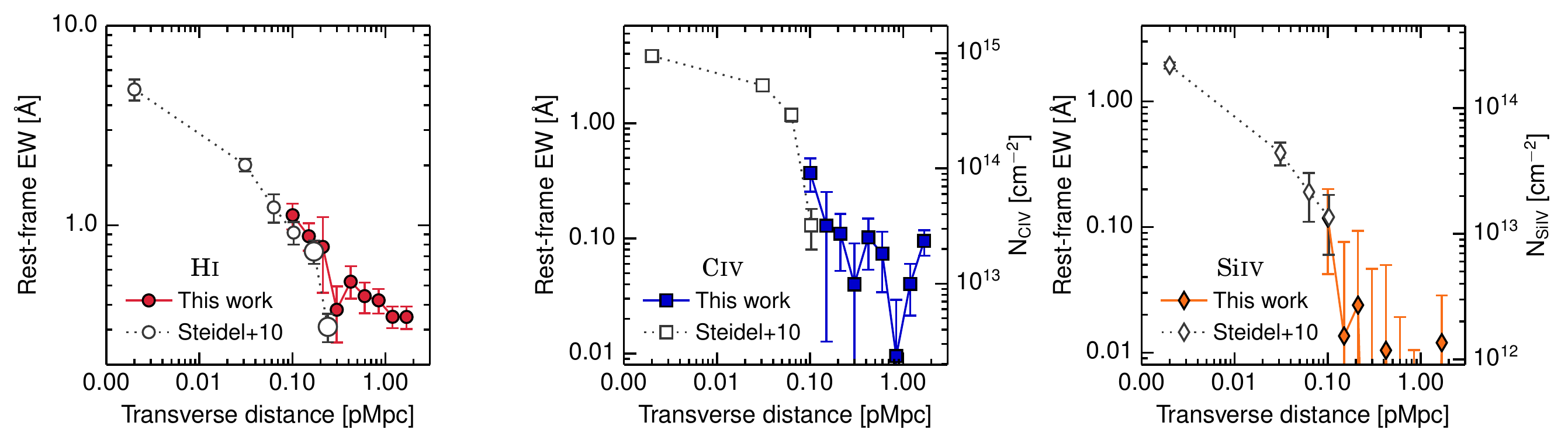}
\caption{Rest-frame EW calculated using the mean spectrum within each
galaxy impact parameter bin (filled points), where each panel shows
a different ion. Here, the average spectrum was divided 
by the mean flux of all spectral pixels and the flux decrement was 
integrated over $\pm500$~\kmps. 
Also plotted are EWs calculated using galaxy pairs from low-resolution spectra 
(open points) as well as \hone\ EWs from galaxy-QSO pairs (left panel, large open circles
at 128--280~pkpc) taken from \citet{steidel10}. 
Note that for clarity, each panel has a different range along the $y$-axis. 
EWs were also calculated for 
an average spectrum created from 1000 random $\pm500$~\kmps\ regions, and the 
result was negative for all three ions. For reference, on the right-hand y-axis
of the two rightmost panels
we show the corresponding column density calculated from Equation~\ref{eq:cdens},
which assumes that the absorbers are on the linear portion of the curve-of-growth.
At the point of overlap, $\sim10^2$~pkpc, our points agree with \citet{steidel10},
and although we do not see the sharp drop off at larger impact 
parameters predicted by their models, we do find consensus 
in the measured \hone\ EWs at large impact parameters.}
\label{fig:ewB}
\end{figure*}

\begin{table*}
\caption{Median EWs and $1\sigma$ errors (in \AA) calculated by integrating 
the normalised flux decrement over a $\pm500$~\kmps\ region
 as a function of transverse distance for \hone, \cfour\ and \sifour. These values
 are plotted in Figure~\ref{fig:ewB}. We also show two different column density estimates, 
 $\log_{10}$~\newa\ and  $\log_{10}$~\newb, where N is in cm$^{-2}$. \newa\ was
 computed by applying Equation~\ref{eq:cdens} to the EWs from the row above.
 \newb\ was measured by performing the EW integration on the recovered optical depths converted back to flux, 
  rather than on the observed flux as was done for \newa, and then applying Equation~\ref{eq:cdens}.  
  Empty values correspond to negative column densities (due to negative EWs), and u.l. denotes $1\sigma$ upper limits. }
\label{tab:ew}
\begin{tabular}{@{}lcrrrrrrrrr}
\hline\hline
\multicolumn{2}{r}{$D_{\rm trans}$ (pMpc)} & 	0.04--0.13 & 	0.13--0.18 & 	0.18--0.25 & 	0.25--0.36 & 	0.36--0.50 & 	0.50--0.71 & 	0.71--1.00 & 	1.00--1.42 & 	1.42--2.00\\
Ion & & & & & & & & & & \\\hline
\hone & EW	& $1.12_{-0.17}^{+0.16}$	& $0.88_{-0.14}^{+0.14}$	& $0.78_{-0.32}^{+0.32}$	& $0.38_{-0.12}^{+0.12}$	& $0.52_{-0.09}^{+0.10}$	& $0.44_{-0.08}^{+0.08}$	& $0.42_{-0.06}^{+0.06}$	& $0.35_{-0.04}^{+0.04}$	& $0.35_{-0.05}^{+0.04}$ \\ 
\hline 
  & EW	& $0.37_{-0.12}^{+0.13}$	& $0.13_{-0.12}^{+0.12}$	& $0.11_{-0.06}^{+0.05}$	& $0.04_{-0.05}^{+0.05}$	& $0.10_{-0.05}^{+0.05}$	& $0.07_{-0.04}^{+0.04}$	& $0.01_{-0.02}^{+0.02}$	& $0.04_{-0.02}^{+0.02}$	& $0.10_{-0.02}^{+0.02}$ \\ 
\cfour & $\log_{10}$ \newa	& $13.62_{-0.18}^{+0.12}$	& $13.16_{-1.38}^{+0.28}$	& $13.09_{-0.29}^{+0.18}$	& $12.66_{-\infty}^{+0.36}$	& $13.06_{-0.27}^{+0.17}$	& $12.92_{-0.35}^{+0.19}$	& $12.03_{-\infty}^{+0.49}$	& $12.66_{-0.29}^{+0.17}$	& $13.03_{-0.12}^{+0.10}$ \\ 
  & $\log_{10}$ \newb	& $13.55_{-0.17}^{+0.11}$	& $12.66_{-\infty}^{+0.46}$	& $12.66_{-0.97}^{+0.30}$	& $12.80_{-1.37}^{+0.29}$	& $12.70_{-0.61}^{+0.25}$ & ${12.28}$ (u.l.) & \ldots & \ldots	& $12.40_{-0.42}^{+0.23}$ \\ 
\hline 
  & EW	& $0.12_{-0.08}^{+0.08}$	& $0.01_{-0.07}^{+0.06}$	& $0.02_{-0.06}^{+0.07}$	& $-0.01_{-0.05}^{+0.05}$	& $0.01_{-0.04}^{+0.04}$	& $-0.01_{-0.03}^{+0.03}$	& $-0.01_{-0.02}^{+0.02}$	& $-0.02_{-0.01}^{+0.01}$	& $0.01_{-0.02}^{+0.02}$ \\ 
\sifour & $\log_{10}$ \newa	& $13.12_{-0.53}^{+0.22}$	& $12.19_{-\infty}^{+0.77}$	& $12.43_{-\infty}^{+0.57}$ & ${12.71}$ (u.l.)	& $12.07_{-\infty}^{+0.67}$ & ${12.28}$ (u.l.) & ${12.09}$ (u.l.) & \ldots	& $12.13_{-\infty}^{+0.38}$ \\ 
  & $\log_{10}$ \newb	& $12.83_{-\infty}^{+0.33}$ & ${12.62}$ (u.l.) & \ldots & \ldots & \ldots & \ldots & \ldots & \ldots & \ldots \\ 
\hline 
\end{tabular}
\end{table*}

\subsection{Equivalent widths}
\label{sec:ew}

In studies where individual absorption lines are examined, 
EW is often used to parametrise their strength independently of 
the line shapes, and can also be computed using spectral stacking.
Another advantage of employing EW as a metric is that it can be applied to
spectra with low S/N or resolution. In particular, to measure absorption at very small 
impact parameters (i.e., less than our smallest value of 35~pkpc),
the usage of QSO-galaxy pairs is currently not viable, and instead galaxy-galaxy
pairs (with much lower resolution spectra) must be used. By stacking such spectra
to create mean flux profiles,
\citet{steidel10} measured EW as a function of 
galaxy impact parameter (see their Figure~21 and their Table~4). 
We have plotted their results for the ions which are also in our sample
(\hone, \cfour, and \sifour) in Figure~\ref{fig:ewB}. In addition to measuring 
EWs for small impact parameters by using galaxy-galaxy pairs, they also used galaxy-quasar
pairs to calculate \hone\ EWs for impact parameters of 128--280 pkpc, which we denote
by the large black open circles. We note that the galaxy-quasar 
sample from \citet{steidel10} is comprised of the same QSOs 
and mostly the same galaxies as the sample used in this work. 

We calculate typical EWs as a function of 
galaxy impact parameter by shifting the absorption 
spectrum (i.e. the associated region of the intervening QSO) 
of every galaxy to the rest-frame wavelength of the ion in question
and computing the mean spectrum within the transverse distance bin.
To simulate the effect of a suppressed continuum,
as appropriate for the low-resolution spectra used in \citet{steidel10}, 
we follow \citet{rakic12} and
calculate the mean flux level of all pixels in all spectra
within the wavelength region probed for a particular ion. 
We then divide the mean flux profile by this value before integrating
the flux decrement over $\pm500$~\kmps\ to compute the EW. 
The results are given in Table~\ref{tab:ew} and Figure~\ref{fig:ewB}.
We have verified that the conclusions presented here hold if we use 
150, 300 and 600~\kmps\ intervals instead.

For overlapping impact parameter bins, we see good agreement between the 
two samples for \hone\ and \sifour. Additionally, \cfour\ shows
good agreement when we compare the outer transverse distance 
bin of \citet{steidel10} with our second-smallest bin. 
The models from \citet{steidel10} predict
that when one extrapolates EWs to larger impact parameters, there is
sharp drop in EW around impact parameters of $\sim10^2$~pkpc for the metal ions
and around $\sim200$~pkpc for \hone. We do not see such an effect in our data,
where the EW values drop off relatively slowly with
increasing transverse distance. 
However, the models were only intended to explain the behaviour of the 
EWs down to 0.1~\AA, and were insensitive to any plateau at smaller EW values.  
On the other hand, we find good agreement between
both sets of measured \hone\ EW values at large impact parameters, in spite
of the differences in the galaxy sample, bin size, and EW measurement technique.

To ease comparison with other studies, we convert EW to column density as in \citet{savage91}, assuming 
the linear curve-of-growth regime (i.e. $\tau<1$): 
\begin{equation}
\begin{split}
 {\rm N} &= 4 \pi \epsilon_0 \dfrac{m_e c^2}{\pi e^2} \dfrac{\rm EW}{ f_{Z,k} \lambda_{Z,k}^2} \\
	&= 1.13\times10^{20} \dfrac{1}{f_{Z,k}} \left(\dfrac{1 \textup{\AA}}{\lambda_{Z,k} }\right)^2 
	 \left(\dfrac{\rm EW}{1 \textup{\AA}}\right)\, {\rm cm}^{-2}
\end{split}
\label{eq:cdens}
\end{equation}
where $\epsilon_0$ is the permittivity of free space, $m_e$ is the electron mass,
$e$ is the electric charge, $c$ is the speed of light, and
$\lambda_{Z,k}$ and $f_{Z,k}$ are the rest-frame wavelength and 
oscillator strength of the $k$th transition of ion $Z$.
Figure~\ref{fig:odhist} demonstrates that the linear regime is not a valid
assumption for \hone\ since a substantial number of pixels have $\tau>1$, however,
we can still apply Equation~\ref{eq:cdens} to \cfour\ and \sifour.\footnote{ 
Examining the curve-of-growth and assuming a Doppler parameter $b_D=10$~\kmps, 
we find that the linear regime is valid for N$_{\cfourm}\lesssim8\times10^{13}$~cm$^{-2}$ 
and N$_{\sifourm}\lesssim3\times10^{13}$~cm$^{-2}$. Furthermore, from visual inspection 
we conclude that the spectral regions around the 30 smallest impact parameter galaxies
show no evidence for saturation in \cfour\ and \sifour.}
We show this conversion as a secondary vertical axis in Figure~\ref{fig:ewB}. In addition,
in Table~\ref{tab:ew} we give the column densities computed from Equation~\ref{eq:cdens} 
applied to the given EWs (\newa). Furthermore, instead of calculating EW
directly from the observed spectrum, we convert the recovered optical depths
back to flux values, and integrate over those to determine column densities, which
are also presented in Table~\ref{tab:ew} (\newb).

\subsection{Galaxy redshift measurements}
\label{sec:var_gal_sample}

\begin{figure*}
\includegraphics[width= \textwidth,clip=true,trim = 0 0mm 0 0mm]{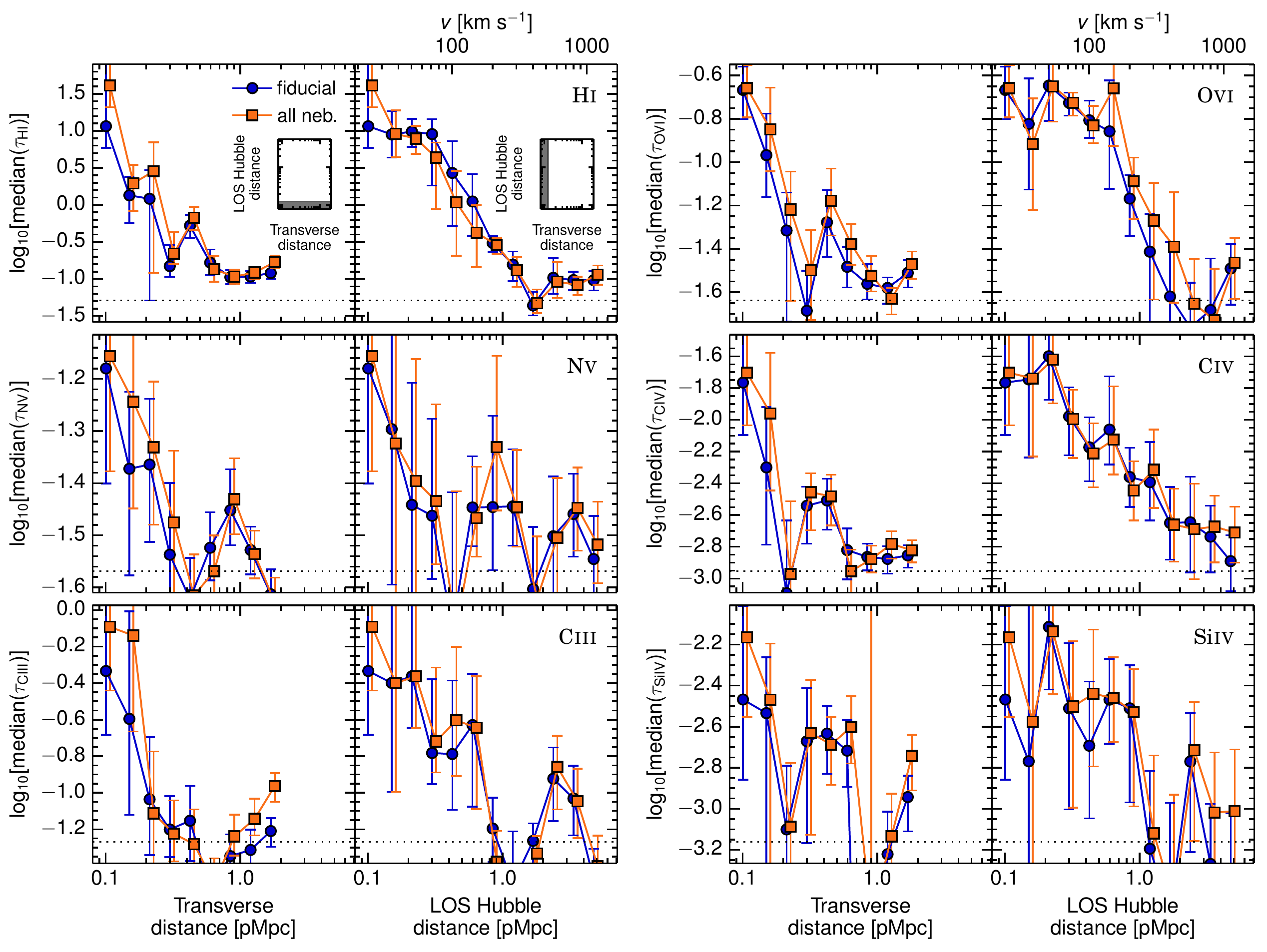}
\caption{The median log optical depth \taumedk\, where each set of two horizontal panels
shows a different ion. The left and right panels in each set of two show points along 
the transverse and LOS directions, respectively.
The cut sizes, shown in the insets, are the same as in Figure~\ref{fig:cuts_a}, i.e., 0--30~\kmps\ and 
0.04--0.18~pMpc for the points along the transverse and LOS directions, respectively. 
The results from the fiducial sample used throughout this work, which consists
of galaxy redshifts measured using a mix of rest-frame UV lines and nebular emission lines, 
are shown by the blue circles (835 galaxies), 
while the orange squares indicate the medians from all galaxies that have 
their redshift measured using nebular emission lines (354 galaxies). 
Points have been offset horizontally by 0.03~dex for clarity. The elongation of the 
optical depth enhancement along the LOS does not decrease when the sample only 
contains galaxies with nebular redshifts, which indicates that redshift measurement 
errors do not contribute significantly to this effect. 
}
\label{fig:maps_onlynebular}
\end{figure*}

As discussed earlier, the various nebular and rest-frame UV-based 
galaxy redshifts suffer from different levels of uncertainty. 
In general, there is significantly more uncertainty associated with 
those redshifts determined from interstellar absorption or \lya\
emission lines than with nebular redshifts measured by 
NIRSPEC or MOSFIRE. Since the elongation of the absorption
enhancement along the LOS could be caused at least in part by redshift
errors (e.g., \citealt{rakic13}),  reducing them helps disentangle this effect from that
of peculiar velocities. 
Currently, our galaxy sample has \ngalneb\ galaxies with redshifts 
measured from nebular emission lines, and \ngalmf\ of those are derived
from MOSFIRE observations, which we estimate have errors of
$\Delta v \approx 18$~\kmps\ (compared to \ngalns\ galaxies with 
NIRSPEC measured redshifts that have an estimated error of 
$\Delta v \approx 60$~\kmps). The measurement
errors from these instruments are much smaller than the $\sim240$~\kmps\ extent
of elongation seen along the LOS.

Therefore, we experiment with using different subsamples of KBSS galaxies.
In Figure~\ref{fig:maps_onlynebular}, we again show the (unsmoothed)
median optical depths from cuts taken along the maps in Figure~\ref{fig:all_maps_b},
using the innermost bins and plotting points along the transverse direction (left columns) 
and LOS direction (right columns).
 The blue circles show the cuts constructed using the full galaxy sample 
 and are identical to the points in Figure~\ref{fig:cuts_a}.
We also consider the results when only galaxies whose
redshifts were measured from nebular emission lines using NIRSPEC or MOSFIRE are used
(\ngalneb\ galaxies or \ngalnebpercent\% of the full sample for \hone, orange squares). 

In general, we do not see any significant differences between the cuts taken from
the fiducial sample and the one using only nebular redshifts. For the points along
the transverse direction, one might expect that more accurate redshift measurements would 
result in an increase in the observed absorption. Although a small
effect can be seen, that is, in most cases along the transverse direction
the orange squares are at slightly higher optical depths than the blue circles,
the result is not significant for any fixed transverse distance
(although the increase in optical depths could be significant 
if integrated over the full impact parameter range). 

We note that discrepancies between the two samples may be due to the different
galaxies used to construct each optical depth profile.  
To test this, in Appendix~\ref{sec:galsample_appendix}
we take the \ngalbothnebanduv\ galaxies that have had \emph{both} their nebular 
and rest-frame UV redshifts measured,
and show the results determined using only one type of redshift 
measurement at a time in Figure~\ref{fig:maps_onlynebular_appendix}.
The slight enhancement along the transverse direction is no longer apparent
in this comparison, so it may be a consequence of the different
galaxies used in the nebular-only sample.
However, it is important to note that in Figure~\ref{fig:maps_onlynebular_appendix}
the number of galaxies in the sample is relatively small
and hence that subtle differences may be hidden in the noise.
Furthermore, in Figure~\ref{fig:oljaA} 
we directly plot our results against those from
\citet{rakic12}, where only 10\% of the galaxy sample had 
redshifts measured from nebular emission lines. We find that 
the \hone\ optical depths along the LOS agree between the two 
sample for the innermost bin, although the redshift error effects
become apparent in larger impact parameter bins.

The most striking result from Figure~\ref{fig:maps_onlynebular} 
is that we do not observe significant differences between the LOS-direction 
optical depth profiles of the two galaxy samples. 
The lack of change in the extent of the optical depth enhancement
when galaxy redshift errors are reduced from $\sim150$ to $\sim10$~\kmps\ implies 
that the small-scale anisotropy detected for \hone\ by \citet{rakic12} and 
\citet{rudie12}, and for \hone\ and metals in this work,
originates from gas peculiar velocities rather than from redshift errors.

Finally, as in \citet{rudie12}, we would like to extract the peculiar velocity field
by subtracting the redshift errors ($\approx18$~\kmps) and transverse extent
of the absorption (180~pkpc or $\approx42$~\kmps) in quadrature from the 
observed velocity enhancement; doing so results in $\sim\pm240$~\kmps, and
is in good agreement with the peculiar velocities measured in \citet{rudie12}
of $\sim\pm260$~\kmps. 
The magnitudes of the peculiar velocities we observe are 
consistent with what would be expected from infall, outflows, and
virial velocities, so although it is currently difficult to disentangle 
the exact nature of their origin, it is likely that we are detecting
motions of the gas around LBGs.

\section{Summary And Conclusions}
\label{sec:conclusion}

We have studied metals in the CGM around \ngaltotal\ $z\approx2\text{--}2.8$ 
galaxies taken from the KBSS. 
The sample  contains \ngalneb\ galaxies that have been
observed with NIRSPEC and/or MOSFIRE,
allowing us to measure their redshifts using nebular emission lines 
with estimated errors of only $\Delta v\approx60$~\kmps\ and 
$\Delta v\approx18$~\kmps\ for the two instruments, respectively. 
The galaxies studied lie in the fields of 15 hyper-luminous
QSOs for which high-quality Keck spectra are available, with galaxy impact parameters ranging from $35$~pkpc
to $2$~pMpc. Using the QSO spectra, 
the optical depth at each pixel in the \lya\ forest redshift range was recovered for 
\hone\ as well as for five metal ions (\osix, \nfive, \cfour, \cthree, and
\sifour). This was done using a slightly modified version of the pixel optical depth 
technique of \citet{aguirre02}, which corrects the 
optical depths for saturation and various forms of contamination.  

The main results are summarised below:
\begin{enumerate}
%%%%%%%%%
 \item We have presented 2-D maps of the median absorption around galaxies in \hone\
(first shown for a smaller sample by \citealt{rakic12}), 
as well as the first maps of \osix, \nfive, \cfour, \cthree,
and \sifour. These maps were created by taking the medians of 
the binned pixel optical depths as a function of the LOS and transverse distances 
from the galaxies in our sample (Figure~\ref{fig:all_maps_b}, \S~\ref{sec:maps}). 
%%%%%%%%
\item For every ion studied except perhaps for \nfive, we measure an enhancement of 
the absorption at small galactocentric radii, 
out to $\sim180$~pkpc in the transverse direction
and $\sim\pm240$~\kmps along the LOS ($\sim1$~pMpc assuming pure Hubble flow).
Inside this region the median optical depth is typically enhanced above that 
of random regions by about
one order of magnitude for the metals and by two order of magnitude 
for \hone (Figures~\ref{fig:all_maps_b} and \ref{fig:zspacedist}, \S~\ref{sec:maps}). 
%%%%%%%%
\item In the transverse direction, \hone\ and \cfour\ show a slight
enhancement out to the maximum impact parameter covered by the survey
(2~pMpc). The non-detection
of an enhancement in the other ions at such large impact parameters does
not necessarily imply that these ions are not present; rather, it may
reflect differences in detection limits 
(Figure~\ref{fig:cuts_a}, \S~\ref{sec:maps}). 
%%%%%%%%
\item The visual impression from the maps is that the enhancement 
of the \hone\ is more extended in the transverse direction than that 
of the metals. However,
normalising the optical depth profiles such that they all have the same
maximum and asymptote, the metal ions do 
not show a significantly different profile shape (i.e. drop-off of
optical depth values) with increasing transverse distance compared
to \hone. In the LOS direction, \cthree\ traces
the normalised \hone\ profile, while \osix, \cfour\ and \sifour\ 
show evidence for more extended absorption (Figure~\ref{fig:cuts_b},
\S~\ref{sec:profiles}). 
%%%%%%%%%
\item Comparing cuts through the median optical depth maps
in the transverse direction to those in
the LOS direction, for all ions except \nfive\ the optical depths 
are higher in the LOS than in the transverse direction out 
to distances of $\sim1$~pMpc (i.e. 240~\kmps) 
(Figure~\ref{fig:zspacedist}, \S~\ref{sec:maps}). 
%%%%%%%%%
\item We examined the median optical depth as a function of 3-D Hubble
distance, which allows us to reduce the noise by combining pixels from both the LOS and 
transverse directions. Focusing on larger distance scales which should be minimally affected by 
redshift-space distortions (i.e. $\gtrsim1$~pMpc), we find the \cfour\
optical depth to be slightly enhanced out to $\sim4$~pMpc. 
(Figure~\ref{fig:hubble_a}, \S~\ref{sec:hubble}). 
%%%%%%%%
\item The pixel optical depth PDFs at all impact parameters 
and for random regions all show similar amounts of scatter.
However, at small impact parameters, particularly in the 
two innermost bins (i.e. $<180$~pkpc), the distributions of optical depths
are typically skewed to higher values
(Figure~\ref{fig:odhist}, \S~\ref{sec:odhist}).  
%%%%%%%%
\item Comparison to the EWs from \citet{steidel10}, which were 
calculated using galaxy pairs and probe smaller impact
parameters than in this work, yields agreement for the 
overlapping impact parameter bins
(Figure~\ref{fig:ewB}, \S~\ref{sec:ew}).
%%%%%%%%
\item The value of the covering fraction, defined as the fraction of galaxies 
for which the median optical depth within $\pm170$~\kmps\ exceeds specific values,
was found to be greater than the covering fraction at a random location 
for galaxy impact parameters below $\sim180$~pkpc, 
for all ions (Figure~\ref{fig:fcover}, \S~\ref{sec:fcover}).
%%%%%%%%%
\item For \hone, \cfour, and in some cases \osix, the covering fraction 
is elevated with respect to the IGM for 
all impact parameters probed, which is consistent with the optical depth
enhancement seen along the transverse distance cuts. However, for high
optical depth threshold values, corresponding to the top 1\% of random
regions, the enhancement is only seen at 
impact parameters $<0.13$~pMpc, which suggests that high optical depth systems
are preferentially found at very small galactocentric distances
 (Figure~\ref{fig:fcover}, \S~\ref{sec:fcover}).
%%%%%%%%
\item Limiting the sample to galaxies with nebular redshifts 
does not impart any significant change to the observed median optical depths
as a function of transverse and LOS distance from the galaxies, even when 
the same sample of galaxies is used to compare the measurement techniques.
This implies that the elongation of the optical depth enhancement along
the LOS direction is due to gas peculiar velocities rather than redshift errors
(Figure~\ref{fig:maps_onlynebular}, \S~\ref{sec:var_gal_sample}).
\end{enumerate}

Thanks to our unique sample, and particularly due to its
unprecedented combination of size and data quality, we were able to
study the 2-D metal distribution around galaxies in a way that 
has not been possible until now. We have presented the first 2-D
maps of metal absorption around galaxies and we have quantified the
enhancement in the absorption signal near $z\approx 2.4$ star-forming
galaxies for 5 different metal ions, as well as for neutral
hydrogen. Observations with MOSFIRE have allowed us to  
significantly reduce the errors on galaxy redshift measurements,
to the point where we now have compelling evidence that 
the redshift-space distortions seen in the 2-D \hone\ and metal 
ion optical depth distributions (i.e. the Finger-of-God effect seen in
the LOS direction) is caused by peculiar motions of the gas. 

As MOSFIRE observations proceed and the KBSS galaxy sample continues to grow, 
improvements in the data will allow
us to split the sample according to various galaxy properties (e.g., mass, 
SFR, etc.) and to study how these characteristics affect the CGM.
Additionally, we plan to examine the ratios between different metal ions
and between metal ions and \hone, and also to compare the results from this 
work to simulations. 

\section*{Acknowledgements}

We are very grateful to Milan Bogosavljevic, Alice Shapley,
Dawn Erb, Naveen Reddy, Max Pettini, Ryan Trainor, and David
Law for their invaluable contributions to the Keck Baryonic
Structure Survey, without which the results presented here
would not have been possible. We also thank Ryan Cooke for his help
with the continuum fitting of QSO spectra, and the anonymous referee
whose valuable comments greatly improved this work.
We gratefully acknowledge support from Marie Curie Training
Network CosmoComp (PITN-GA-2009- 238356) and
from the European Research Council under the European
Union's Seventh Framework Programme (FP7/2007-2013)
/ ERC Grant agreement 278594-GasAroundGalaxies.
CCS, GCR, ALS acknowledge support from grants AST-0908805 
and AST-13131472 from the US National Science Foundation.                            
We thank the W. M. Keck
Observatory staff 
for their assistance with the observations. We also thank the
Hawaiian people, as without their hospitality the observations
presented here would not have been possible.  

%%%%%%%%%%%%%%%%
% Bibliography %
%%%%%%%%%%%%%%%%

\bibliographystyle{mn2e} 
\bibliography{bibliography}

\begin{thebibliography}{64}
\expandafter\ifx\csname natexlab\endcsname\relax\def\natexlab#1{#1}\fi

\bibitem[{{Adelberger} {et~al.}(2005{\natexlab{a}}){Adelberger}, {Shapley},
  {Steidel}, {Pettini}, {Erb}, \& {Reddy}}]{adelberger05b}
{Adelberger} K.~L., {Shapley} A.~E., {Steidel} C.~C., {Pettini} M., {Erb}
  D.~K., {Reddy} N.~A., 2005{\natexlab{a}}, \apj, 629, 636

\bibitem[{{Adelberger} {et~al.}(2005{\natexlab{b}}){Adelberger}, {Steidel},
  {Pettini}, {Shapley}, {Reddy}, \& {Erb}}]{adelberger05a}
{Adelberger} K.~L., {Steidel} C.~C., {Pettini} M., {Shapley} A.~E., {Reddy}
  N.~A., {Erb} D.~K., 2005{\natexlab{b}}, \apj, 619, 697

\bibitem[{{Adelberger} {et~al.}(2004){Adelberger}, {Steidel}, {Shapley},
  {Hunt}, {Erb}, {Reddy}, \& {Pettini}}]{adelberger04}
{Adelberger} K.~L., {Steidel} C.~C., {Shapley} A.~E., {Hunt} M.~P., {Erb}
  D.~K., {Reddy} N.~A., {Pettini} M., 2004, \apj, 607, 226

\bibitem[{{Adelberger} {et~al.}(2003){Adelberger}, {Steidel}, {Shapley}, \&
  {Pettini}}]{adelberger03}
{Adelberger} K.~L., {Steidel} C.~C., {Shapley} A.~E., {Pettini} M., 2003, \apj,
  584, 45

\bibitem[{{Aguirre} {et~al.}(2001){Aguirre}, {Hernquist}, {Schaye}, {Weinberg},
  {Katz}, \& {Gardner}}]{aguirre01}
{Aguirre} A., {Hernquist} L., {Schaye} J., {Weinberg} D.~H., {Katz} N.,
  {Gardner} J., 2001, \apj, 560, 599

\bibitem[{{Aguirre} {et~al.}(2002){Aguirre}, {Schaye}, \& {Theuns}}]{aguirre02}
{Aguirre} A., {Schaye} J., {Theuns} T., 2002, \apj, 576, 1

\bibitem[{{Ajiki} {et~al.}(2002){Ajiki}, {Taniguchi}, {Murayama}, {Nagao},
  {Veilleux}, {Shioya}, {Fujita}, {Kakazu}, {Komiyama}, {Okamura}, {Sanders},
  {Oyabu}, {Kawara}, {Ohyama}, {Iye}, {Kashikawa}, {Yoshida}, {Sasaki},
  {Kosugi}, {Aoki}, {Takata}, {Saito}, {Kawabata}, {Sekiguchi}, {Okita},
  {Shimizu}, {Inata}, {Ebizuka}, {Ozawa}, {Yadoumaru}, {Taguchi}, {Ando},
  {Nishimura}, {Hayashi}, {Ogasawara}, \& {Ichikawa}}]{ajiki02}
{Ajiki} M., {Taniguchi} Y., {Murayama} T., {Nagao} T., {Veilleux} S., {Shioya}
  Y., {Fujita} S.~S., {Kakazu} Y., {Komiyama} Y., {Okamura} S., {Sanders}
  D.~B., {Oyabu} S., {Kawara} K., {Ohyama} Y., {Iye} M., {Kashikawa} N.,
  {Yoshida} M., {Sasaki} T., {Kosugi} G., {Aoki} K., {Takata} T., {Saito} Y.,
  {Kawabata} K.~S., {Sekiguchi} K., {Okita} K., {Shimizu} Y., {Inata} M.,
  {Ebizuka} N., {Ozawa} T., {Yadoumaru} Y., {Taguchi} H., {Ando} H.,
  {Nishimura} T., {Hayashi} M., {Ogasawara} R., {Ichikawa} S.-i., 2002, \apjl,
  576, L25

\bibitem[{{Bahcall} \& {Spitzer}(1969)}]{bahcall69}
{Bahcall} J.~N., {Spitzer} Jr. L., 1969, \apjl, 156, L63

\bibitem[{{Bergeron} \& {Boiss{\'e}}(1991)}]{bergeron91}
{Bergeron} J., {Boiss{\'e}} P., 1991, \aap, 243, 344

\bibitem[{{Booth} {et~al.}(2012){Booth}, {Schaye}, {Delgado}, \& {Dalla
  Vecchia}}]{booth12}
{Booth} C.~M., {Schaye} J., {Delgado} J.~D., {Dalla Vecchia} C., 2012, \mnras,
  420, 1053

\bibitem[{{Chen} {et~al.}(2001){Chen}, {Lanzetta}, \& {Webb}}]{chen01}
{Chen} H.-W., {Lanzetta} K.~M., {Webb} J.~K., 2001, \apj, 556, 158

\bibitem[{{Chen} \& {Mulchaey}(2009)}]{chen09}
{Chen} H.-W., {Mulchaey} J.~S., 2009, \apj, 701, 1219

\bibitem[{{Conroy} {et~al.}(2008){Conroy}, {Shapley}, {Tinker}, {Santos}, \&
  {Lemson}}]{conroy08}
{Conroy} C., {Shapley} A.~E., {Tinker} J.~L., {Santos} M.~R., {Lemson} G.,
  2008, \apj, 679, 1192

\bibitem[{{Cowie} \& {Songaila}(1998)}]{cowie98}
{Cowie} L.~L., {Songaila} A., 1998, \nat, 394, 44

\bibitem[{{Ellison} {et~al.}(2000){Ellison}, {Songaila}, {Schaye}, \&
  {Pettini}}]{ellison00}
{Ellison} S.~L., {Songaila} A., {Schaye} J., {Pettini} M., 2000, \aj, 120, 1175

\bibitem[{{Erb} {et~al.}(2006{\natexlab{a}}){Erb}, {Shapley}, {Pettini},
  {Steidel}, {Reddy}, \& {Adelberger}}]{erb06a}
{Erb} D.~K., {Shapley} A.~E., {Pettini} M., {Steidel} C.~C., {Reddy} N.~A.,
  {Adelberger} K.~L., 2006{\natexlab{a}}, \apj, 644, 813

\bibitem[{{Erb} {et~al.}(2006{\natexlab{b}}){Erb}, {Steidel}, {Shapley},
  {Pettini}, {Reddy}, \& {Adelberger}}]{erb06b}
{Erb} D.~K., {Steidel} C.~C., {Shapley} A.~E., {Pettini} M., {Reddy} N.~A.,
  {Adelberger} K.~L., 2006{\natexlab{b}}, \apj, 647, 128

\bibitem[{{Erb} {et~al.}(2006{\natexlab{c}}){Erb}, {Steidel}, {Shapley},
  {Pettini}, {Reddy}, \& {Adelberger}}]{erb06c}
---, 2006{\natexlab{c}}, \apj, 646, 107

\bibitem[{{Ford} {et~al.}(2013){Ford}, {Oppenheimer}, {Dav{\'e}}, {Katz},
  {Kollmeier}, \& {Weinberg}}]{ford13}
{Ford} A.~B., {Oppenheimer} B.~D., {Dav{\'e}} R., {Katz} N., {Kollmeier} J.~A.,
  {Weinberg} D.~H., 2013, \mnras, 432, 89

\bibitem[{{Franx} {et~al.}(1997){Franx}, {Illingworth}, {Kelson}, {van Dokkum},
  \& {Tran}}]{franx97}
{Franx} M., {Illingworth} G.~D., {Kelson} D.~D., {van Dokkum} P.~G., {Tran}
  K.-V., 1997, \apjl, 486, L75

\bibitem[{{Haas} {et~al.}(2013{\natexlab{a}}){Haas}, {Schaye}, {Booth}, {Dalla
  Vecchia}, {Springel}, {Theuns}, \& {Wiersma}}]{haas13a}
{Haas} M.~R., {Schaye} J., {Booth} C.~M., {Dalla Vecchia} C., {Springel} V.,
  {Theuns} T., {Wiersma} R.~P.~C., 2013{\natexlab{a}}, \mnras, 435, 2931

\bibitem[{{Haas} {et~al.}(2013{\natexlab{b}}){Haas}, {Schaye}, {Booth}, {Dalla
  Vecchia}, {Springel}, {Theuns}, \& {Wiersma}}]{haas13b}
---, 2013{\natexlab{b}}, \mnras, 435, 2955

\bibitem[{{Heckman} {et~al.}(1990){Heckman}, {Armus}, \& {Miley}}]{heckman90}
{Heckman} T.~M., {Armus} L., {Miley} G.~K., 1990, \apjs, 74, 833

\bibitem[{{Heckman} {et~al.}(2000){Heckman}, {Lehnert}, {Strickland}, \&
  {Armus}}]{heckman00}
{Heckman} T.~M., {Lehnert} M.~D., {Strickland} D.~K., {Armus} L., 2000, \apjs,
  129, 493

\bibitem[{{Jones} {et~al.}(2012){Jones}, {Stark}, \& {Ellis}}]{jones12}
{Jones} T., {Stark} D.~P., {Ellis} R.~S., 2012, \apj, 751, 51

\bibitem[{{Kaiser}(1987)}]{kaiser87}
{Kaiser} N., 1987, \mnras, 227, 1

\bibitem[{{Martin}(2005)}]{martin05}
{Martin} C.~L., 2005, \apj, 621, 227

\bibitem[{{McLean} {et~al.}(2012){McLean}, {Steidel}, {Epps}, {Konidaris},
  {Matthews}, {Adkins}, {Aliado}, {Brims}, {Canfield}, {Cromer}, {Fucik},
  {Kulas}, {Mace}, {Magnone}, {Rodriguez}, {Rudie}, {Trainor}, {Wang}, {Weber},
  \& {Weiss}}]{mclean12}
{McLean} I.~S., {Steidel} C.~C., {Epps} H.~W., {Konidaris} N., {Matthews}
  K.~Y., {Adkins} S., {Aliado} T., {Brims} G., {Canfield} J.~M., {Cromer}
  J.~L., {Fucik} J., {Kulas} K., {Mace} G., {Magnone} K., {Rodriguez} H.,
  {Rudie} G., {Trainor} R., {Wang} E., {Weber} B., {Weiss} J., 2012, in Society
  of Photo-Optical Instrumentation Engineers (SPIE) Conference Series, Vol.
  8446, Society of Photo-Optical Instrumentation Engineers (SPIE) Conference
  Series

\bibitem[{{Nielsen} {et~al.}(2013){Nielsen}, {Churchill}, {Kacprzak}, \&
  {Murphy}}]{nielsen13}
{Nielsen} N.~M., {Churchill} C.~W., {Kacprzak} G.~G., {Murphy} M.~T., 2013,
  \apj, 776, 114

\bibitem[{{Oppenheimer} \& {Dav{\'e}}(2008)}]{oppenheimer08}
{Oppenheimer} B.~D., {Dav{\'e}} R., 2008, \mnras, 387, 577

\bibitem[{{Oppenheimer} {et~al.}(2010){Oppenheimer}, {Dav{\'e}}, {Kere{\v s}},
  {Fardal}, {Katz}, {Kollmeier}, \& {Weinberg}}]{oppenheimer10}
{Oppenheimer} B.~D., {Dav{\'e}} R., {Kere{\v s}} D., {Fardal} M., {Katz} N.,
  {Kollmeier} J.~A., {Weinberg} D.~H., 2010, \mnras, 406, 2325

\bibitem[{{Pettini} {et~al.}(2000){Pettini}, {Steidel}, {Adelberger},
  {Dickinson}, \& {Giavalisco}}]{pettini00}
{Pettini} M., {Steidel} C.~C., {Adelberger} K.~L., {Dickinson} M., {Giavalisco}
  M., 2000, \apj, 528, 96

\bibitem[{{Planck Collaboration} {et~al.}(2013){Planck Collaboration}, {Ade},
  {Aghanim}, {Armitage-Caplan}, {Arnaud}, {Ashdown}, {Atrio-Barandela},
  {Aumont}, {Baccigalupi}, {Banday}, \& et~al.}]{planck13}
{Planck Collaboration}, {Ade} P.~A.~R., {Aghanim} N., {Armitage-Caplan} C.,
  {Arnaud} M., {Ashdown} M., {Atrio-Barandela} F., {Aumont} J., {Baccigalupi}
  C., {Banday} A.~J., et~al., 2013, ArXiv e-prints

\bibitem[{{Prochaska} {et~al.}(2011){Prochaska}, {Weiner}, {Chen}, {Mulchaey},
  \& {Cooksey}}]{prochaska11}
{Prochaska} J.~X., {Weiner} B., {Chen} H.-W., {Mulchaey} J., {Cooksey} K.,
  2011, \apj, 740, 91

\bibitem[{{Quider} {et~al.}(2009){Quider}, {Pettini}, {Shapley}, \&
  {Steidel}}]{quider09}
{Quider} A.~M., {Pettini} M., {Shapley} A.~E., {Steidel} C.~C., 2009, \mnras,
  398, 1263

\bibitem[{{Rakic} {et~al.}(2013){Rakic}, {Schaye}, {Steidel}, {Booth}, {Dalla
  Vecchia}, \& {Rudie}}]{rakic13}
{Rakic} O., {Schaye} J., {Steidel} C.~C., {Booth} C.~M., {Dalla Vecchia} C.,
  {Rudie} G.~C., 2013, \mnras, 433, 3103

\bibitem[{{Rakic} {et~al.}(2011){Rakic}, {Schaye}, {Steidel}, \&
  {Rudie}}]{rakic11}
{Rakic} O., {Schaye} J., {Steidel} C.~C., {Rudie} G.~C., 2011, \mnras, 414,
  3265

\bibitem[{{Rakic} {et~al.}(2012){Rakic}, {Schaye}, {Steidel}, \&
  {Rudie}}]{rakic12}
---, 2012, \apj, 751, 94

\bibitem[{{Reddy} {et~al.}(2008){Reddy}, {Steidel}, {Pettini}, {Adelberger},
  {Shapley}, {Erb}, \& {Dickinson}}]{reddy08}
{Reddy} N.~A., {Steidel} C.~C., {Pettini} M., {Adelberger} K.~L., {Shapley}
  A.~E., {Erb} D.~K., {Dickinson} M., 2008, \apjs, 175, 48

\bibitem[{{Rudie} {et~al.}(2012){Rudie}, {Steidel}, {Trainor}, {Rakic},
  {Bogosavljevi{\'c}}, {Pettini}, {Reddy}, {Shapley}, {Erb}, \&
  {Law}}]{rudie12}
{Rudie} G.~C., {Steidel} C.~C., {Trainor} R.~F., {Rakic} O.,
  {Bogosavljevi{\'c}} M., {Pettini} M., {Reddy} N., {Shapley} A.~E., {Erb}
  D.~K., {Law} D.~R., 2012, \apj, 750, 67

\bibitem[{{Rupke} {et~al.}(2005){Rupke}, {Veilleux}, \& {Sanders}}]{rupke05}
{Rupke} D.~S., {Veilleux} S., {Sanders} D.~B., 2005, \apjs, 160, 115

\bibitem[{{Savage} \& {Sembach}(1991)}]{savage91}
{Savage} B.~D., {Sembach} K.~R., 1991, \apj, 379, 245

\bibitem[{{Schaye} {et~al.}(2003){Schaye}, {Aguirre}, {Kim}, {Theuns}, {Rauch},
  \& {Sargent}}]{schaye03}
{Schaye} J., {Aguirre} A., {Kim} T.-S., {Theuns} T., {Rauch} M., {Sargent}
  W.~L.~W., 2003, \apj, 596, 768

\bibitem[{{Schaye} {et~al.}(2000){Schaye}, {Rauch}, {Sargent}, \&
  {Kim}}]{schaye00}
{Schaye} J., {Rauch} M., {Sargent} W.~L.~W., {Kim} T.-S., 2000, \apjl, 541, L1

\bibitem[{{Shapley} {et~al.}(2003){Shapley}, {Steidel}, {Pettini}, \&
  {Adelberger}}]{shapley03}
{Shapley} A.~E., {Steidel} C.~C., {Pettini} M., {Adelberger} K.~L., 2003, \apj,
  588, 65

\bibitem[{{Shen} {et~al.}(2013){Shen}, {Madau}, {Guedes}, {Mayer}, {Prochaska},
  \& {Wadsley}}]{shen13}
{Shen} S., {Madau} P., {Guedes} J., {Mayer} L., {Prochaska} J.~X., {Wadsley}
  J., 2013, \apj, 765, 89

\bibitem[{{Simcoe}(2011)}]{simcoe11}
{Simcoe} R.~A., 2011, \apj, 738, 159

\bibitem[{{Songaila}(1998)}]{songaila98}
{Songaila} A., 1998, \aj, 115, 2184

\bibitem[{{Steidel} {et~al.}(1999){Steidel}, {Adelberger}, {Giavalisco},
  {Dickinson}, \& {Pettini}}]{steidel99}
{Steidel} C.~C., {Adelberger} K.~L., {Giavalisco} M., {Dickinson} M., {Pettini}
  M., 1999, \apj, 519, 1

\bibitem[{{Steidel} {et~al.}(2003){Steidel}, {Adelberger}, {Shapley},
  {Pettini}, {Dickinson}, \& {Giavalisco}}]{steidel03}
{Steidel} C.~C., {Adelberger} K.~L., {Shapley} A.~E., {Pettini} M., {Dickinson}
  M., {Giavalisco} M., 2003, \apj, 592, 728

\bibitem[{{Steidel} {et~al.}(2010){Steidel}, {Erb}, {Shapley}, {Pettini},
  {Reddy}, {Bogosavljevi{\'c}}, {Rudie}, \& {Rakic}}]{steidel10}
{Steidel} C.~C., {Erb} D.~K., {Shapley} A.~E., {Pettini} M., {Reddy} N.,
  {Bogosavljevi{\'c}} M., {Rudie} G.~C., {Rakic} O., 2010, \apj, 717, 289

\bibitem[{{Steidel} {et~al.}(1996){Steidel}, {Giavalisco}, {Pettini},
  {Dickinson}, \& {Adelberger}}]{steidel96}
{Steidel} C.~C., {Giavalisco} M., {Pettini} M., {Dickinson} M., {Adelberger}
  K.~L., 1996, \apjl, 462, L17

\bibitem[{{Steidel} {et~al.}(2014){Steidel}, {Rudie}, {Strom}, {Pettini},
  {Reddy}, {Shapley}, {Trainor}, {Erb}, {Turner}, {Konidaris}, {Kulas}, {Mace},
  {Matthews}, \& {McLean}}]{steidel14}
{Steidel} C.~C., {Rudie} G.~C., {Strom} A.~L., {Pettini} M., {Reddy} N.~A.,
  {Shapley} A.~E., {Trainor} R.~F., {Erb} D.~K., {Turner} M.~L., {Konidaris}
  N.~P., {Kulas} K.~R., {Mace} G., {Matthews} K., {McLean} I.~S., 2014, ArXiv
  e-prints

\bibitem[{{Steidel} {et~al.}(2004){Steidel}, {Shapley}, {Pettini},
  {Adelberger}, {Erb}, {Reddy}, \& {Hunt}}]{steidel04}
{Steidel} C.~C., {Shapley} A.~E., {Pettini} M., {Adelberger} K.~L., {Erb}
  D.~K., {Reddy} N.~A., {Hunt} M.~P., 2004, \apj, 604, 534

\bibitem[{{Stocke} {et~al.}(2006){Stocke}, {Penton}, {Danforth}, {Shull},
  {Tumlinson}, \& {McLin}}]{stocke06}
{Stocke} J.~T., {Penton} S.~V., {Danforth} C.~W., {Shull} J.~M., {Tumlinson}
  J., {McLin} K.~M., 2006, \apj, 641, 217

\bibitem[{{Trainor} \& {Steidel}(2012)}]{trainor12}
{Trainor} R.~F., {Steidel} C.~C., 2012, \apj, 752, 39

\bibitem[{{Tremonti} {et~al.}(2007){Tremonti}, {Moustakas}, \&
  {Diamond-Stanic}}]{tremonti07}
{Tremonti} C.~A., {Moustakas} J., {Diamond-Stanic} A.~M., 2007, \apjl, 663, L77

\bibitem[{{Tumlinson} {et~al.}(2011){Tumlinson}, {Thom}, {Werk}, {Prochaska},
  {Tripp}, {Weinberg}, {Peeples}, {O'Meara}, {Oppenheimer}, {Meiring}, {Katz},
  {Dav{\'e}}, {Ford}, \& {Sembach}}]{tumlinson11}
{Tumlinson} J., {Thom} C., {Werk} J.~K., {Prochaska} J.~X., {Tripp} T.~M.,
  {Weinberg} D.~H., {Peeples} M.~S., {O'Meara} J.~M., {Oppenheimer} B.~D.,
  {Meiring} J.~D., {Katz} N.~S., {Dav{\'e}} R., {Ford} A.~B., {Sembach} K.~R.,
  2011, Science, 334, 948

\bibitem[{{Tummuangpak} {et~al.}(2013){Tummuangpak}, {Shanks}, {Bielby},
  {Crighton}, {Francke}, {Infante}, \& {Theuns}}]{tummuangpak13}
{Tummuangpak} P., {Shanks} T., {Bielby} R., {Crighton} N.~H.~M., {Francke} H.,
  {Infante} L., {Theuns} T., 2013, ArXiv e-prints

\bibitem[{{Weiner} {et~al.}(2009){Weiner}, {Coil}, {Prochaska}, {Newman},
  {Cooper}, {Bundy}, {Conselice}, {Dutton}, {Faber}, {Koo}, {Lotz}, {Rieke}, \&
  {Rubin}}]{weiner09}
{Weiner} B.~J., {Coil} A.~L., {Prochaska} J.~X., {Newman} J.~A., {Cooper}
  M.~C., {Bundy} K., {Conselice} C.~J., {Dutton} A.~A., {Faber} S.~M., {Koo}
  D.~C., {Lotz} J.~M., {Rieke} G.~H., {Rubin} K.~H.~R., 2009, \apj, 692, 187

\bibitem[{{Wiersma} {et~al.}(2010){Wiersma}, {Schaye}, {Dalla Vecchia},
  {Booth}, {Theuns}, \& {Aguirre}}]{wiersma10}
{Wiersma} R.~P.~C., {Schaye} J., {Dalla Vecchia} C., {Booth} C.~M., {Theuns}
  T., {Aguirre} A., 2010, \mnras, 409, 132

\bibitem[{{Wiersma} {et~al.}(2011){Wiersma}, {Schaye}, \& {Theuns}}]{wiersma11}
{Wiersma} R.~P.~C., {Schaye} J., {Theuns} T., 2011, \mnras, 415, 353

\bibitem[{{Zhu} {et~al.}(2014){Zhu}, {M{\'e}nard}, {Bizyaev}, {Brewington},
  {Ebelke}, {Ho}, {Kinemuchi}, {Malanushenko}, {Malanushenko}, {Marchante},
  {More}, {Oravetz}, {Pan}, {Petitjean}, \& {Simmons}}]{zhu14}
{Zhu} G., {M{\'e}nard} B., {Bizyaev} D., {Brewington} H., {Ebelke} G., {Ho} S.,
  {Kinemuchi} K., {Malanushenko} V., {Malanushenko} E., {Marchante} M., {More}
  S., {Oravetz} D., {Pan} K., {Petitjean} P., {Simmons} A., 2014, \mnras

\bibitem[{{Zibetti} {et~al.}(2005){Zibetti}, {M{\'e}nard}, {Nestor}, \&
  {Turnshek}}]{zibetti05}
{Zibetti} S., {M{\'e}nard} B., {Nestor} D., {Turnshek} D., 2005, \apjl, 631,
  L105

\end{thebibliography}

%%%%%%%%%%%%	
% Appendix %
%%%%%%%%%%%%

\appendix

\section{The pixel optical depth method}
\label{sec:pod_details}

In this appendix, we describe our implementation of the pixel optical
depth technique. Our version is based on \citet{aguirre02}, which 
was itself based on \citet{cowie98, ellison00, schaye00}, and we will
point out where our method differs from that of \citet{aguirre02}. 

Using the redshift ranges given in \S~\ref{sec:od_recovery}, 
we calculate the optical depth for 
each species $Z$ and each multiplet component $k$ (if applicable) as:
\begin{equation}
 \tau_{Z,k}(z) \equiv -\ln[F_{Z,k}(z)]
\end{equation}
where $F_{Z,k}(z)$ is the normalised flux at redshift $z$, obtained from 
\begin{equation}
 F_{Z,k}(z) \equiv F(\lambda = \lambda_{Z,k} [1 + z])
\end{equation}
where $\lambda_{Z,k}$ is the transition's rest wavelength. 
For all optical depths other than $\tau_{\honem,\lyam}(z)$, linear interpolation of the flux
is used so that all discrete $z$ values are the same as for $\tau_{\honem,\lyam}(z)$.
Before beginning the correction, we search for saturated pixels, 
defined as those pixels satisfying
$F(\lambda)\leq N_{\sigma}\sigma(\lambda)$,
where $\sigma(\lambda)$ is the normalised noise array, 
and we set $N_{\sigma} = 3$. Since the optical depths of the saturated pixels
are not reliable, but will be high compared to the unsaturated pixels,
we set their values to $\tau_{Z,k}(z) = 10^{6}$. 
Using such an extreme optical depth to flag the saturated pixels will 
not skew our results provided median statistics are used throughout the analysis. 

First we perform the recovery for \hone, since we will need the recovered optical depths
to correct metals which are contaminated by higher order Lyman lines.  
For \hone, the main source of error comes from the saturation of the \lya\ lines. 
Therefore, for all \lya\ pixels previously flagged as saturated, we look to all available 
higher order Lyman lines (in practice we have used $N=16$ higher 
order components). We only use higher order pixels which satisfy
$N_\sigma \sigma(\lambda)\leq F(\lambda)\leq1-N_\sigma \sigma(\lambda)$
at $\lambda = \lambda_{\honem,\lynm} [1 + z]$, to ensure that they are 
not poorly detected.  Scaling the higher order optical depths to the \lya\ component strength,
we set the recovered \lya\ optical depth, $\tau_{\honem,\lyam}^{\rm rec}(z)$ 
to be the minimum of all well-detected higher order pixels:
\begin{equation}
\tau^{\rm rec}_{\honem,\lyam}(z) \equiv \min \left[\dfrac{\tau_{\honem,\lynm}(z)g_{\honem,\lyam}}{g_{\honem,\lynm}}\right].
\end{equation}
Here,  $g_{\honem,\lynm} \equiv f_{\honem,\lynm} \lambda_{\honem,\lynm}$ where
$f_{\honem,\lynm}$ is the oscillator strength.

In addition to the above correction for saturated pixels
which is identical to the one used in \citet{ellison00} and  \citet{aguirre02},
we implement a procedure to search for and flag contaminated \lya\ pixels.
We consider a \lya\ pixel to be contaminated if
\begin{equation}
\begin{split}
&F_{\honem,\lynm}(z) - N_\sigma\sigma(\lambda = \lambda_{\honem,\lynm} [1 + z]) > \\
   &\max \left[ F_{\honem,\lyam}(z), N_\sigma \sigma(\lambda_{\honem,\lyam} [1 + z])  \right]^{g_{\honem,\lynm}/g_{\honem,\lyam}}.
\end{split}
\end{equation}
Specifically, for unsaturated \lya\ pixels, we scale the \lya\ flux to obtain the flux expected at 
higher-order transition $n$; if the observed flux at transition $n$ is 
significantly greater than would be expected from the \lya\ flux, the 
\lya\ pixel is likely contaminated. In order to apply this algorithm 
to saturated \lya\ pixels (for which the flux estimate is likely to 
be unreliable), we instead use the \lya\ noise array to set the \lya\ flux. 
All \lya\ pixels found to be contaminated in this way are flagged and 
discarded for the remainder of the analysis.

The recovered \hone\ \lya\ optical depths are then used to correct 
unsaturated pixels of metals contaminated by Lyman series lines.
For the recovery of \osix\ and \cthree, which are located in the \lyb\ forest,
the procedure involves the subtraction of the Lyman lines starting with \lyb\ (i.e., $n=2$):
\begin{equation}
 \tau_{Z,k}(z) := \tau_{Z,k}(z) - \sum_{n=2}^{N} \dfrac{g_{\honem,\lynm}}{g_{\honem,\lyam}} \tau^{\rm rec}_{\honem,\lyam} 
 \left(\dfrac{\lambda \lambda_{\honem,\lyam}}{ \lambda_{\honem,\lynm}}\right)
\label{eq:honesub}
\end{equation}
where $\lambda=\lambda_{Z,k}[1+z]$ and we take $N=5$ (the effects of varying
$N$ are explored in Appendix~\ref{sec:o6_var_pod}). 
For \osix, the \hone\ optical depth subtraction procedure
is performed for both multiplet components $k$.

Something else that we have implemented in the \hone\ subtraction
procedure  which was not done in \citet{aguirre02},
is the treatment of saturated metal pixels. 
More specifically, since the true optical depth of saturated pixels
cannot be measured accurately in the presence of noise, 
we cannot make a reliable \hone\ subtraction. 
Hence, we do not apply  Equation~\ref{eq:honesub} to pixels for which 
$F(\lambda)\leq N_{\sigma}\sigma(\lambda)$. 
Rather, if the sum of contaminating \hone\ optical depths is sufficient to saturate
the absorption, i.e. if
\begin{equation}
 \exp\left[-\sum_{n=2}^{N} \dfrac{g_{\honem,\lynm}}{g_{\honem,\lyam}} \tau^{\rm rec}_{\honem,\lyam} 
\left(\dfrac{\lambda \lambda_{\honem,\lyam}}{ \lambda_{\honem,\lynm}}\right)\right] <  N_{\sigma}\sigma(\lambda),
\end{equation}
then the pixel cannot be used to estimate the metal optical depth. 
For \cthree, this pixel is discarded immediately. 
Since \osix\ is a doublet and we are able to compare the two components in the next
 contamination correction step, it is not as crucial to discard all contaminated pixels
right away. Instead, we perform this flagging procedure on each of the
doublet components and only discard those redshifts for which both components are contaminated. 

As mentioned above, the next step is to correct the optical depths
of metals which have two multiplet components $k$ (\osix, \nfive, and \sifour)
by taking the minimum of the the optical depths, where the optical depth 
of the weaker component is scaled to match that of the stronger one. 
Specifically, we take:
\begin{equation}
\tau_{Z,1}^{\rm rec}(z) = \min \left[\tau_{Z,1}(z), \dfrac{g_{Z,1}}{g_{Z,2}} \tau_{Z,2}(z)\right].
\end{equation}
However, it may not be a good idea to use the minimum optical depth
at every pixel. This is because, particularly
in the case where both pixels are not contaminated, doing so will
result in an optical depth value that is biased low.
To combat this,  \citet{aguirre02} would only take the optical depth of the weaker 
component if it were positive, i.e. $\tau_{Z,2}(z) > 0$. 
Here we use a different approach, where we take the noise level into account,
and only use the optical depth of the weaker component if:
\begin{equation}
\begin{split}
 &\left[\exp[-\tau_{Z,2}(z)] - N_\sigma \sigma(\lambda = \lambda_{Z,2}[1+z]) \right]^{\frac{g_{Z,1}}{g_{Z,2}}}  > \\
&\exp[-\tau_{Z,1}(z)]  + N_\sigma \sigma(\lambda = \lambda_{Z,1}[1+z]).
\end{split}
\label{eq:doubmin}
  \end{equation}
The left hand side of this condition contains the optical depth of the 
weaker component expressed in terms of flux, with the noise term subtracted,
and then scaled to match the strength of the stronger component optical depth,
which is on the right-hand side. 
This condition states that the scaled optical depth of the weaker component must be 
\textit{significantly} lower than that of the stronger one in order 
that it be used for the recovered optical depth value.

Finally, the rest wavelength of \cfour\ puts the recovery region redwards of the
\lya\ forest, and the majority of the contamination therefore consists of 
self-contamination from its own doublet.
We correct for this self-contamination as follows \citet{aguirre02}. 
First, every pixel is checked for contamination from other
ions, which is done by testing to see if its optical depth is higher
than what would be expected from self-contamination alone. 
  Specifically, we define $\tau_{\cfourm,1}(\lambda)$ as $\tau_{\cfourm,1}(z)$ 
where $z = \lambda / \lambda_{\cfourm,1} - 1$.
If the optical depth at wavelength $\lambda$ comes from the weaker component,
we want to know the optical depth at the location of the strong component
 $\lambda_s = [\lambda_{\cfourm,1} / \lambda_{\cfourm,2}] \lambda$. We also 
want to consider the case where the optical depth at $\lambda$ comes from the 
strong component, and therefore need the optical depth at the wavelength of the 
weak component, $\lambda_w = \lambda_{\cfourm,2} [1 + z]$. 
We then scale these optical depths to test if the optical depth at $\lambda$ 
is higher than expected from self-contamination, which gives the condition:
\begin{equation}
\begin{split}
&\exp\left[-\tau_{\cfourm,1}(\lambda)\right] + N_\sigma \bar\sigma(\lambda) <  \\
&\exp\left[-\dfrac{g_{\cfourm,2}}{g_{\cfourm,1}} \tau_{\cfourm,1}(\lambda_s) - 
   \dfrac{g_{\cfourm,1}}{g_{\cfourm,2}}\tau_{\cfourm,1}(\lambda_w)\right]
\end{split}
\end{equation}
where $\bar\sigma^2(\lambda)=\sigma^2(\lambda)+\sigma^2(\lambda_s)+\sigma^2(\lambda_w)$; 
pixels meeting this condition are probably contaminated with absorption from other ions, 
and are therefore not used for the correction or any subsequent analysis.
Then, an iterative doublet subtraction algorithm is used to remove self-contamination,
where the scaled optical depth of the strong component at $\lambda_s$ 
is subtracted from the optical depth at $\lambda$. 
Before doing this, for those pixels with $\tau_{\cfourm,1}(z)< 0$,we set 
$\tau_{\cfourm,1}(z) = 10^{-4}$ 
so that negative optical depth values won't affect the subtraction procedure.
After first setting $\tau_{\cfourm,1}^{\rm rec}(\lambda) = \tau_{\cfourm,1}(\lambda)$,
the following is repeated until convergence occurs (about 5 iterations):
\begin{equation}
 \tau_{\cfourm,1}^{\rm rec}(\lambda):=\tau_{\cfourm,1}(\lambda)-\dfrac{g_{\cfourm,2}}{g_{\cfourm,1}}
 \tau_{\cfourm,1}^{\rm rec}\left(\lambda_s\right).
\end{equation}

Here we note the implementation of an automated continuum fitting procedure which we apply
to the spectral regions with wavelengths greater than that of their quasar's \lya\ emission. 
The purpose of this is to homogenise the continuum fitting errors, and it is performed
as follows. This region of each spectrum is divided into bins of size $\Delta \lambda$ in the rest-frame, 
each with central wavelength $\lambda_i$ and median flux $\bar{f_i}$. We then interpolate 
a B-spline through $\bar{f_k}$ and discard any pixels with flux values that are $N_{\sigma}^{cf}\times\sigma$ 
below the interpolated flux values. $\bar{f_k}$ is recalculated without the discarded pixels, 
and the procedure is repeated until convergence is reached. 
In our implementation, we use $N_{\sigma}^{cf}=2$ and $\Delta \lambda = 20$~\AA\ as in \citet{schaye03}.

Finally, we discuss our masking procedure. As mentioned in \S~\ref{sec:gal_sample}, six
of the QSOs have DLAs located in their \lya\ forest regions. 
For the recovery of \hone\ and \nfive, we use the spectra that have had the damping
wings of these DLAs fitted out and the saturated region masked. However, for the recovery 
of \osix\ and \cthree, we want to subtract as much \hone\ contamination as possible. Therefore,
we also perform an \hone\ recovery without the \lya\ forest region DLAs fitted out and masked,
and use these values for the subtraction of contamination from the \osix\ and \cthree\ regions. 
We found that the results for \osix\ and \cthree\ obtained from using the unmasked DLA \hone\ recovery
are similar to those obtained from masking out the higher order components of the DLA by hand.
Lastly, for all recoveries we have masked out two more DLAs which are bluewards
of their QSO's \lya\ forest region, 
i.e. saturated Lyman continuum absorption associated with strong \hone\ absorbers,
 and we have also masked out all Lyman break regions,
which occurs for three of our QSOs.

\def\appwid{0.35\textwidth}
\def\appwidh{0.35\textwidth}

\section{Variation in pixel optical depth recovery}
\label{sec:var_pod}

In this section we explore the sensitivity of our results
to variations in the method used to recover the pixel optical depths. 
For this purpose we will compare the enhancement in the median optical depth
relative to a random location as a function of 3-D Hubble distance. 
Note that corrections may change the median \taumedk\ even if they do 
not change the median \tauk/\taumedk\ plotted in this section. As we will 
demonstrate, our results are not particularly sensitive to any of the 
corrections that we apply in the recovery of pixel optical depths.

\begin{figure}
\includegraphics[width=\appwid]{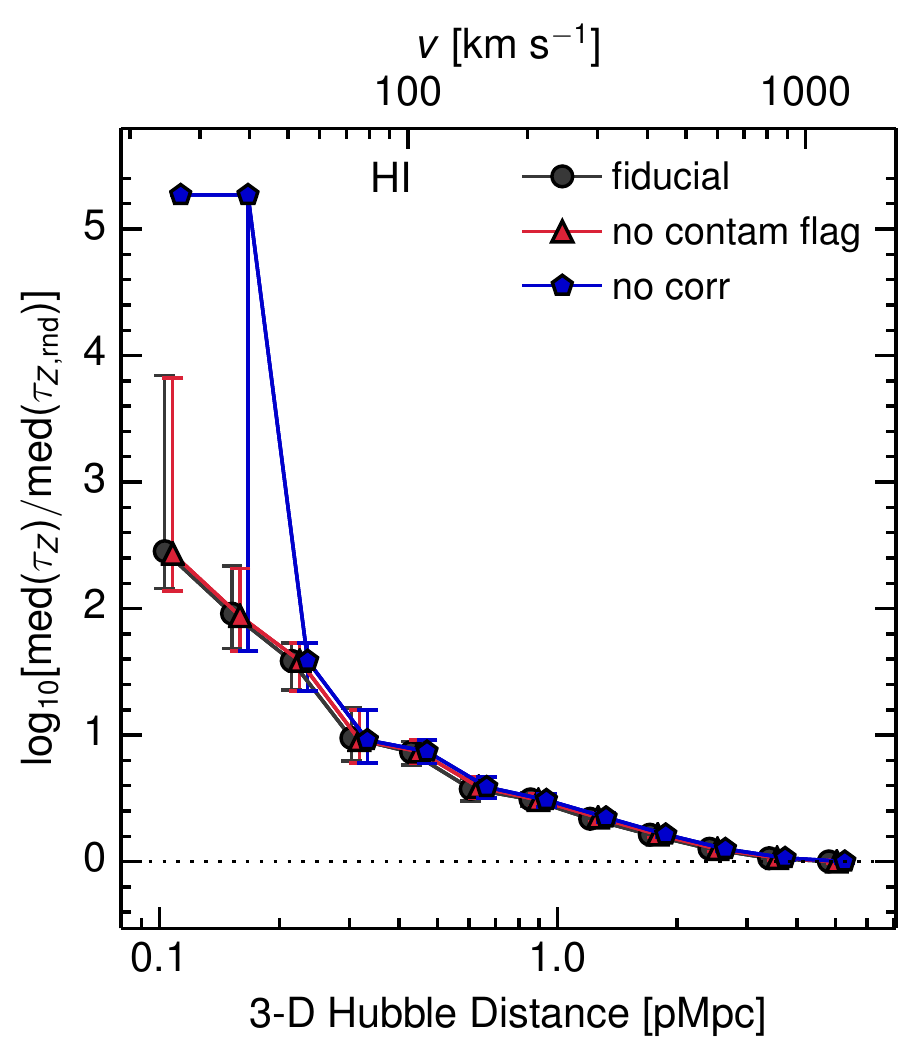}
\caption{Similar to Figure~\ref{fig:hubble_a}, 
but only for \hone, dividing by \taumedk, and modifying the optical depth recovery method.
The modifications are: not flagging pixels determined to be contaminated from their
 higher-order \hone\ flux (``\textit{no contam flag}''), and doing no correction at all 
 (i.e. not using higher-order \hone\ lines to correct saturated \hone\ \lya\ pixels, 
  ``\textit{no corr}''). The points are offset 
horizontally by 0.02~dex for clarity.}
\label{fig:o6d}
\end{figure}

\subsection{\hone}

We begin by examining one of the changes made to the \hone\ \lya\ recovery
algorithm with respect to that used in \citet{aguirre02}, which is the 
use of higher-order \hone\ flux to flag contaminated \hone\ \lya\ pixels.
We show the resulting median optical depth divided by \taumedk\ as a function 
of Hubble distance in Figure~\ref{fig:o6d} for the fiducial \hone\ recovery 
as well as one performed without flagging contaminated pixels, and we can 
see that this change in the algorithm has almost no impact on the results.
We have also plotted the median optical depths obtained when no \hone\ correction
is done (i.e. saturated \hone\ \lya\ pixels are not corrected using higher-order
\hone\ components). This correction is most important for small galactocentric distances,
where the median optical depths actually reach the optical depth value
used to flag saturated pixels. However, the absence of this correction would not 
affect our overall conclusions.

\begin{figure}
\includegraphics[width=\appwid]{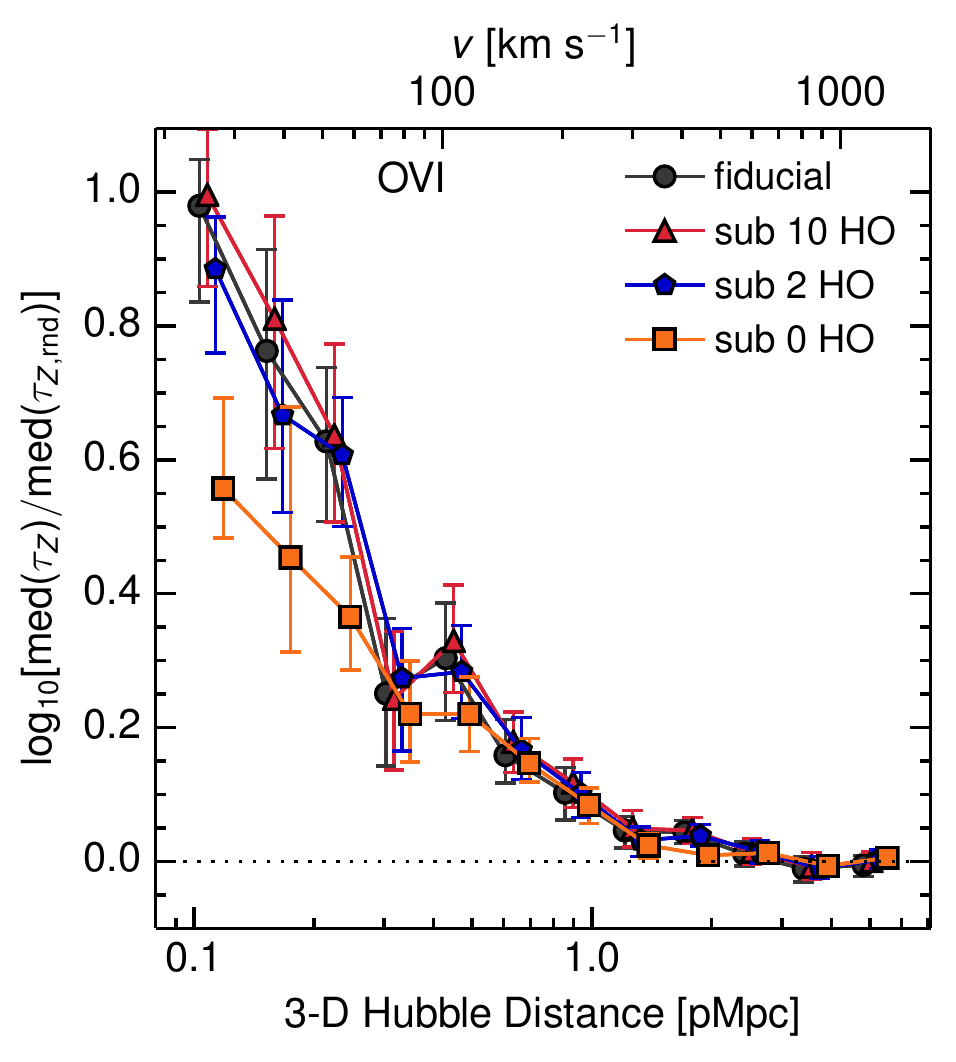}
\caption{Similar to Figure~\ref{fig:o6d}, 
but for \osix, and modifying the optical depth recovery method
with respect to the fiducial one (which invokes the subtraction of 
five higher order \hone\ lines). The modifications are: subtracting 10, 2, and 0 higher-order 
(HO) \hone\ lines, where the doublet minimum is taken in every case. }
\label{fig:o6a}
\end{figure}

\begin{figure}
\includegraphics[width=\appwid]{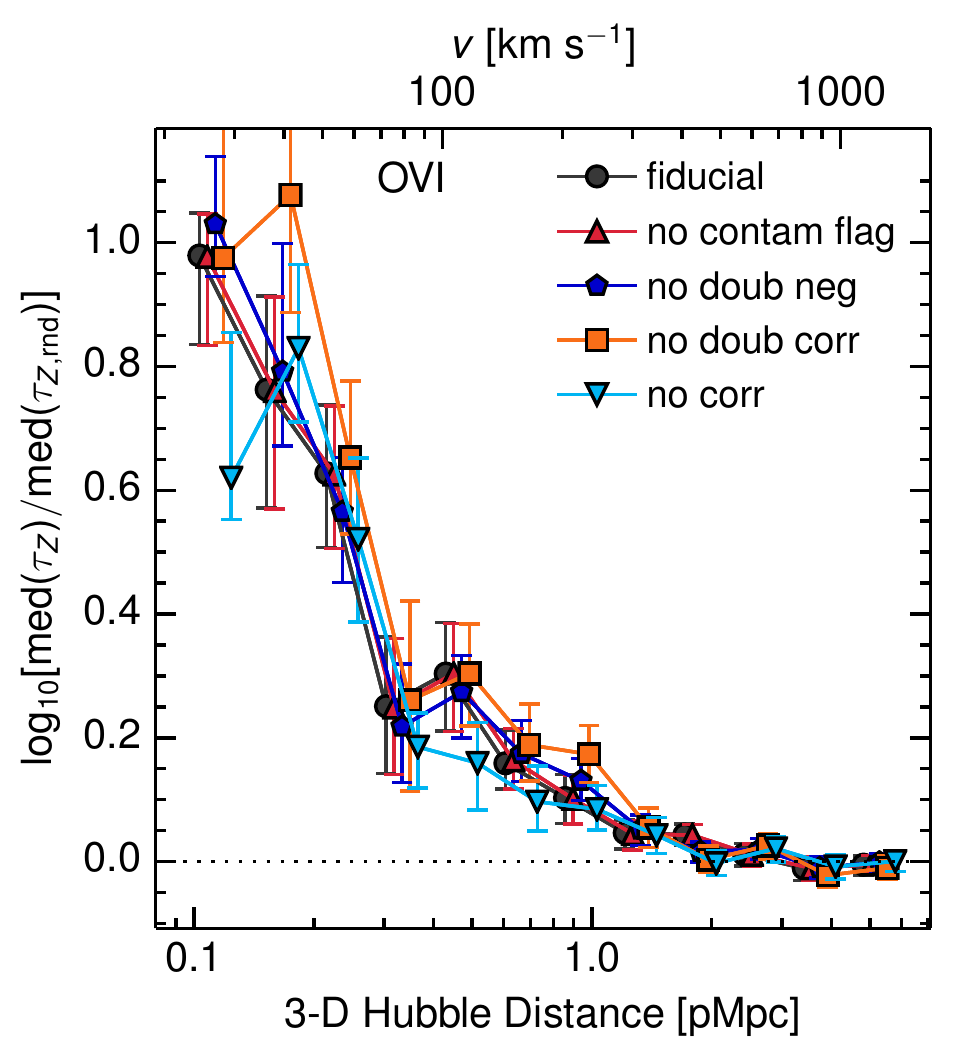}
\caption{Similar to Figure~\ref{fig:o6d}, but for the following modifications with respect
to the fiducial optical depth recovery: 
pixels determined to be saturated with contaminating \hone\ are not flagged (``\textit{no contam flag}'');
the weaker doublet component is not considered if it has a negative optical depth (``\textit{no doub neg}'');
no doublet minimum taken (``\textit{no doub corr}''); 
and no correction at all (i.e., neither \hone\ subtraction nor doublet minimum, ``\textit{no corr}''). 
 }
\label{fig:o6b}
\end{figure}

\subsection{\osix}

\label{sec:o6_var_pod}	

For \osix, we first vary the number of higher order \hone\ Lyman lines that are subtracted
from the \osix\ region. The fiducial number that we use is five, and 
we have also tried
subtracting ten, two, and zero higher order \hone\ lines, 
the outcomes of which are shown in Figure~\ref{fig:o6a}. In the recoveries where
at least two higher-order \hone\ lines are subtracted, the curves are almost identical. 
The main   difference (which cannot be seen from this figure)
is that the median optical depth in random locations (\taumedk) increases as
fewer \hone\ lines are subtracted. 

Next we investigate the effects of the pixel optical depth 
recovery modifications with respect to \citet{aguirre02}. 
Specifically, in Figure~\ref{fig:o6b}, we show the median
optical depth that results from 
not discarding pixels for
which the contaminating \hone\ absorption is thought to be saturated based on 
the sum of corresponding higher-order \hone\ optical depths;
and also using the doublet minimum condition from \citet{aguirre02}: that is,
instead of using Equation~\ref{eq:doubmin}, the doublet minimum is taken for 
every pixel, except for pixels where the weaker component has a negative optical depth.
Neither of these changes has any impact on the resulting median optical depth.
In Figure~\ref{fig:o6b}, we also look at the median optical depth from recoveries where 
we did not perform the doublet correction (that is, we still subtract
five higher order \hone\ lines, but then do not take the minimum
of the \osix\ doublet at each redshift); 
finally where we did not perform any correction at all. 
Although not taking the doublet minimum does not have a significant
affect on the outcome, not performing any correction at all 
does reduce the dynamic range with respect to the fiducial case. 

\begin{figure*}
\includegraphics[width=\appwid]{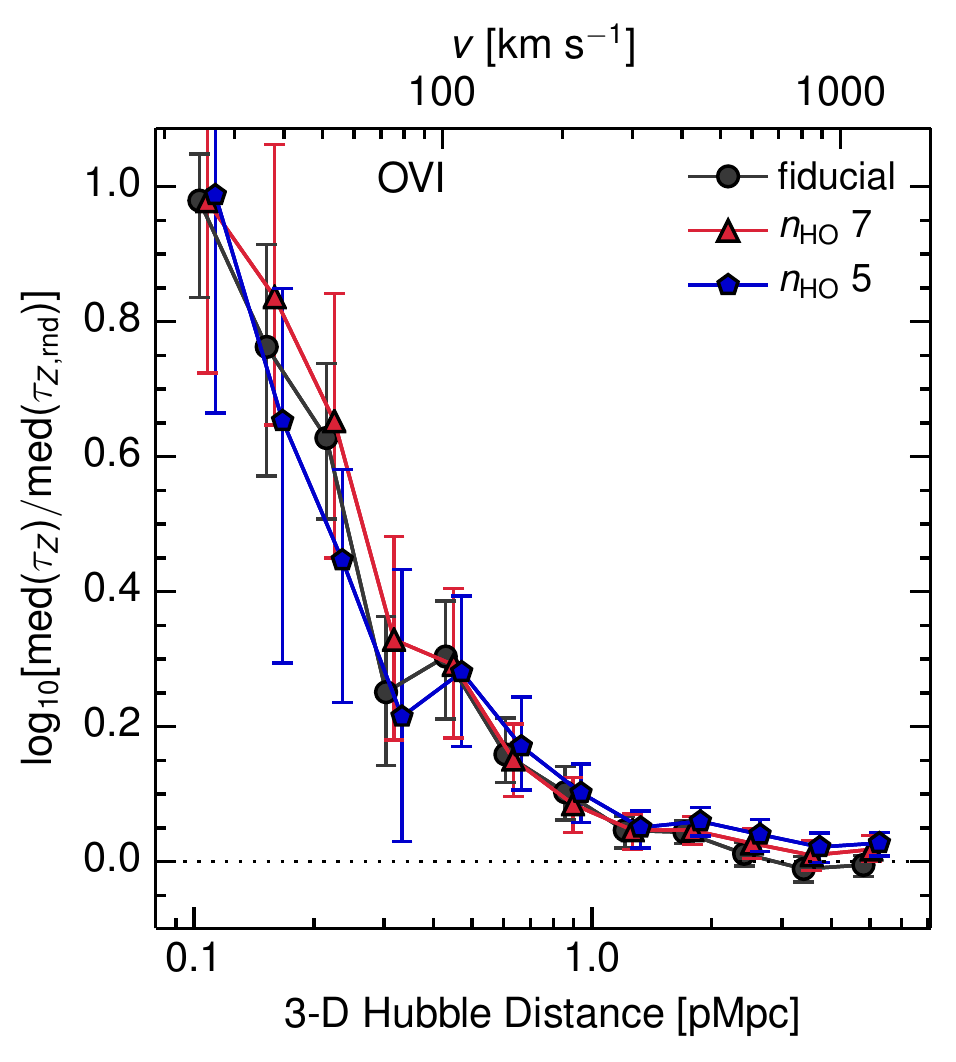}
\includegraphics[width=0.28\textwidth]{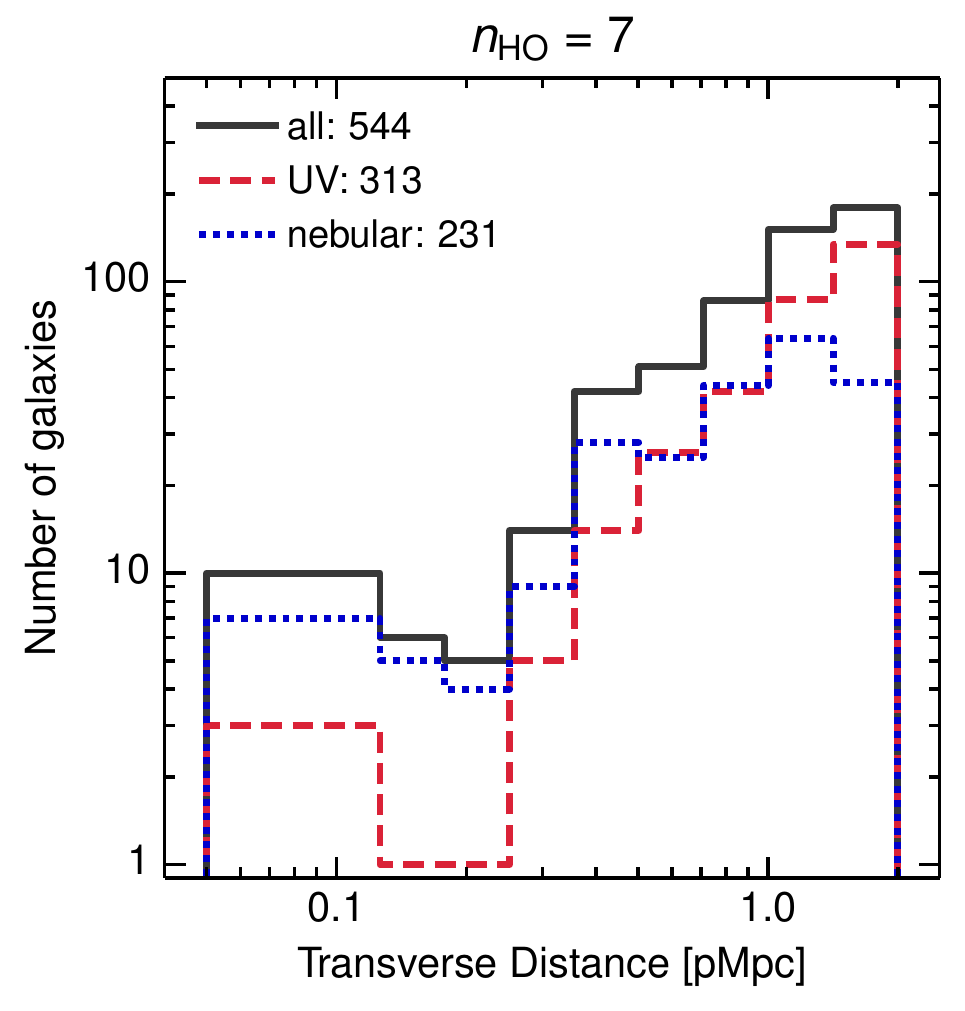}
\includegraphics[width=0.28\textwidth]{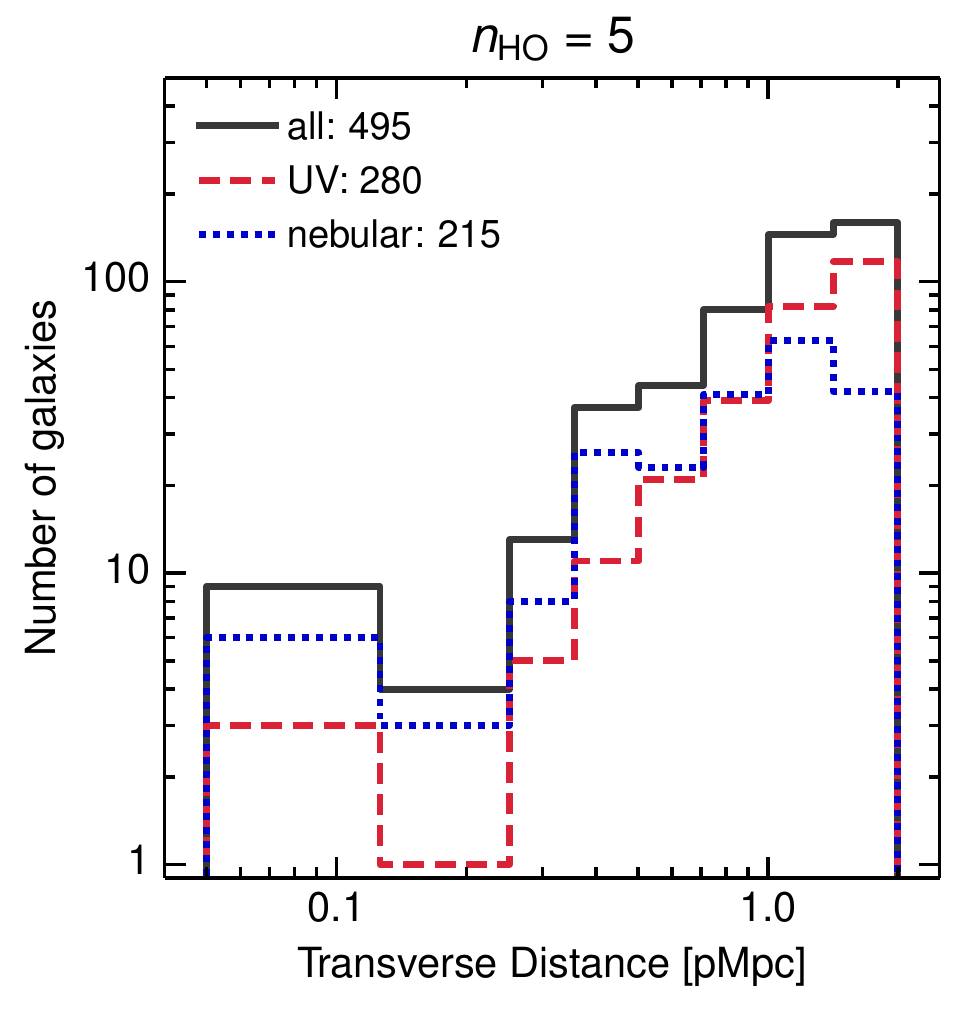}\\
\caption{\emph{Left}: Similar to Figure~\ref{fig:o6d}, 
but limiting the spectral region to that contaminated by:
any number (fiducial), seven and five higher order \hone\ lines. 
\emph{Centre and right}: Histograms of the number of galaxies per impact parameter bin for 
each of the spectral region cuts shown above. For reference, the fiducial number 
of galaxies in each bin is \ngaltotal\ galaxies in total, with \ngaluv\ and 
\ngalneb\ having their redshifts measured from rest-frame UV features and nebular emission
lines, respectively.  }
\label{fig:o6e}
\end{figure*}

Another technique which can be applied to the \osix\ recovery is to make redshift cuts such that those wavelength ranges
 contaminated by more higher order \hone\ lines are removed from the analysis.
For example, one can choose to limit the recovery to  the region in 
the spectrum where only three \hone\ lines (\lya, \lyb, \lyg, and no further higher order 
lines) are present. In the top row Figure~\ref{fig:o6e} we show
the result of limiting the spectral region to that contaminated by all, seven,
and five higher order \hone\ lines, respectively.
For the five higher order \hone\ line limit, there are so few galaxies in the innermost
impact parameter bins (right panel of Figure~\ref{fig:o6e} 
that the error bars are likely not reliable.
In general, we find these cuts
slightly increase the dynamic range of the optical depths, but for 
up to 3 higher order lines, i.e. our most restrictive cut, the dynamic
range in $\tau_{\osixm}$ is significantly smaller.
Given the low number of galaxies at the smallest impact parameters where the optical
depth is enhanced with respect to the median (and the even lower numbers resulting from 
reducing the sample size), such differences are probably 
due to small number statistics.

\begin{figure*}
\includegraphics[width=\appwid]{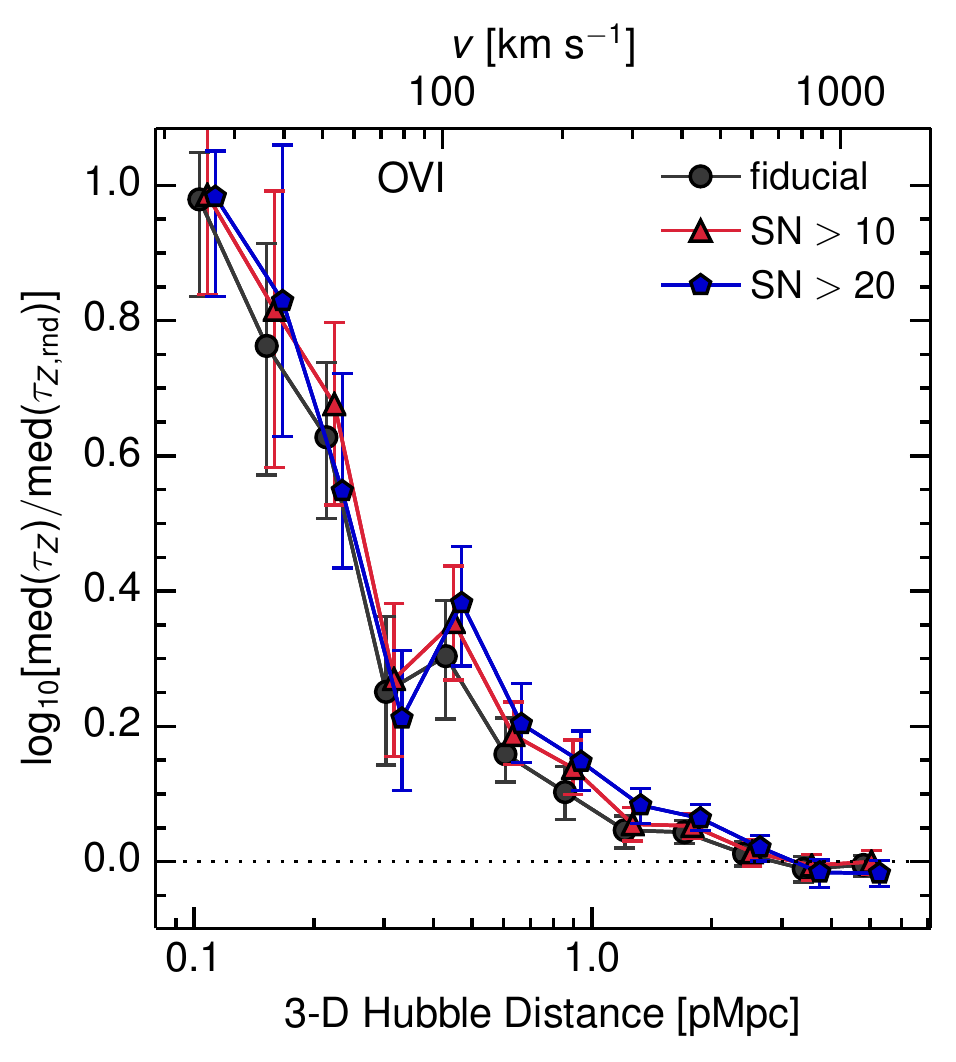} 
\includegraphics[width=0.28\textwidth]{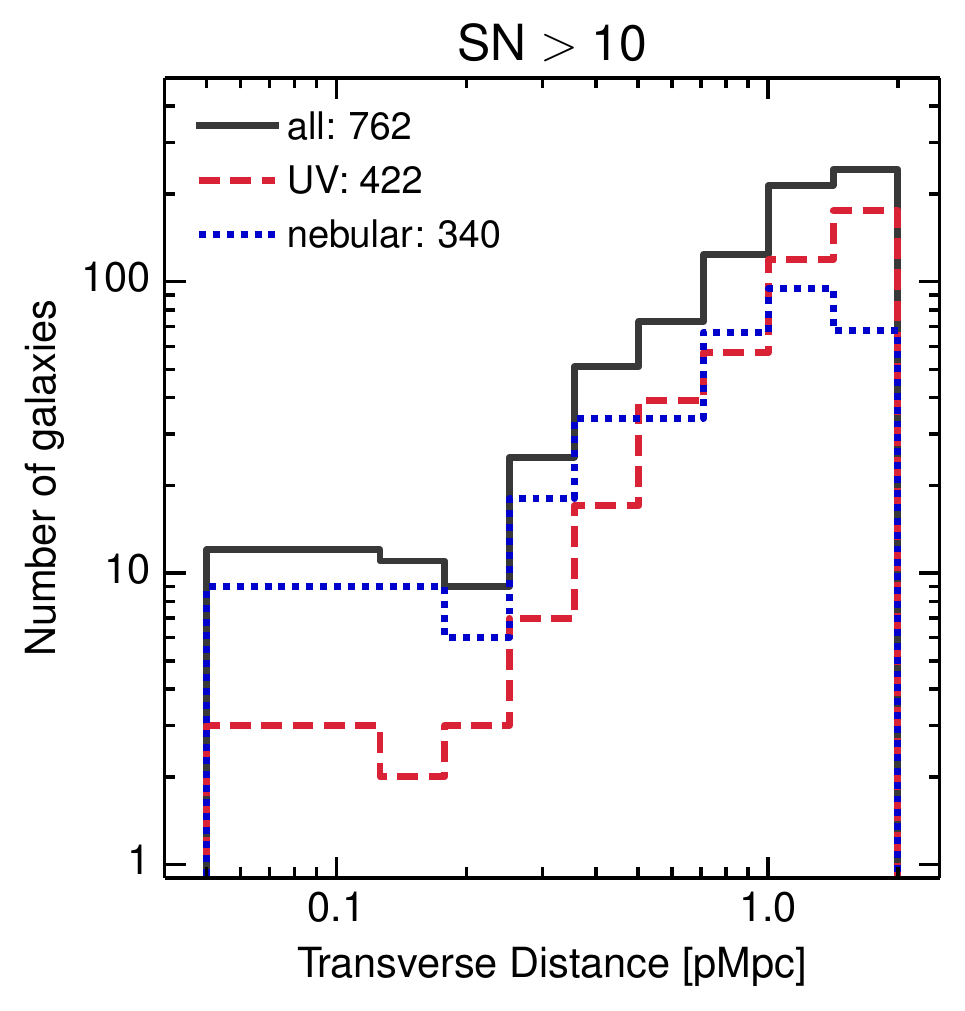} 
\includegraphics[width=0.28\textwidth]{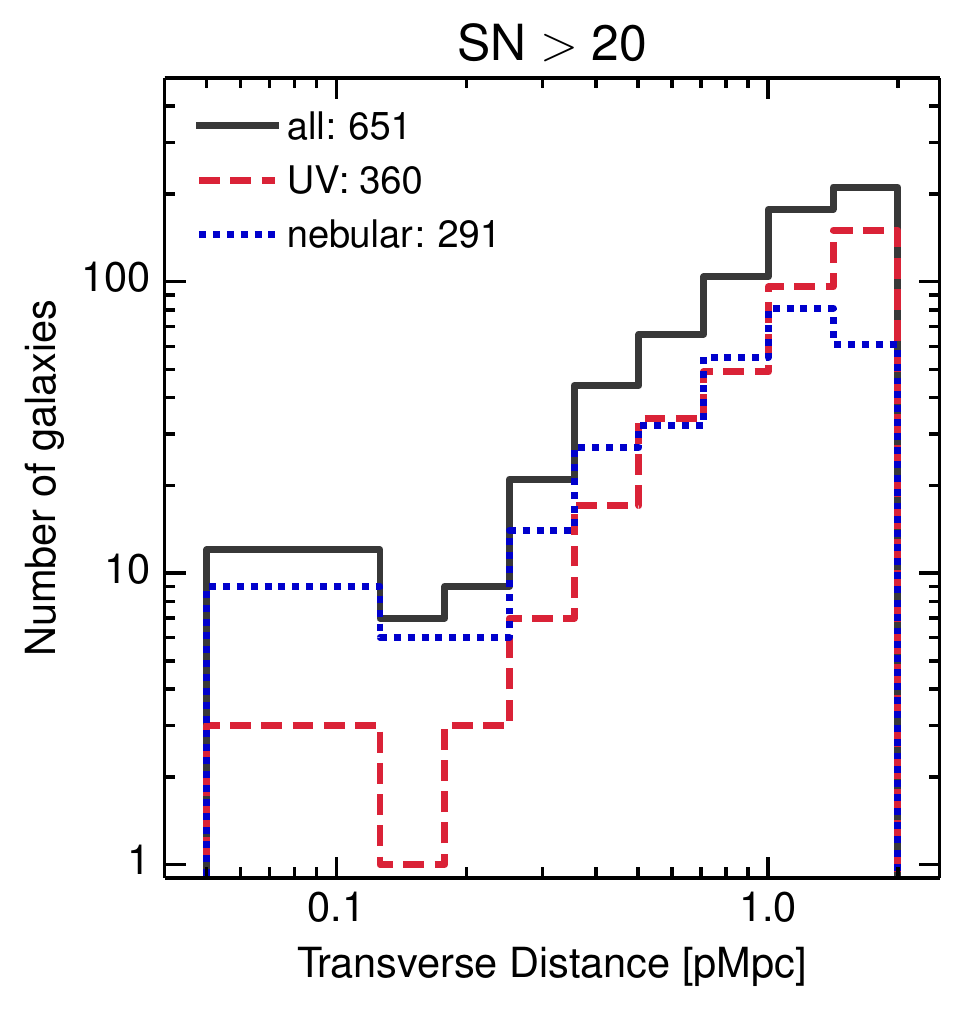} \\
\caption{\emph{Right}: Similar to Figure~\ref{fig:o6d}, 
but limiting the used spectral region to that having
S/N greater than 0 (fiducial), 10 and 20. 
\emph{Centre and right}: Histograms of the number of galaxies per impact 
parameter bin for each of the spectral region S/N cuts shown above.
For reference, the fiducial number 
of galaxies in each bin is \ngaltotal\ galaxies in total, with \ngaluv\ and 
\ngalneb\ having their redshifts measured from rest-frame UV features and nebular emission
lines, respectively.  }
\label{fig:o6f}
\end{figure*}

Additionally, one could optimise the \osix\ recovery by making cuts in the 
S/N of the HIRES spectra, since it is at the lowest wavelengths of Keck (those of the \osix\ region)
where the S/N declines quite rapidly. We try excluding individual regions with S/N less than 
10 and 20 (using a higher S/N cut results in $<10$ galaxies in the smallest 
impact parameter bin), and show the results in the top row of Figure~\ref{fig:o6f}.
Indeed, such cuts do appear to enhance the median central absorption relative
to that in a random location. However,
as was the case for the higher order \hone\ line cuts, this could certainly
be due to the varying and small number of galaxies in the
smallest impact parameter bins (see the histograms in the bottom row of Figure~\ref{fig:o6f}).

\begin{figure}
\includegraphics[width=\appwid]{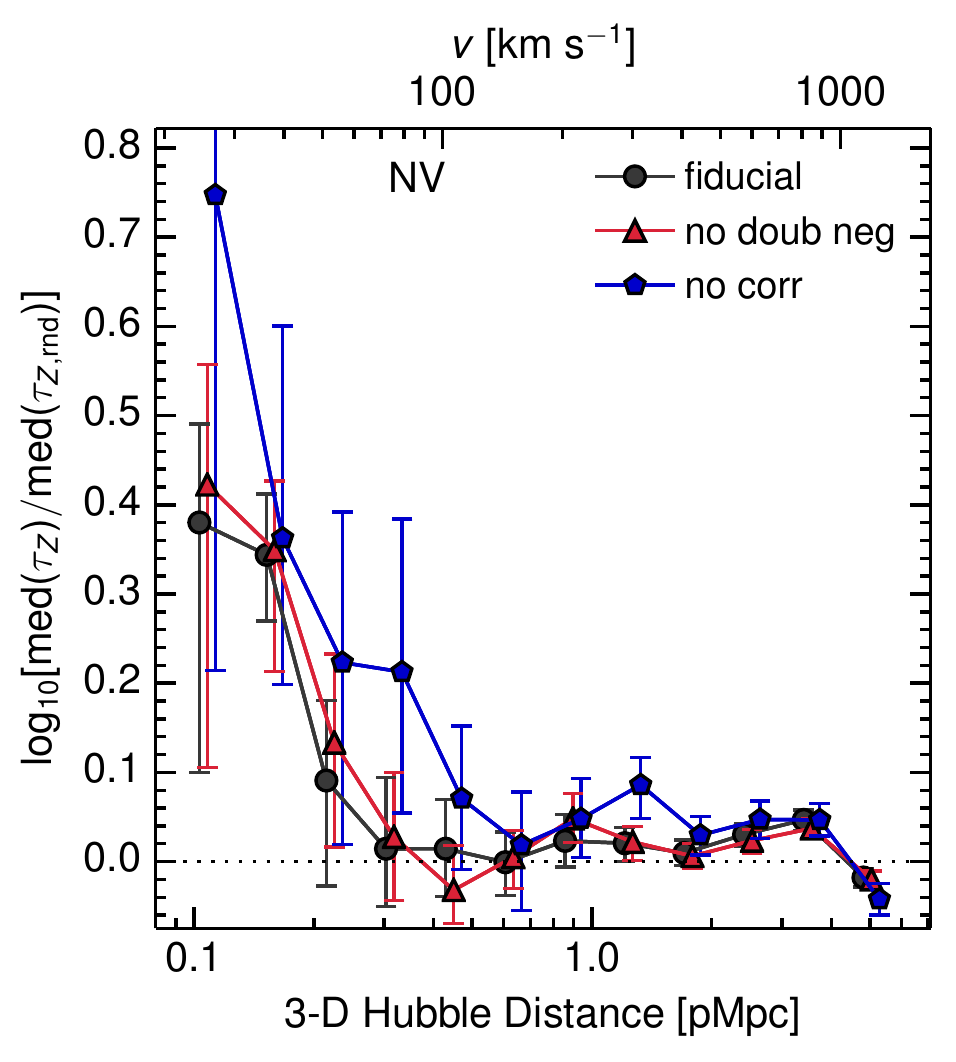}
\caption{Similar to Figure~\ref{fig:o6d}, 
but for \nfive\ and comparing the use of the fiducial optical
depth recovery method (taking the doublet minimum) to recoveries where
the weaker doublet component is not considered if it has a negative optical depth (``\textit{no doub neg}''),
and where no correction is done at all (``\textit{no corr}'').}
\label{fig:n5a}
\end{figure}

\subsection{\nfive}
\label{sec:n5_var_pod}

The optical depth correction for \nfive\ consists only of examining both doublet components at each redshift 
and taking the minimum of the two optical depths. In Figure~\ref{fig:n5a}, we compare the 
3-D-Hubble distance curves determined both with and without this correction, 
as well the curve resulting from using the doublet minimum condition from \citet{aguirre02}
(where instead of using Equation~\ref{eq:doubmin}, the doublet minimum is taken for 
all pixels except those where the weaker component has a negative optical depth). 
Neither of these changes have a significant impact on the resulting median optical depth. 
Not using the doublet correction yields a larger dynamic range; however, 
the errors become much larger. 
Hence, the significance with which the 
enhancement is detected is typically smaller without the correction.
We also note that since the points here are correlated,
the fact that the first five green points are above the black ones 
may not be a significant effect. Indeed, the differences between the
two recovery methods  is no longer seen when we use cuts along the transverse 
direction of the 2-D optical depth maps
(where the points are independent) instead of the 3-D Hubble distances.

\begin{figure*}
\includegraphics[width=\appwid]{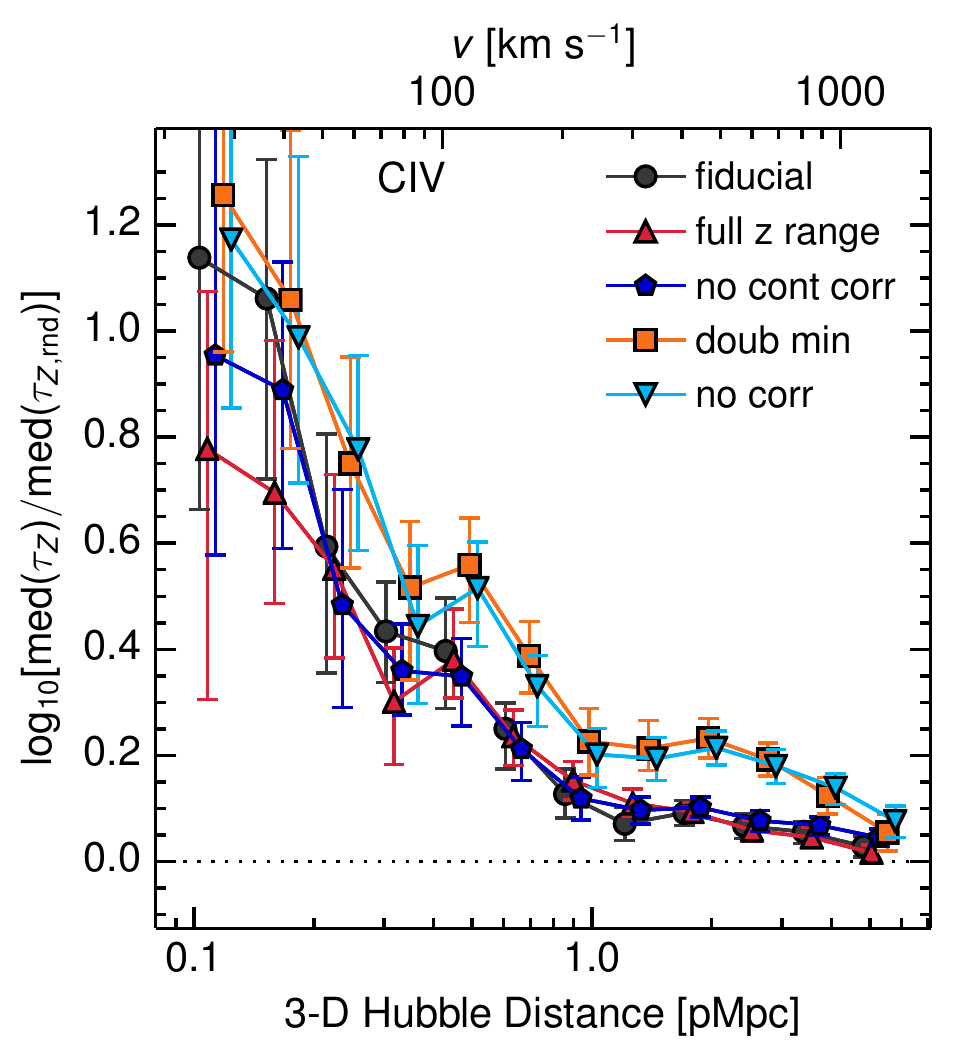}
\includegraphics[width=0.56\textwidth]{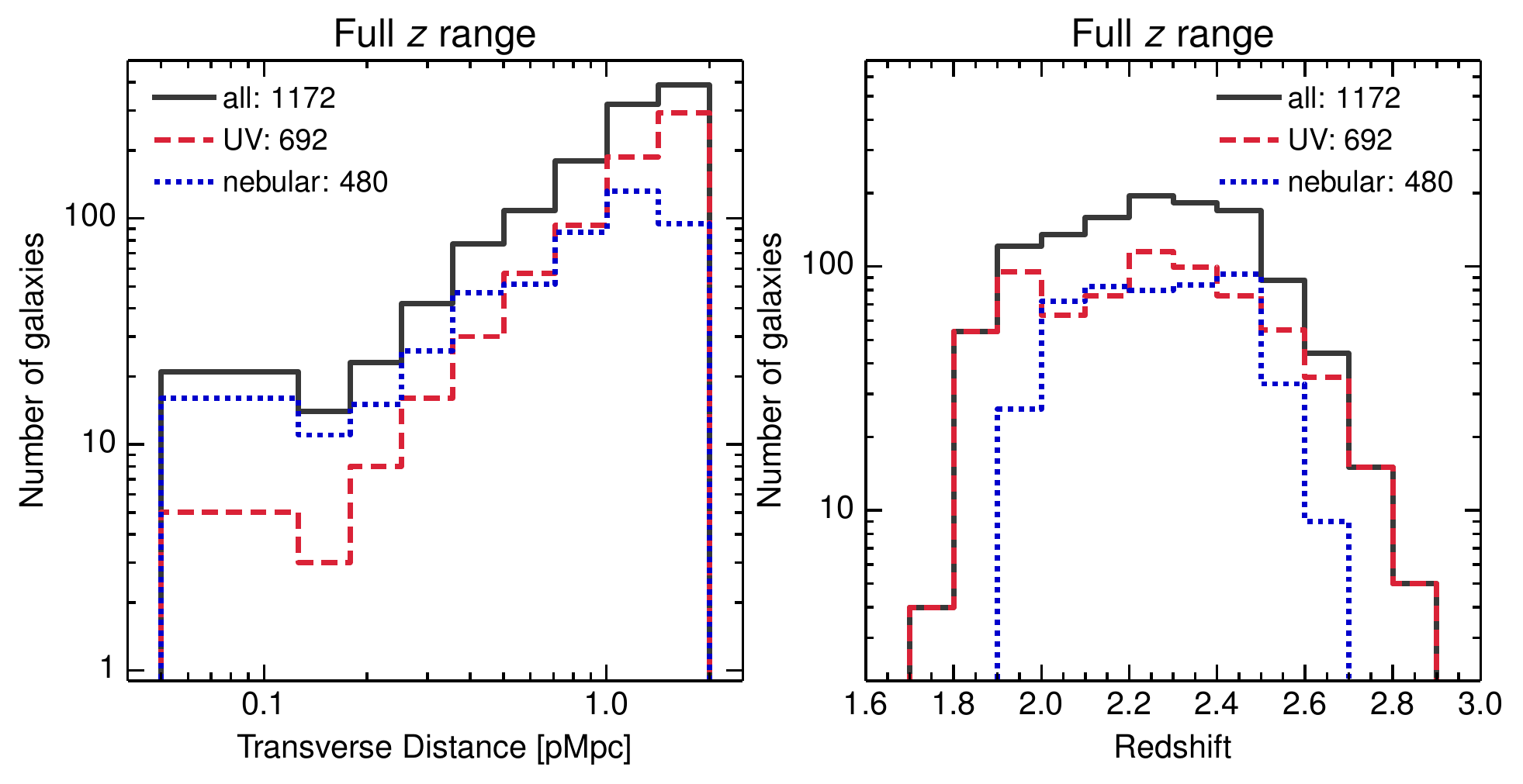}
\caption{\emph{Left}: Similar to Figure~\ref{fig:o6d}, 
but plotting different \cfour\ recoveries.
These are: self-contamination correction (``\textit{fiducial}'');  
using the full redshift range(``\textit{full z range}''); 
without the automatic continuum correction (``\textit{no cont corr}'');
taking the doublet minimum instead 
of the self-contamination correction (``\textit{doub min}'');
and without any correction (``\textit{no corr}''). 
\emph{Centre and right}: The histograms show the number of galaxies 
per impact parameter (centre) and redshift
bin (right) obtained when the full \cfour\ redshift range is used.
For reference, the fiducial number 
of galaxies in each bin is \ngaltotal\ galaxies in total, with \ngaluv\ and 
\ngalneb\ having their redshifts measured from rest-frame UV features and nebular emission
lines, respectively.  
}

\label{fig:c4a}
\end{figure*}

\subsection{\cfour}
\label{sec:c4_var_pod}	

As described in \S~\ref{sec:od_recovery}, we normally perform an automated continuum fit 
to any regions in the spectrum redwards of the \lya\ emission. However,
we find that this only slightly boosts the dynamic range in the recovered median 
optical depths, as can be seen in  Figure~\ref{fig:c4a}.

We have also examined the effect of 
not applying the self-contamination correction, but rather taking the doublet
minimum, as well as not doing any correction at all.
In both instances, the dynamic range probed is actually larger than 
in the self-contamination correction case. 
However, this increase in dynamic range is probably due to 
self-contamination. The \cfour\ doublet separation is 
$\sim500$~\kmps. As this is smaller than the scale over which the
absorption is enhanced, self-contamination will boost the enhancement 
in the apparent \cfour\ absorption near galaxies. 

Finally, our fiducial \cfour\ recovery only uses pixels
within the redshift range of the \lya\ forest of the quasar, 
i.e. down to the redshift for which \lya\ absorption 
coincides with the \lyb\ emission line. 
However, the region where \cfour\ can be recovered accurately (i.e., redwards
of the QSO \lya\ emission) extends to lower redshifts than this range,
and so we experiment with performing the recovery down to the redshift
for which \cfour\ absorption coincides with the quasar's 
\lya\ emission line. 

The result is shown as the red points Figure~\ref{fig:c4a}.
Extending the recovered redshift range increases the number 
of galaxies that fall within the spectral coverage range 
(see the galaxy histograms in the right two panels of Figure~\ref{fig:c4a}). 
However, the signal is slightly reduced, although the difference
is not significant. The galaxy sample
now contains, and the median redshift is now 2.27 instead of 
\medz\ for the fiducial sample. This could skew the
sample to lower masses, since at lower redshift these galaxies are more easily observed. 
Alternatively, the small decrease in the enhancement may reflect
small number statistics since it is not significant.

\begin{figure}
\includegraphics[width=\appwid]{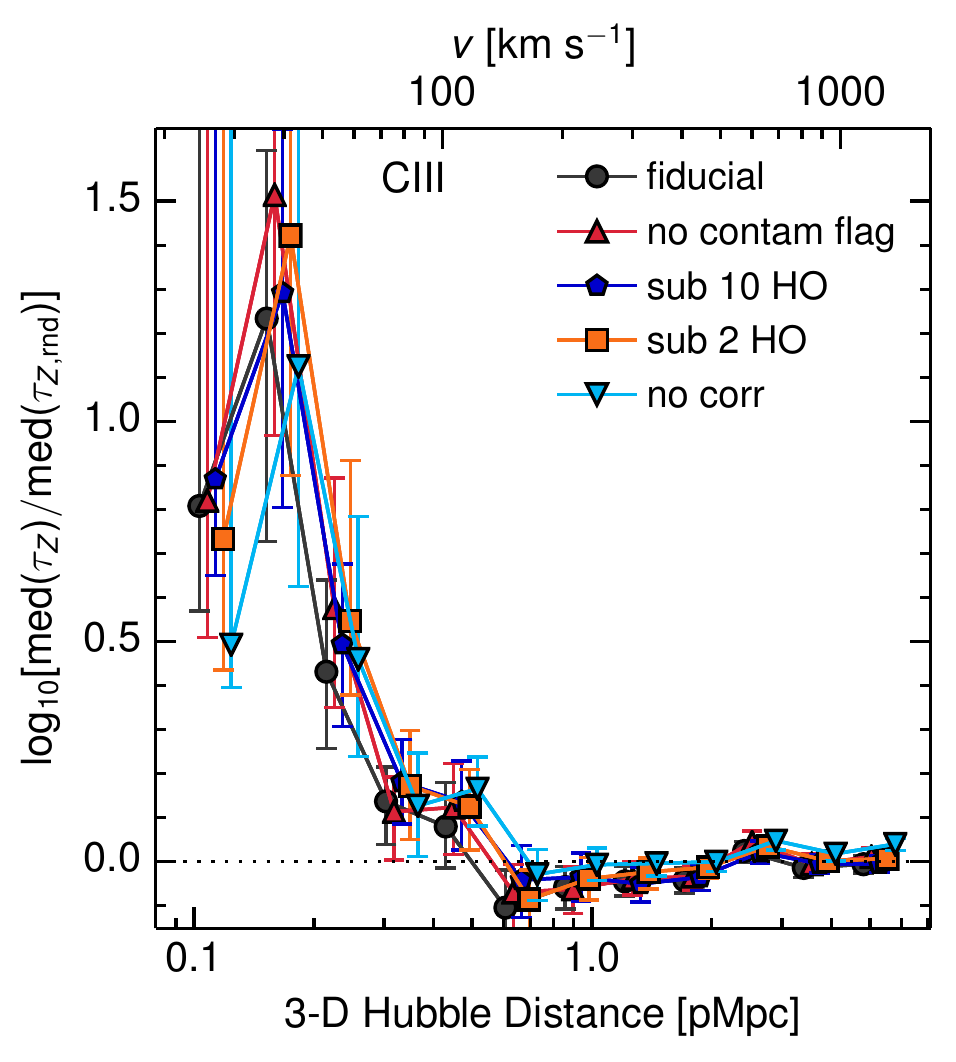}
\caption{Similar to Figure~\ref{fig:o6d}, 
but for \cthree\ and modifying the optical depth recovery method
with respect to the fiducial one (which invokes the subtraction of 
five higher order \hone\ lines). The modifications are: 
not flagging pixels determined to be contaminated from their
 higher-order \hone\ flux (``\textit{no contam flag}'');
and subtracting 10 (``\textit{sub 10 HO}''), 2 (``\textit{sub 2 HO}''), 
and 0 (``\textit{no corr}'') higher-order \hone\ lines. }
\label{fig:c3a}
\end{figure}

\subsection{\cthree}

Since the correction for \cthree\ involves the subtraction of higher-order contaminating \hone\ lines, 
we repeat the procedure done for \osix\ where we test the effect of removing different 
numbers of higher order \hone\ lines (the fiducial number being 5). In Figure~\ref{fig:c3a}
we show the result of removing 10, 5, 2, and no higher order lines (no correction). 
We also show the effect of not flagging pixels determined to be saturated due to \hone\ contamination. 
We find almost no change
except for the case of no correction, where the optical depth dynamic range is slightly reduced.

\begin{figure}
\includegraphics[width=\appwid]{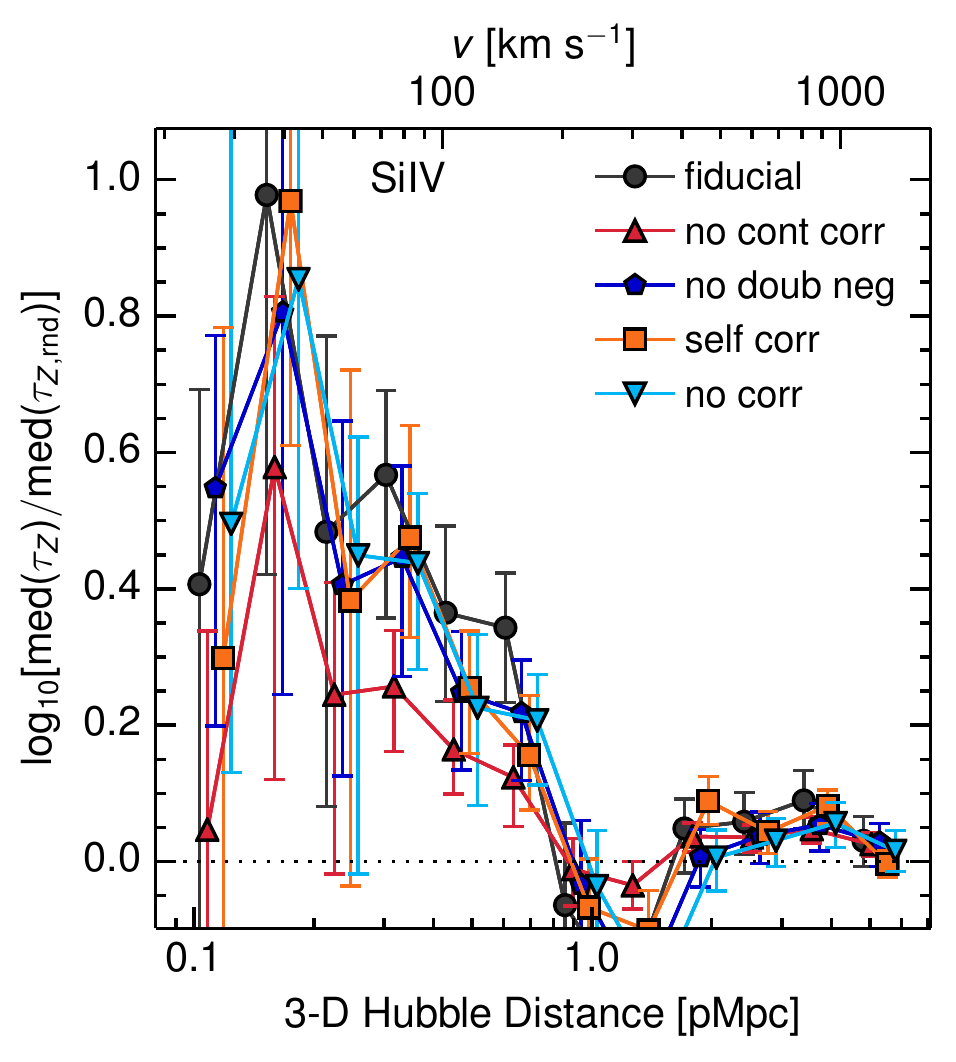}
\caption{Similar to Figure~\ref{fig:o6d}, 
but only for \sifour\ and for different optical depth recovery methods.
These are: taking the doublet minimum (``\textit{fiducial}'');  
without the automatic continuum correction (``\textit{no cont corr}''); 
the weaker doublet component is not considered if it has a negative optical depth (``\textit{no doub neg}'');
performing a self-contamination correction instead of taking the 
doublet minimum (``\textit{self corr}''); 
and without any correction (``\textit{no corr}''). }
 \label{fig:si4a}
\end{figure}

\subsection{\sifour}

In Figure~\ref{fig:si4a},  we find that for \sifour\ (unlike for
 \cfour) the automated correction to the continuum fit substantially 
increases the dynamic range probed. Since \sifour\ is a relatively
weak transition and its optical depths are very close to the noise
level, it is likely quite sensitive to the continuum fit.

Using the doublet minimum condition from \citet{aguirre02}
(where instead of using Equation~\ref{eq:doubmin}, the doublet minimum is taken for 
all pixels except those where the weaker component has a negative optical depth),
employing the self-contamination correction instead 
of the doublet minimum, or doing no correction at all 
does not significantly alter the dynamic range of 
the recovered optical depths, although we note that using the 
self-contamination correction results in larger errors. This may be due
to the fact that \sifour\ is a relatively weak line (compared to \cfour) and 
the greatest source of contamination likely comes from \cfour\ rather 
than from its own doublet.

\section{Covering fraction for EW thresholds}
\label{sec:fcover_ew} 

  \begin{figure*}
  \includegraphics[width=0.9\textwidth]{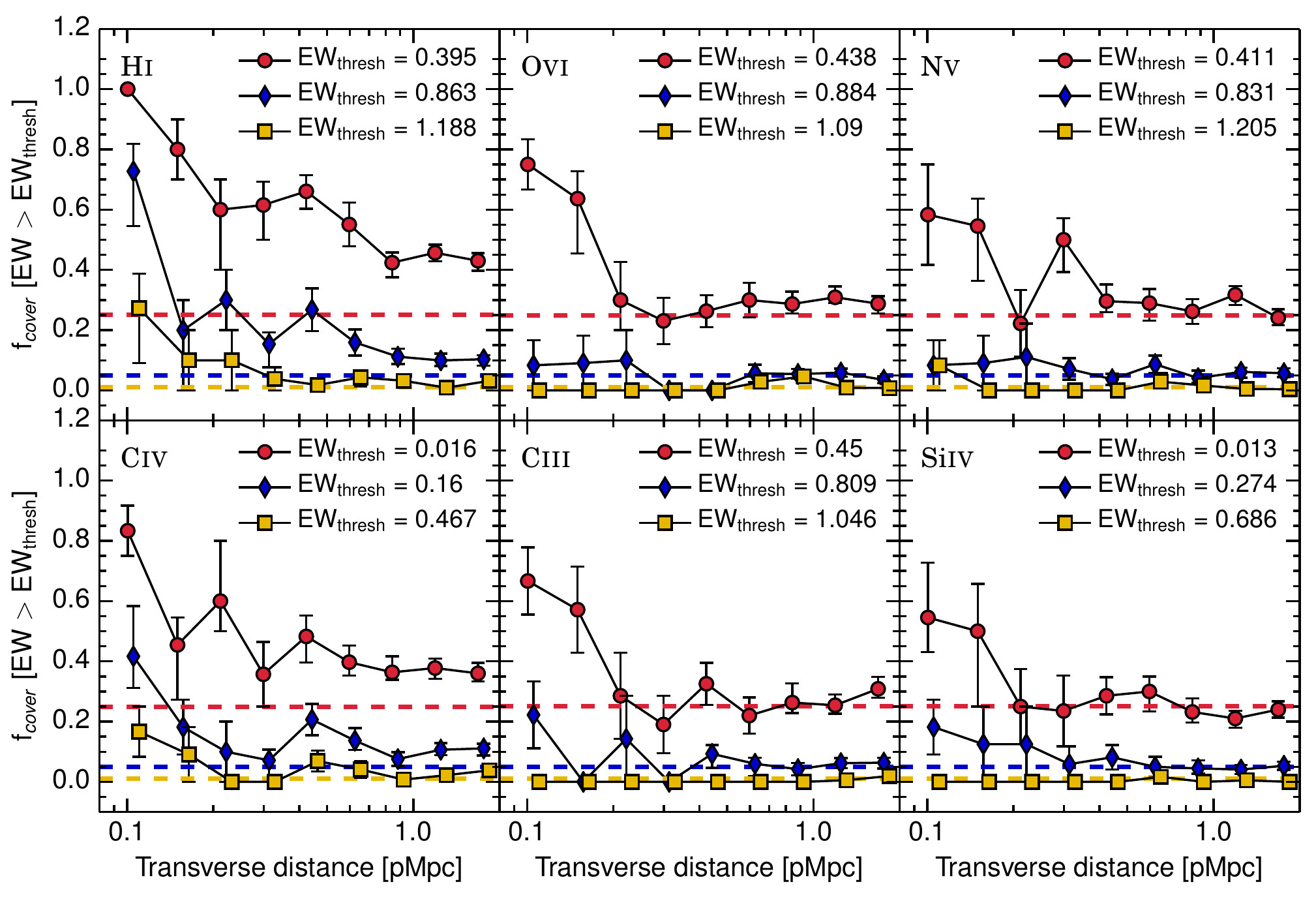}
  \caption{Covering fraction for each ion, defined as the fraction of galaxies within some 
  impact parameter bin (plotted along the y-axis) that have an EW, 
  calculated by integrating the flux decrement over $\pm170$~\kmps,
  above some threshold value (in \AA). These threshold values are set by some multiple 
  of the median EW for 1000 randomly drawn 
  regions integrated over $\pm170$~\kmps. These regions are then used to
  determine the covering fraction of the \ewthresh\ values, 
  which are denoted by the dotted horizontal lines. 
  Points determined using different \tauthresh\ values have been offset
 horizontally by 0.2~dex for clarity.}
  \label{fig:fcoverew}
  \end{figure*}

\begin{table*}
\caption{Covering fraction and 1-$\sigma$ errors as a function of transverse distance (top row),
 which is defined as the fraction of galaxies in each impact parameter bin that
  have an EW within $\pm170$~\kmps\ above some threshold value. 
  The threshold values are set by the EWs at which the covering fraction 
  of 1000 random $\pm170$~\kmps\, regions, $f_{\rm IGM}$,  are equal to 
   0.25, 0.05, and 0.01 (second column),
    and the results are plotted in Figure~\ref{fig:fcoverew}.}
\label{tab:fcoverew}
\begin{tabular}{@{}lcccccccccc}
\hline\hline
\multicolumn{2}{r}{$D_{\rm trans}$ (pMpc)} & 	0.04--0.13 & 	0.13--0.18 & 	0.18--0.25 & 	0.25--0.36 & 	0.36--0.50 & 	0.50--0.71 & 	0.71--1.00 & 	1.00--1.42 & 	1.42--2.00\\
Ion & $f_{\rm IGM}$ & & & & & & & & & \\\hline
  & 0.25	& $1.00_{-0.00}^{+0.00}$	& $0.80_{-0.10}^{+0.10}$	& $0.60_{-0.20}^{+0.10}$	& $0.62_{-0.12}^{+0.08}$	& $0.66_{-0.06}^{+0.05}$	& $0.55_{-0.07}^{+0.07}$	& $0.42_{-0.05}^{+0.03}$	& $0.46_{-0.03}^{+0.03}$	& $0.43_{-0.03}^{+0.02}$ \\ 
\hone & 0.05	& $0.73_{-0.18}^{+0.09}$	& $0.20_{-0.20}^{+0.10}$	& $0.30_{-0.10}^{+0.10}$	& $0.15_{-0.08}^{+0.04}$	& $0.27_{-0.07}^{+0.07}$	& $0.16_{-0.04}^{+0.04}$	& $0.11_{-0.02}^{+0.03}$	& $0.10_{-0.02}^{+0.02}$	& $0.10_{-0.02}^{+0.01}$ \\ 
  & 0.01	& $0.27_{-0.18}^{+0.11}$	& $0.10_{-0.10}^{+0.10}$	& $0.10_{-0.10}^{+0.10}$	& $0.04_{-0.04}^{+0.04}$	& $0.02_{-0.02}^{+0.02}$	& $0.04_{-0.03}^{+0.01}$	& $0.03_{-0.02}^{+0.02}$	& $0.01_{-0.00}^{+0.01}$	& $0.03_{-0.01}^{+0.01}$ \\ 
\hline 
  & 0.25	& $0.75_{-0.08}^{+0.08}$	& $0.64_{-0.18}^{+0.09}$	& $0.30_{-0.10}^{+0.13}$	& $0.23_{-0.08}^{+0.08}$	& $0.26_{-0.05}^{+0.05}$	& $0.30_{-0.06}^{+0.06}$	& $0.29_{-0.03}^{+0.04}$	& $0.31_{-0.02}^{+0.04}$	& $0.29_{-0.03}^{+0.02}$ \\ 
\osix & 0.05	& $0.08_{-0.08}^{+0.08}$	& $0.09_{-0.09}^{+0.09}$	& $0.10_{-0.10}^{+0.10}$	& $0.00_{-0.00}^{+0.00}$	& $0.00_{-0.00}^{+0.00}$	& $0.06_{-0.03}^{+0.03}$	& $0.05_{-0.02}^{+0.02}$	& $0.06_{-0.01}^{+0.01}$	& $0.04_{-0.01}^{+0.01}$ \\ 
  & 0.01	& $0.00_{-0.00}^{+0.00}$	& $0.00_{-0.00}^{+0.00}$	& $0.00_{-0.00}^{+0.00}$	& $0.00_{-0.00}^{+0.00}$	& $0.00_{-0.00}^{+0.00}$	& $0.03_{-0.01}^{+0.01}$	& $0.05_{-0.02}^{+0.02}$	& $0.01_{-0.00}^{+0.00}$	& $0.01_{-0.00}^{+0.00}$ \\ 
\hline 
  & 0.25	& $0.58_{-0.17}^{+0.17}$	& $0.55_{-0.18}^{+0.09}$	& $0.22_{-0.11}^{+0.11}$	& $0.50_{-0.11}^{+0.07}$	& $0.30_{-0.04}^{+0.06}$	& $0.29_{-0.06}^{+0.05}$	& $0.26_{-0.04}^{+0.04}$	& $0.32_{-0.03}^{+0.03}$	& $0.24_{-0.02}^{+0.03}$ \\ 
\nfive & 0.05	& $0.08_{-0.08}^{+0.08}$	& $0.09_{-0.09}^{+0.09}$	& $0.11_{-0.11}^{+0.11}$	& $0.07_{-0.04}^{+0.04}$	& $0.04_{-0.02}^{+0.02}$	& $0.09_{-0.03}^{+0.03}$	& $0.04_{-0.02}^{+0.02}$	& $0.06_{-0.02}^{+0.01}$	& $0.06_{-0.01}^{+0.02}$ \\ 
  & 0.01	& $0.08_{-0.08}^{+0.08}$	& $0.00_{-0.00}^{+0.00}$	& $0.00_{-0.00}^{+0.00}$	& $0.00_{-0.00}^{+0.00}$	& $0.00_{-0.00}^{+0.00}$	& $0.03_{-0.01}^{+0.03}$	& $0.02_{-0.02}^{+0.01}$	& $0.00_{-0.00}^{+0.00}$	& $0.00_{-0.00}^{+0.00}$ \\ 
\hline 
  & 0.25	& $0.83_{-0.08}^{+0.08}$	& $0.45_{-0.18}^{+0.09}$	& $0.60_{-0.10}^{+0.20}$	& $0.36_{-0.11}^{+0.11}$	& $0.48_{-0.09}^{+0.07}$	& $0.40_{-0.04}^{+0.05}$	& $0.36_{-0.02}^{+0.05}$	& $0.38_{-0.04}^{+0.03}$	& $0.36_{-0.03}^{+0.03}$ \\ 
\cfour & 0.05	& $0.42_{-0.10}^{+0.17}$	& $0.18_{-0.09}^{+0.09}$	& $0.10_{-0.10}^{+0.10}$	& $0.07_{-0.07}^{+0.04}$	& $0.21_{-0.05}^{+0.05}$	& $0.14_{-0.03}^{+0.04}$	& $0.08_{-0.02}^{+0.02}$	& $0.11_{-0.02}^{+0.02}$	& $0.11_{-0.02}^{+0.02}$ \\ 
  & 0.01	& $0.17_{-0.08}^{+0.08}$	& $0.09_{-0.09}^{+0.09}$	& $0.00_{-0.00}^{+0.00}$	& $0.00_{-0.00}^{+0.00}$	& $0.07_{-0.03}^{+0.03}$	& $0.04_{-0.03}^{+0.03}$	& $0.01_{-0.01}^{+0.01}$	& $0.02_{-0.01}^{+0.01}$	& $0.04_{-0.01}^{+0.01}$ \\ 
\hline 
  & 0.25	& $0.67_{-0.11}^{+0.11}$	& $0.57_{-0.14}^{+0.14}$	& $0.29_{-0.14}^{+0.14}$	& $0.19_{-0.10}^{+0.10}$	& $0.33_{-0.07}^{+0.07}$	& $0.22_{-0.06}^{+0.06}$	& $0.26_{-0.03}^{+0.06}$	& $0.25_{-0.03}^{+0.04}$	& $0.31_{-0.03}^{+0.04}$ \\ 
\cthree & 0.05	& $0.22_{-0.11}^{+0.11}$	& $0.00_{-0.00}^{+0.00}$	& $0.14_{-0.14}^{+0.14}$	& $0.00_{-0.00}^{+0.00}$	& $0.09_{-0.05}^{+0.03}$	& $0.06_{-0.04}^{+0.02}$	& $0.04_{-0.02}^{+0.02}$	& $0.06_{-0.02}^{+0.02}$	& $0.06_{-0.02}^{+0.02}$ \\ 
  & 0.01	& $0.00_{-0.00}^{+0.00}$	& $0.00_{-0.00}^{+0.00}$	& $0.00_{-0.00}^{+0.00}$	& $0.00_{-0.00}^{+0.00}$	& $0.00_{-0.00}^{+0.00}$	& $0.00_{-0.00}^{+0.00}$	& $0.00_{-0.00}^{+0.00}$	& $0.01_{-0.01}^{+0.01}$	& $0.02_{-0.01}^{+0.01}$ \\ 
\hline 
  & 0.25	& $0.55_{-0.11}^{+0.18}$	& $0.50_{-0.25}^{+0.16}$	& $0.25_{-0.12}^{+0.12}$	& $0.24_{-0.12}^{+0.12}$	& $0.29_{-0.06}^{+0.06}$	& $0.30_{-0.07}^{+0.05}$	& $0.23_{-0.04}^{+0.04}$	& $0.21_{-0.03}^{+0.03}$	& $0.24_{-0.03}^{+0.03}$ \\ 
\sifour & 0.05	& $0.18_{-0.09}^{+0.09}$	& $0.12_{-0.12}^{+0.12}$	& $0.12_{-0.12}^{+0.12}$	& $0.06_{-0.06}^{+0.06}$	& $0.08_{-0.04}^{+0.04}$	& $0.05_{-0.03}^{+0.03}$	& $0.04_{-0.02}^{+0.03}$	& $0.04_{-0.02}^{+0.02}$	& $0.05_{-0.01}^{+0.01}$ \\ 
  & 0.01	& $0.00_{-0.00}^{+0.00}$	& $0.00_{-0.00}^{+0.00}$	& $0.00_{-0.00}^{+0.00}$	& $0.00_{-0.00}^{+0.00}$	& $0.00_{-0.00}^{+0.00}$	& $0.02_{-0.02}^{+0.02}$	& $0.00_{-0.00}^{+0.00}$	& $0.01_{-0.01}^{+0.01}$	& $0.00_{-0.00}^{+0.00}$ \\ 
\hline 
\hline
\end{tabular}
  \end{table*}

Here we investigate the outcome of using an alternate threshold for the
 covering fraction, where we use the EW (rather than the median optical 
depth) within $\pm170$~\kmps\ of every galaxy, to facilitate comparison 
with low-quality data. 
 The EW within $\pm170$~\kmps  is also computed
 for 1000 random regions within the spectra. We define the covering fraction 
 as the fraction of galaxies within an impact parameter bin with an EW above
 \ewthresh, where we take \ewthresh\ values as those where the covering
 fractions for random regions are 0.25, 0.05 and 0.01. The results are shown in 
Figure~\ref{fig:fcoverew} and Table~\ref{tab:fcoverew}. 
 
 In general, the results are similar to those obtained using our first
 covering fraction definition: for \hone, the covering fraction is 
 above that for random regions out to the largest
 impact parameter in our sample (2~pMpc), while for metals
 the covering fraction is only elevated for the smallest transverse
 distance bins. One difference to note is that for the EW thresholds, 
 a signal is seen for the covering fraction of \nfive\ (which was
 not the case using median optical depth thresholds).

\section{Galaxy redshift measurements}
\label{sec:galsample_appendix} 

  \begin{figure*}
  \includegraphics[width=\textwidth]{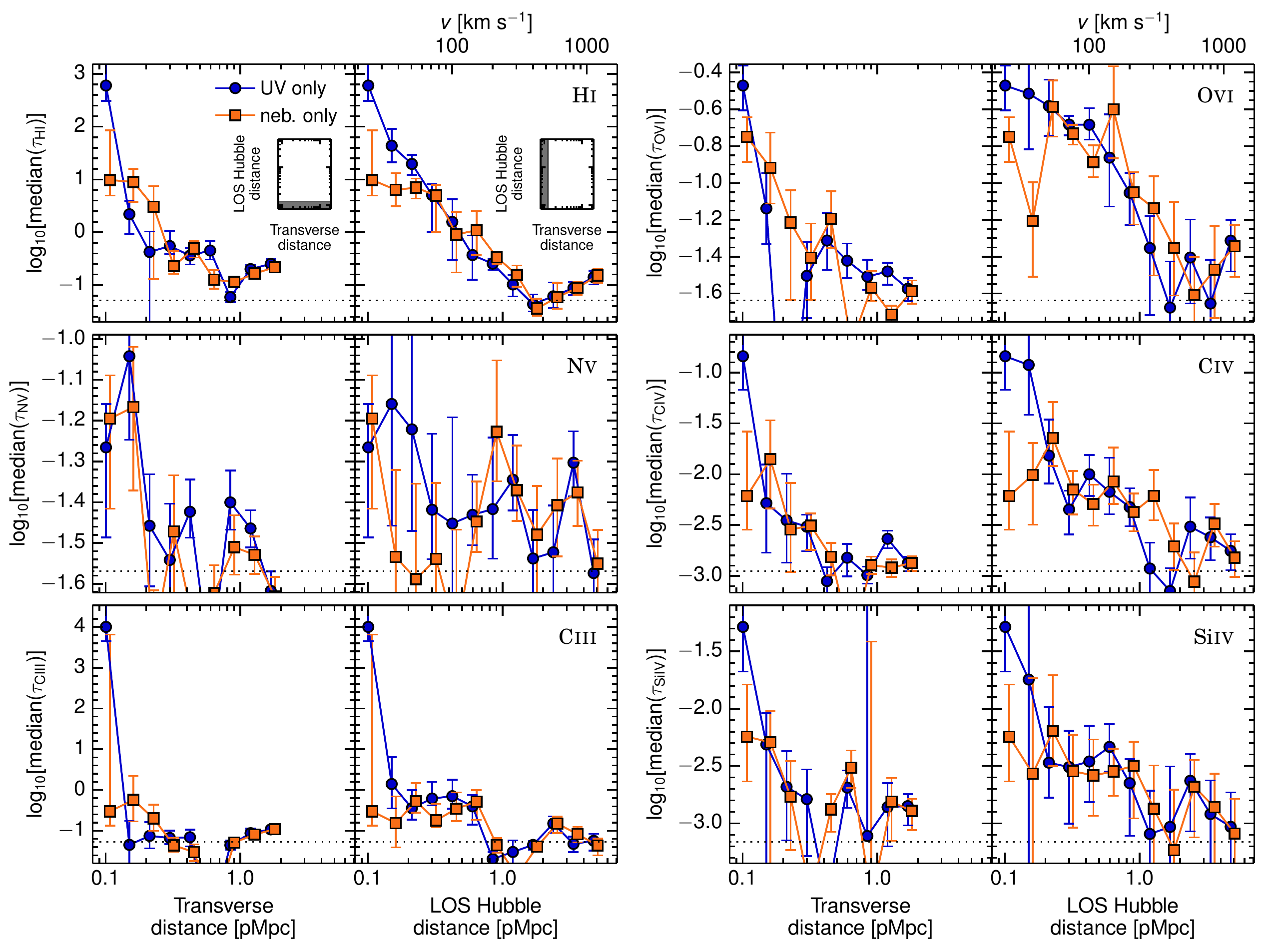}
  \caption{As Figure~\ref{fig:maps_onlynebular}, but for the same 
   \ngalbothnebanduv\ galaxies, using either their redshifts measured from
    rest-frame UV features (blue circles) or from nebular emission lines (orange squares). 
  Except for the first bin of \cfour, there are no significant differences between 
the results based on nebular and rest-frame UV redshifts.
 }
  \label{fig:maps_onlynebular_appendix}
  \end{figure*}

\begin{figure*}
\includegraphics[width=0.7\textwidth]{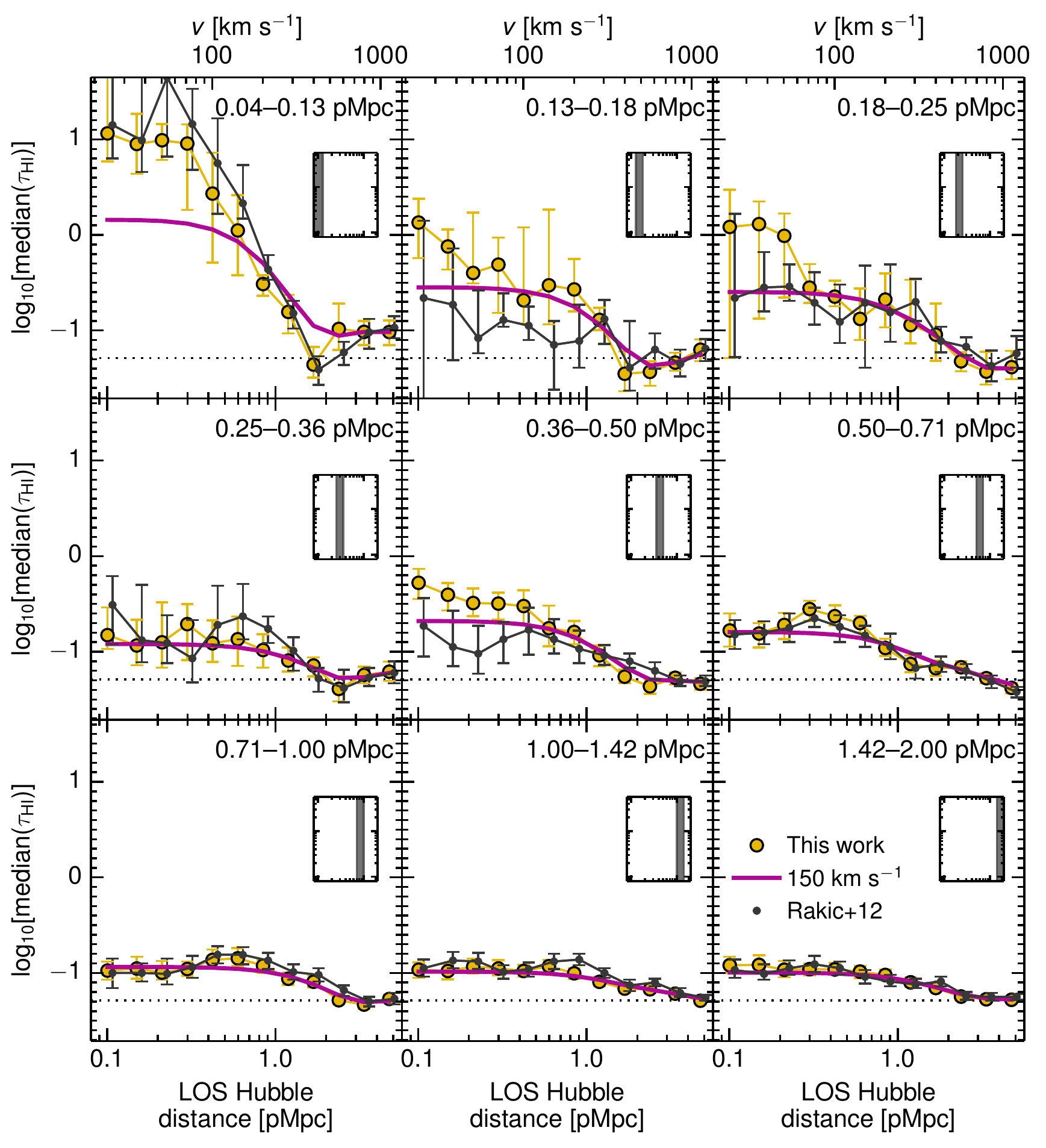}
\caption{Cuts along the LOS through the unsmoothed \hone\ map from Figure~\ref{fig:all_maps_b}
(using the values from Table~\ref{tab:cuts}), 
where each panel shows a different transverse distance bin as denoted by the label and shown
by the inset. The yellow circles show the data from this work, and the magenta
lines give the values after smoothing with a $\sigma=150$~\kmps\ Gaussian,
to mimic the effect of larger redshift errors.
For comparison, we display the points from Figure~6 and Table~2 of \citet{rakic12},
whose galaxy sample only had 10\% nebular redshifts. Although we find no difference 
between the two data sets in the first panel (0.04--0.13~pMpc), 
the second and third panels (0.13--0.18 and
0.18--0.25~pMpc) show that the results from \citet{rakic12} are consistent with 
the current data after convolution with 150~\kmps\ redshift errors.}
\label{fig:oljaA}
\end{figure*}

Figure~\ref{fig:maps_onlynebular} investigated 
if there are any differences in the absorption profiles using
the full galaxy sample (which contains galaxy redshifts measured 
from a mix of rest-frame UV features and nebular emission lines)
to a sample of galaxies with redshifts measured only using nebular
emission lines. However, the comparison was complicated by the 
fact that these two samples contain different galaxies (and also 
that the latter group has less than half as many galaxies as the former).

To remove galaxy sample effects, we now use only the  \ngalbothnebanduv\ galaxies
that have redshifts measured using both techniques, and directly compare
the results in Figure~\ref{fig:maps_onlynebular_appendix}. 
The slight enhancement at large LOS distances in the optical depth of
the nebular only sample that was visible in Figure~\ref{fig:maps_onlynebular} for 
\osix\ and \cfour\ is less apparent here. 
Overall there is only one significant difference between
the two samples, for 
the smallest impact parameter / LOS bin, the nebular-only optical depths
are consistently lower than those measured from rest-frame UV features. 
This may suggest that the peak in the 
absorption is systematically offest from the galaxy's systemic
redshift. 

Furthermore, our galaxy sample has changed significantly from that of \citet{rakic12}. 
Particularly in the innermost impact parameter bins, many more galaxies now have
redshifts measured using MOSFIRE. This presents a problem: since the redshift errors in \citet{rakic12}
were so large, then would we not expect to 
see a difference in the LOS extent of \hone\ optical depths between the two samples? 
Specifically, the results from \citet{rakic12} should essentially show the effect of 
being smoothed on 150~\kmps\ scales with respect to the current data. 

To explore this, we show our
median \hone\ optical depth results
(yellow circles) alongside those from \citet[][grey points]{rakic12}
in Figure~\ref{fig:oljaA}. Specifically, we plot the unsmoothed data taken from cuts
along the LOS through the \hone\ map in our Figure~\ref{fig:all_maps_b} and
Figure~5 of \citet{rakic12}. We also show how the current data look when smoothed
with a $\sigma=150$~\kmps\ Gaussian as the magenta line. Naively, we would expect
the points from \citet{rakic12} to be consistent with this curve, and indeed, except for the 
smallest transverse cut (upper left panel), they are in relatively good agreement. 

As for the innermost transverse distance bin, it is still 
curious that \hone\ optical depths
from this work and \citet{rakic12} are in near perfect agreement.
A possible explanation comes from the fact that galaxies with low
impact parameters
were preferentially targeted with NIRSPEC -- already in \citet{rakic12},
25\% of galaxies with impact parameters less than 250~kpc had nebular
redshifts (compared to 9\% for the remainder of the sample), reducing
the amount of change we would expect to see between this study and the
previous one.

It is worth mentioning that the latest data resolve some tensions with theory. 
Figure~7 of \citet{rakic13} demonstrates that the observed LOS optical depth values 
tend to be lower then those seen in simulations for impact parameters
0.013-0.25~pMpc, which is no longer 
the case with the updated KBSS galaxy sample.

\label{lastpage}
\end{document}